\documentclass[12pt,english]{article}
\usepackage{a4}
\usepackage{color}
\usepackage{times}
\usepackage{amssymb}
\usepackage{amsmath}
\usepackage{stmaryrd}
\usepackage{amscd}
\usepackage{latexsym}
\usepackage{bbm}
\usepackage{epsfig}
\usepackage{graphicx}
\usepackage{amsthm}
\usepackage{bm}
\usepackage{amsfonts}
\usepackage{dcolumn}
\usepackage{babel}
\def\un#1{\relax\ifmmode\@@underline#1\else
        $\@@underline{\hbox{#1}}$\relax\fi}

\let\du=\du                     % dot-under
\def\a{\alpha}
\def\b{\beta}
\def\c{\chi}
\def\d{\delta}

\def\f{\phi}
\def\g{\gamma}
\def\h{\eta}

\def\j{\psi}
\def\k{\kappa}
\def\l{\lambda}
\def\m{\mu}
\def\n{\nu}
\def\o{\omega}
\def\p{\pi}
\def\q{\theta}
\def\r{\rho}
\def\s{\sigma}

\def\x{\xi}

\def\D{\Delta}
\def\F{\Phi}
\def\G{\Gamma}
\def\J{\Psi}
\def\L{\Lambda}
\def\O{\Omega}
\def\P{\Pi}

\def\S{\Sigma}

\def\ve{\varepsilon}
\def\vf{\varphi}

\def\vq{\vartheta}

\def\ce{{\cal E}}

\def\cg{{\cal G}}

\def\car{{\cal R}}

\def\cv{{\cal V}}
\def\cw{{\cal W}}

\def\cy{{\cal Y}}
\def\cz{{\cal Z}}

\def\slpa{\slash{\pa}}                            % slashed partial derivative
\def\bo{{\raise-.3ex\hbox{\large$\Box$}}}               % D'Alembertian
\def\pa{\partial}                                       % curly d
\def\de{\nabla}                                         % del
\def\TH{{\raise.2ex\hbox{$\displaystyle \bigodot$}\mskip-4.7mu \llap H \;}}
\def\face{{\raise.2ex\hbox{$\displaystyle \bigodot$}\mskip-2.2mu \llap {$\ddot
        \smile$}}}                                      % happy face
\def\dg{\sp\dagger}                                     % hermitian conjugate
\def\sp#1{{}^{#1}}                              % superscript (unaligned)
\def\slash#1{\rlap{\hbox{$\mskip 1 mu /$}}#1}      % good slash for lower case
\def\Hat#1{\widehat{#1}}                        % big hat
\def\Bar#1{\overline{#1}}                       % big bar
\def\VEV#1{\left\langle #1\right\rangle}        % < >
\def\abs#1{\left| #1\right|}                    % | |
\def\leftrightarrowfill{$\mathsurround=0pt \mathord\leftarrow \mkern-6mu
        \cleaders\hbox{$\mkern-2mu \mathord- \mkern-2mu$}\hfill
        \mkern-6mu \mathord\rightarrow$}
\def\dvec#1{\vbox{\ialign{##\crcr
        \leftrightarrowfill\crcr\noalign{\kern-1pt\nointerlineskip}
        $\hfil\displaystyle{#1}\hfil$\crcr}}}           % <--> accent
\def\dt#1{{\buildrel {\hbox{\LARGE .}} \over {#1}}}     % dot-over for sp/sb
\def\frac#1#2{{\textstyle{#1\over\vphantom2\smash{\raise.20ex
        \hbox{$\scriptstyle{#2}$}}}}}                   % fraction
\def\sfrac#1#2{{\vphantom1\smash{\lower.5ex\hbox{\small$#1$}}\over
        \vphantom1\smash{\raise.4ex\hbox{\small$#2$}}}} % alternate fraction
\def\bfrac#1#2{{\vphantom1\smash{\lower.5ex\hbox{$#1$}}\over
        \vphantom1\smash{\raise.3ex\hbox{$#2$}}}}       % "
\def\afrac#1#2{{\vphantom1\smash{\lower.5ex\hbox{$#1$}}\over#2}}    % "
\def\[{\lfloor{\hskip 0.35pt}\!\!\!\lceil}
\def\]{\rfloor{\hskip 0.35pt}\!\!\!\rceil}
\def\Lag{{\cal L}}
\def\du#1#2{_{#1}{}^{#2}}

\def\fracm#1#2{\hbox{\large{${\frac{{#1}}{{#2}}}$}}}
\def\ha{{\fracmm12}}

\def\un{\underline}
\def\fracmm#1#2{{{#1}\over{#2}}}

\def\low#1{{\raise -3pt\hbox{${\hskip 0.75pt}\!_{#1}$}}}

\def\dt#1{\Dot{#1}}

\def\DDot#1{\buildrel{_{_{\hskip 0.01in}\bullet\bullet}}\over{#1}}
\def\ddt#1{\DDot{#1}}

\def\Hat#1{\widehat{#1}}

\newskip\humongous \humongous=0pt plus 1000pt minus 1000pt
\def\caja{\mathsurround=0pt}
\def\eqalign#1{\,\vcenter{\openup2\jot \caja
        \ialign{\strut \hfil$\displaystyle{##}$&$
        \displaystyle{{}##}$\hfil\crcr#1\crcr}}\,}
\newif\ifdtup

\newcommand{\be}{\begin{equation}}
\newcommand{\ee}{\end{equation}}
\newcommand{\nbe}{\begin{equation*}}
\newcommand{\nee}{\end{equation*}}

\newcommand{\lb}{\label}

\newcommand{\R}{\mathbb R}

\numberwithin{equation}{section}

\begin{document}

\thispagestyle{empty}

\noindent
\vskip1.0cm
\begin{center}

{\large\bf Modified Supergravity and Early Universe: the Meeting Point 
                      of Cosmology and High-Energy Physics}
\vglue.1in

Sergei V. Ketov~${}^{a,b}$  
\vglue.1in
${}^a$~Department of Physics, Tokyo Metropolitan University, 
Minami-ohsawa 1-1 \\ Hachioji-shi, Tokyo 192-0397, Japan \\
${}^b$~Kavli Institute for the Physics and Mathematics of the Universe (IPMU), The 
University of Tokyo, Kashiwanoha 5-1-5, Kashiwa-shi, Chiba 277-8568, Japan \\
\vglue.1in
ketov@phys.se.tmu.ac.jp
\end{center}

\begin{center}
{\Large\bf Abstract}
\end{center}

\noindent We review the new theory of modified supergravity, dubbed the $F({\cal R})$
supergravity, and some of its recent applications to inflation and reheating in the early 
universe cosmology. The $F({\cal R})$ supergravity is the $N=1$ locally supersymmetric 
extension of the $f(R)$ gravity in four space-time dimensions.  A manifestly supersymmetric 
formulation of the $F({\cal R})$ supergravity exist in terms of $N=1$  superfields, by using the 
(old) minimal Poincar\'e supergravity in curved superspace.  We find the conditions for stability, 
the absence of ghosts  and tachyons. Three models of the $F({\cal R})$ supergravity are studied.
The first example is devoted to a recovery of the standard (pure) $N=1$ supergravity
 with a negative cosmological constant from the $F(\car)$ supergravity. As the second 
example, a generic $\car^2$ supergravity is investigated, and the existence of the AdS 
bound on the scalar curvature is found. As the third (and most important) example, 
a simple viable realization of chaotic inflation in supergravity is found. Our approach is {\it minimalistic} 
since it does not introduce new exotic fields or new interactions, beyond those already present in 
(super)gravity. The universal reheating mechanism is automatic. We establish the consistency of our approach 
and also apply it to preheating and  reheating after inflation. The Higgs inflation and its correspondence to the 
Starobinsky inflation are established in the context of supergravity. We briefly review other relevant issues such as 
non-Gaussianity, $CP$-violation, origin of baryonic asymmetry, lepto- and baryo-genesis. The $F(\car)$ supergravity 
has promise for possible solutions to those outstanding problems too.
\vglue.4in



\newpage

\begin{center}
{\large\bf CONTENTS}

\vglue.3in

\end{center}
\begin{enumerate}
\item Introduction and Motivation \hfill 3 
\item Starobinsky Approach to Inflation \hfill 5  
\item $f(R)$ Gravity  \hfill 6
\item Inflationary Theory and Observations \hfill 8 
\item Supergravity and Superspace \hfill 11 
\item $F(\car)$ Supergravity in Superspace \hfill 13
\item No-scale $F(\car)$ Supergravity \hfill 17
\item Fields from Superfields in $F(\car)$ Supergravity \hfill 18
\item Generic $\car^2$ Supergravity, and AdS Bound \hfill 20
\item Inflationary Model in $F(\car)$ Supergravity \hfill 25
\item More about Inflationary Dynamics in our Model \hfill 29 
\item Facing Observational Tests \hfill 32
\item Effective Scalar Potential for Preheating \hfill 36
\item Preheating after Inflation \hfill 39
\item Current Status of our Model \hfill 44
\item Cosmological Constant in $F(\car)$ Supergravity \hfill 45
\item Nonmimimal Scalar-Curvature Coupling in Gravity 
and Supergravity, \\ and Higgs Inflation \hfill 49
\item Quantum Particle Production (Reheating) \hfill 56
\item Conclusion \hfill 59
\item Outlook: $CP$-violation, Baryonic Asymmetry,
Lepto- and Baryo-genesis,\\ Non-Gaussianity, and Experimental Tests  \hfill 61
\end{enumerate}

Acknowledgements  \hfill 63

Appendix: Scalar Potential in Generic $F(\car)$ Supergravity \hfill 64

References \hfill 65

\newpage

\section{Introduction and Motivation} 

A brief history of our universe in pictures is nicely summarized in the NASA website
of the Wilkinson Microwave Anisotropy Probe (WMAP) satellite mission \cite{nasa}. Very
recently (March 2013) more data about the observational constraints on inflation has 
become available from the PLANCK satellite mission \cite{planck}. In this review paper
we focus on a field-theoretical description of the inflationary phase of early 
universe and its post-inflationary dynamics (pre-heating and re-heating) in the
context of modified supergravity proposed and studied in 
Refs.~\cite{us1,us2,us3,us4,us5,us6,us7,us8,us9,us10,us11,us12,us13,us14}.

Cosmological inflation (a phase of `rapid' quasi-exponential accelerated 
expansion of universe) \cite{star1,star2,a1,stein,lindef} predicts homogeneity of our 
universe at large scales, its spatial flatness, large size and entropy, and the almost 
scale-invariant spectrum of cosmological perturbations, in good agreement 
with the ongoing WMAP and PLANCK measurements of the CMB radiation spectrum 
\cite{wmap,wmap9,planck2}. Inflation is also the only known way to generate structure 
formation in the universe via amplifying quantum fluctuations in vacuum. See, e.g., 
Refs.~\cite{b1,b2,bte1,bte2,bte3} for a comprehensive review of inflationary physics
and mathematics.

However, inflation is just the cosmological paradigm, not a theory! The known 
field-theoretical mechanisms of inflation use a slow-roll scalar field  $\f$ 
(called {\it inflaton}) with proper scalar potential $V(\f)$.

The scale of inflation is well beyond the electro-weak scale, ie. it is 
well beyond the Standard Model of Elementary Particles! Thus the inflationary 
stage in the early universe is the most powerful High-Energy Physics
(HEP) accelerator in Nature (up to ${}10^{10}~TeV$). Therefore, inflation is 
the great and unique window to very HEP!

The nature of inflaton and the origin of its scalar potential are the big 
mysteries.

In this paper the units $\hbar=c=1$ and $M_{\rm Pl}=\k^{-1}=
\fracmm{1}{\sqrt{8\p G_{\rm N}}}=2.4\times 10^{18}$ GeV, and the spacetime signature
$(+,-,-,-)$ are used. See ref.~\cite{ll} for our use of Riemann geometry 
of a curved spacetime. 

The CMB radiation picture from the WMAP and PLANCK are the main source of data about 
early universe. Deciphering it in terms of density perturbations, gravity wave 
polarization, power spectrum and its various indices is a formidable task. It also 
requires the heavy mathematical formalism based on General Relativity 
\cite{bte1,bte2,bte3}. Fortunately, we do not need much of that formalism for our 
purposes, since the relevant indices can also be introduced in terms of the inflaton 
scalar potential (Sec.~4). We assume that inflation did happen. There exist many 
inflationary models --- see eg., Ref.~\cite{b2} for their description and comparison 
(without supersymmetry). Our aim is a viable theoretical description of inflation in the 
context of supergravity and its relation to HEP of elementary particles beyond the SM.

The main Cosmological Principle of a spatially homogeneous and isotropic 
$(1+3)$-dimensional universe (at large scales) gives rise to the FLRW
metric 
\be \lb{frw1} 
 ds_{\rm FLRW}^2 = 
 dt^2 - a^2(t)\left[ \fracmm{dr^2}{1-kr^2} +r^2d\Omega^2\right]
  \ee
where the function $a(t)$ is known as the scale factor in  `cosmic' (comoving)
coordinates $(t,r,\theta,\phi)$, and $k$ is the FLRW topology index, 
$k=(-1,0,+1)$. The FLRW metric (\ref{frw1}) admits the six-dimensional 
isometry group $G$ that is either $SO(1,3)$, $E(3)$ or $SO(4)$, acting on the 
orbits $G/SO(3)$, with  the spatial three-dimensional sections $H^3$, $E^3$ or 
$S^3$, respectively. The Weyl tensor of any FLRW metric vanishes, 
\be \lb{weylv}
C^{\rm FLRW}_{\m\n\l\r}=0
\ee
where $\m,\n,\l,\r =0,1,2,3$. The early universe inflation (acceleration) 
means 
\be  \lb{idef} 
\ddt{a}(t)>0~,~{\rm or~ equivalently~,}~~
 \fracmm{d}{dt}\left( \fracmm{H^{-1}}{a}\right)<0 
\ee
where $H=\dt{a}/a$ is called Hubble function, and $\fracmm{H^{-1}}{a}$ is called Hubble
radius. The latter describes the causally connected region whose size is {\it decreasing}
 during inflation. We take $k=0$ for simplicity.  The amount of inflation 
(called the {\it e-foldings} number) is given by 
\be \lb{efol}
N_e =\ln \fracmm{a(t_{\rm end})}{a(t_{\rm start})}= 
\int^{t_{\rm end}}_{t_{\rm start}} 
 H~dt   \approx \fracmm{1}{M^2_{\rm Pl}}\int_{\f_{\rm end}}^{\f}
\fracmm{V}{V'}\,d\f  \ee 

It is well recognized by now that one has to go beyond the Einstein-Hilbert action for 
gravity, both from the experimental viewpoint (because of dark energy) and from the 
theoretical viewpoint (because of the UV incompleteness of quantized Einstein gravity
and the need of its unification with the Standard Model of Elementary Particles).

In our approach the origin of inflation is {\it geometrical} or {\it gravitational}, ie. 
is closely related to space-time and gravity. It can be technically accomplished by 
taking into account the higher-order curvature terms on the left-hand-side of Einstein 
equations (modified gravity), and extending gravity to supergravity. The higher-order 
curvature terms are supposed to appear in the gravitational effective action of Quantum 
Gravity. Their derivation from Superstring Theory may be possible too. The true problem 
is a {\it selection} of those high-order curvature terms that are physically relevant 
and/or derivable from a fundamental theory of Quantum Gravity.

There are many phenomenological models of inflation in the literature, which
usually employ some new fields and new interactions. It is, therefore, quite
reasonable and meaningful to search for the {\it minimal} inflationary model 
building, by getting the most economical and viable inflationary scenarios. We 
are going to use the approach proposed the long time ago by Starobinsky 
\cite{star1,star2}, which is also known as the (chaotic) $R^2$-inflation. We assume 
that the general 
coordinate invariance in spacetime is fundamental, and it should not be 
sacrificed. Moreover, it should be extended to the more fundamental, {\it local} 
supersymmetry that is known to imply the general coordinate invariance. It thus
leads us to supergravity which, in addition, automatically has several viable 
candidates for {\it Dark Matter} particle (see Sec.~20 for more).

On the theoretical side, the available inflationary models may be also 
evaluated  with respect to their {\it ``cost''}, ie. against what one gets from 
a given model in relation to what one puts in! Our approach does {\it not} 
introduce new fields, beyond those already present in gravity and supergravity.
We also exploit (super)gravity interactions {\it only}, ie. do {\it not}
 introduce new interactions, in order to describe inflation.

Before going into details, let us address two common prejudices and objections.

The higher-order curvature terms are usually expected to be relevant near the 
spacetime curvature singularities. It is also quite possible that some 
higher-derivative gravity, subject to suitable constraints, could be in the 
effective action of a quantized theory of gravity,~\footnote{To the best of our
knowledge, this proposal was first formulated by A.D. Sakharov in 1967 \cite{saha}.} 
like eg., in String Theory. However, there are also some common doubts 
 against the higher-derivative terms, in principle.

First, it is often argued that all higher-derivative field theories, 
including the  higher-derivative gravity theories, have ghosts (i.e. are 
unphysical), because of Ostrogradski theorem (1850) in Classical Mechanics. 
As a matter of fact, though the presence of ghosts is a generic feature of 
the higher-derivative theories indeed, it is not always the case, while many 
explicit examples are known (Lovelock gravity, Euler densities, some $f(R)$ 
gravity theories, etc.) --- see eg., ref.~\cite{florida} for more details. In 
our approach, the absence of ghosts and tachyons is required, while it is also 
considered as one of the main physical selection criteria for the ``good''
higher-derivative field theories.

Another common objection against the higher-derivative gravity theories 
is due to the fact that all the higher-order curvature terms in the action
are to be suppressed by the inverse powers of $M_{\rm Pl}$ on dimensional 
reasons and, therefore, they seem to be `very small and negligible'. Though it 
is generically true, it does not mean that all the higher-order curvature 
terms are irrelevant at all scales much less than $M_{\rm Pl}$. For instance, 
it appears that the {\it quadratic} curvature terms have {\it dimensionless} 
couplings, while they can easily describe the early universe inflation (in the
high-curvature regime). A non-trivial function of $R$ in the effective gravitational 
action may also `explain' dark dnergy in the present universe 
\cite{fgrev1,fgrev2,fgrev3}.

\section{Starobinsky approach to inflation}

The Starobinsky models were the first inflationary models introduced  as early as 1980 
\cite{star1,star2}. Remarkably, they are still viable, being consistent with all 
cosmological observations at present. To say more, they are currently {\it preferred}
 by the most recent WMAP9 and PLANCK observational data \cite{wmap9,planck2}. In this 
section we approach the Starobinsky models from the very different (formal) perspective.

It can be argued that it is the {\it scalar} curvature-dependent part 
of the gravitational effective action that is most relevant to the large-scale 
dynamics $H(t)$. Here are some simple arguments.

In four spacetime dimensions all the independent {\it quadratic} curvature invariants
are  $R^{\m\n\l\r}R_{\m\n\l\r}$~, $R^{\m\n}R_{\m\n}$ and $R^2$. However, the Gauss-Bonnet
combination 
\be \lb{td1}
 \int d^4x\, \sqrt{-g} \left( R^{\m\n\l\r}R_{\m\n\l\r} 
-4R^{\m\n}R_{\m\n}
 +R^2\right)
\ee
is topological (ie. a total derivative) for any metric, while
\be \lb{td2}
\int d^4x\, \sqrt{-g} \left( 3R^{\m\n}R_{\m\n} -R^2\right)
\ee
is also topological for any FLRW metric, because of eq.~(\ref{weylv}). Hence, 
the FLRW-relevant quadratically-generated gravity action is 
$(8\p G_N=1)$
\be \lb{qua}
 S= - \ha \int d^4x\, \sqrt{-g} 
\left( R - R^2/M^2 \right)
\ee
This action is known as the simplest Starobinsky model \cite{star1,star2}. Its 
equations of motion allow a stable inflationary solution, and it is an {\it 
attractor!} When $H\gg M$, one finds 
\be \lb{starinf}
 H\approx \left( \fracmm{M}{6}\right)^2 \left(t_{\rm end}-t\right)
\ee
It is the particular realization of chaotic inflation  (ie. with chaotic 
initial conditions) \cite{linde1} with a Graceful Exit.

In the case of a generic gravitational action with the higher-order curvature 
terms, the Weyl dependence can be excluded due to eq.~(\ref{weylv}) again. A
dependence upon the Ricci tensor may also be excluded since, otherwise, it 
would lead to the extra propagating massless spin-2 degree of freedom (in 
addition to a metric) described by the field $\pa\Lag/\pa R_{\m\n}$. The higher
 derivatives of the scalar curvature in the gravitational Lagrangian $\Lag$ 
just lead to more propagating scalars \cite{wands}, so we simply ignore them for
 simplicity in what follows.
 
\section{$f(R)$ Gravity}

The Starobinsky model (\ref{qua}) is the special case of the $f(R)$ gravity 
theories \cite{fgrev1,fgrev2,fgrev3} having the action 
\be \lb{fgra}
 S_f = -\fracmm{1}{16\p G_N} \int d^4x\, \sqrt{-g}\,\tilde{f}(R) 
\ee 
In the absence of extra matter, the gravitational (trace) equation of motion is
of the fourth order with respect to the time derivative,
\be \lb{fgrem1}
 \fracmm{3}{a^3} \fracmm{d}{dt} \left( a^3 \fracmm{d\tilde{f}'(R)}{dt}\right)
 +R\tilde{f}'(R)-2\tilde{f}(R)=0 
\ee
where we have used $H=\fracmm{\dt{a}}{a}$ and $R=-6(\dt{H}+ 2H^2)$. The primes
 denote the derivatives with respect to $R$, and the dots denote the 
derivative with respect to $t$. Static de-Sitter 
solutions correspond to the roots of the equation \cite{stab1}
$$ \lb{mull}
R\tilde{f}'(R)=2\tilde{f}(R)$$

The $00$-component of the gravitational equations is of the third order with
respect to the time derivative,
\be \lb{fgrem2}
 3H \fracmm{d\tilde{f}'(R)}{dt} -3(\dt{H} +H^2)\tilde{f}'(R)
 -\fracmm{1}{2}\tilde{f}(R)=0 
\ee

The (classical and quantum) {\it stability} conditions in $f(R)$ gravity 
are well known \cite{fgrev1,fgrev2}, and are given by (in our notation)  
\be \lb{stab} 
\tilde{f}'(R)>0 \qquad {\rm and}\qquad  \tilde{f}''(R)<0
\ee
respectively. The first condition (\ref{stab}) is needed to get a physical 
(non-ghost) graviton, while the second condition (\ref{stab}) is needed to get 
a physical (non-tachyonic) scalaron (see Sec.~9 for more).

Any $f(R)$ gravity is known to be classically {\it equivalent} to the certain 
scalar-tensor gravity having an (extra) propagating scalar field 
\cite{eq1,eq2,eq3}. The formal equivalence can be established via the
Legendre-Weyl transformation to be described below.

First, the $f(R)$-gravity action (\ref{fgra}) can be rewritten to the form
\be \lb{let}
 S_A = \fracmm{-1}{2\k^2}\int d^4x\,\sqrt{-g}\,\left\{ AR-Z(A)\right\} 
\ee
where the real scalar (or Lagrange multiplier) $A(x)$ is related to the 
scalar curvature $R$ by the Legendre-like transformation:
\be\lb{let2}
  R=Z'(A) \qquad{\rm and}\qquad \tilde{f}(R)=RA(R)-Z(A(R)) 
\ee
with $\k^2=8\p G_{N}=M_{\rm Pl}^{-2}$.

Next, a Weyl transformation of the metric,
\be \lb{weylt}
g_{\m\n}(x)\to \exp \left[ \fracmm{2\k\f(x)}{\sqrt{6}} \right] g_{\m\n}(x) 
\ee
with arbitrary field parameter $\f(x)$ yields
\be \lb{let3} 
\sqrt{-g}\,R \to \sqrt{-g}\, \exp \left[ \fracmm{2\k\f(x)}{\sqrt{6}} \right]
\left\{ R -\sqrt{\fracmm{6}{-g}}\pa_{\m}\left(\sqrt{-g}g^{\m\n}\pa_{\n}
\f\right)\k -\k^2g^{\m\n}\pa_{\m}\f\pa_{\n}\f\right\} 
\ee
Therefore, when choosing
\be \lb{let4}  
 A(\k\f) = \exp \left[ \fracmm{-2\k\f(x)}{\sqrt{6}} \right]  
\ee
and ignoring a total derivative in the Lagrangian, we can rewrite the 
action to the form 
\be \lb{stand}
\eqalign{
S[g_{\m\n},\f] = ~&~ \int d^4x\, \sqrt{-g}\left\{ \fracmm{-R}{2\k^2}
+\fracmm{1}{2}g^{\m\n}\pa_{\m}\f\pa_{\n}\f  \right. \cr
 ~&~ \left.  + \fracmm{1}{2\k^2}
\exp \left[ \fracmm{4\k\f(x)}{\sqrt{6}}\right] Z(A(\k\f)) \right\}
\cr} 
\ee
in terms of the physical (and canonically normalized) scalar field $\f(x)$,
without any higher derivatives and ghosts. As a result, one arrives at
 the standard action of the real dynamical scalar field $\f(x)$ 
{\it minimally} coupled to Einstein gravity and having the  scalar 
potential
\be \lb{spot1}
 V(\f) = -\fracmm{M^2_{\rm Pl}}{2}\exp \left\{
 \fracmm{4\f}{M_{\rm Pl}\sqrt{6}}\right\}
Z\left( \exp \left[ \fracmm{-2\f}{M_{\rm Pl}\sqrt{6}} \right] \right) 
\ee

In the context of the inflationary theory, the {\it scalaron} 
(= scalar part of spacetime metric) $\f$ can be identified with inflaton. 
This inflaton has the clear origin as the spin-$0$ part of spacetime metric, 
and may also be understood as the conformal mode of the metric in Minkowski or 
(A)dS vacuum.

In the Starobinsky case, $\tilde{f}(R)=R-R^2/M^2$, 
the inflaton scalar potential  reads 
\be \lb{starpo}
 V(y) = V_0\left( e^{-y}-1 \right)^2 
\ee
where we have introduced the notation
 \be \lb{not1}
 y=\sqrt{\fracmm{2}{3}}\fracmm{\f}{M_{\rm Pl}} \qquad {\rm and}\qquad 
V_0=\fracmm{1}{8}M^2_{\rm Pl}M^2
\ee
It is worth noticing here the appearance of the inflaton vacuum energy 
$V_0$ driving inflation. The end of inflation (Graceful Exit) is also clear:
the scalar potential (\ref{starpo}) has a very  flat (slow-roll) `plateau', 
ending with a `waterfall' towards the minimum (Fig.~1). 

\begin{figure}[t]
\begin{center}
\vglue.1in
\makebox{
\epsfxsize=3in
\epsfbox{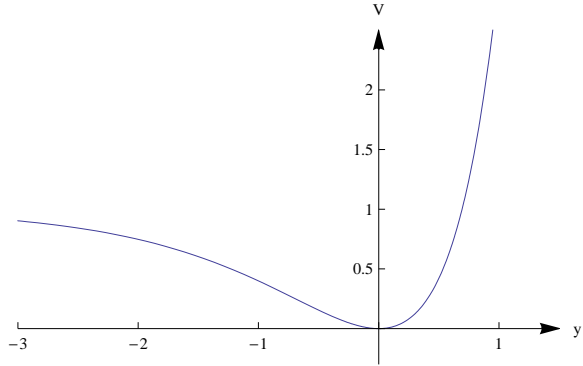}
}
\caption{\small The inflaton scalar potential $v(x)=(e^{y}-1)^2$ in the 
Starobinsky  model, after $y\to -y$}
\end{center}
\end{figure}

It is worth emphasizing that the inflaton (scalaron) scalar potential 
(\ref{starpo}) is derived here by merely assuming the existence of the $R^2$ 
term in the gravitational action. The Newton (weak gravity) limit is not 
applicable to early universe (including its inflationary stage), so that
the dimensionless coefficient in front of the $R^2$ term does not have to be
very small at early time. It distinguishes the primordial `Dark Energy' driving 
inflation in the early Universe from the `Dark Energy' responsible for the present 
universe acceleration.
 
\section{Inflationary Theory and Observations}

The {\it slow-roll} inflation parameters are defined by 
\be \lb{slp}
\ve(\f) = \fracmm{1}{2} M^2_{\rm Pl} \left( \fracmm{V'}{V}\right)^2 \qquad
{\rm and}\qquad \h (\f) = M^2_{\rm Pl} \fracmm{V''}{V} 
\ee
A necessary condition for the slow-roll approximation is the
smallness of the inflation parameters
\be \lb{smallp}
 \ve(\f)\ll 1 \qquad {\rm and} \qquad \abs{\h(\f)}\ll 1 
\ee
The first condition implies $\ddt{a}(t)>0$. The second one guarantees that 
inflation lasts long enough, via domination of the {\it friction} term in the 
inflaton equation of motion, $3 H\dt{\f} =- V'$. 

The CMB temperature fluctuations \cite{nasa,wmap} have the scale $\d T/T \approx 10^{-5}$
at the WMAP normalization of $500$ Mpc. Actually, the scalar $(\r_s)$ and tensor  
$(\r_t)$ perturbations of metric do {\it decouple}. The scalar perturbations couple to 
the density of matter and radiation, so they are responsible for the inhomogeneities and 
anisotropies in the universe. The tensor perturbations (or gravity waves) also contribute
 to the CMB, while their experimental detection would tell us much more about inflation. 
The CMB raditation is expected to be {\it polarized} due to Compton scattering 
at the time of decoupling \cite{dodeb,weinb}. 

The primordial (Zeldovich-Harrison) spectrum is proportional to $k^{n-1}$, in terms of 
the comoving wave number $k$ and the spectral index $n$, in the 2-point function 
(observable) 
\be \lb{tempr}
\VEV{  \fracmm{\d T(x)}{T} \fracmm{ \d T(y)}{T} }\propto 
\int \fracmm{d^3k}{k^3} e^{ik(x-y)} k^{n-1} 
\ee 

In theory, the slope $n_s$ of the {\it scalar} power spectrum, associated with the 
density perturbations, $\left( \fracmm{\d\r}{\r}\right)^2\propto k^{n_s-1}$,  is given
by $n_s  = 1+2\h -6\ve$, the slope of the {\it tensor} primordial spectrum, 
associated with gravitational waves, is $ n_t= -2\ve$, and the 
{\it tensor-to-scalar ratio} is $r = \d\r_s/ \d\r_t= 16\ve$ (see eg., ref.~\cite{b2}).

It is straightforward to calculate those indices in any inflationary model with
a given inflaton scalar potential. In the case of the Starobinsky model and
its scalar potential (\ref{starpo}), one finds \cite{mchi,chi2,us6} 
\be \lb{chi}
 n_s =1 - \fracmm{2}{N_e} + \fracmm{3\ln N_e}{2N_e^2} -\fracmm{2}{N^2_e}
+{\cal O}\left( \fracmm{\ln^2 N_e}{N^3_e}\right)  
\ee
and
\be \lb{rind}
 r\approx \fracmm{12}{N_e^2}\approx 0.004
\ee
with $N_e\approx 55$. The very small value of $r$ is the sharp prediction of
the Starobinsky inflationary model towards $r$-measurements in a future.

Those theoretical values are to be compared to the observed values of the CMB radiation 
For instance, the WMAP7 observations \cite{wmap} yield 
\be \lb{wmap7} 
n_s=0.963\pm 0.012 \qquad {\rm and} \qquad  r<0.24
\ee
with the $95$ \% level of confidence. 

The most recent PLANCK data yields \cite{planck2}
\be \lb{pla2}
n_s=0.960\pm 0.007 \qquad {\rm and} \qquad  r<0.11
\ee
also with the $95$ \% level of confidence. 

The amplitude of the initial perturbations, $\D^2_R=M^4_{\rm Pl}V/(24\p^2\ve)$,
 is also the physical observable whose experimental value is known since 1992 due to
the Cosmic Background Explorer (COBE) satellite mission \cite{cobe}:
\be \lb{cobe}
\left(\fracmm{V}{\ve}\right)^{1/4} =0.027\,M_{\rm Pl}=
6.6\times 10^{16}~{\rm GeV}
\ee

It determines the normalization of the $R^2$-term in the action (\ref{qua})
\be \lb{norma}
 \fracmm{M}{M_{\rm Pl}}= 4\cdot \sqrt{\fracmm{2}{3}}\cdot (2.7)^2
\cdot \fracmm{e^{-y}}{(1-e^{-y})^2} \cdot 10^{-4} \approx 
(3.5\pm 1.2)\cdot 10^{-6} 
\ee
The inflaton mass is given by $M_{\rm inf}=M/\sqrt{6}$, and there are no free parameters left.

\begin{figure}[t]
\begin{center}
\vglue.1in
\makebox{
\epsfxsize=3in
\epsfbox{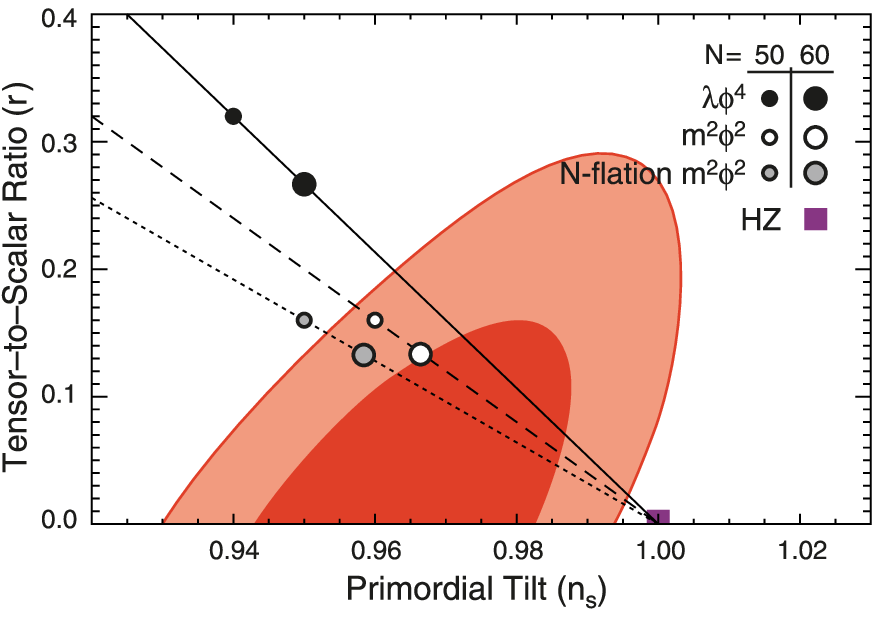}
 }
\caption{\small Starobinsky inflation vs. $m^2\f^2/2$ and $\l\f^4$ \cite{lnw}}
\end{center}
\end{figure}

The main theoretical lessons we can draw from that are:\\
(i) the main {\it discriminants} amongst all inflationary models are given by
the values of $n_s$ and $r$;\\
(ii) the Starobinsky model (1980) of chaotic inflation is very simple and 
economic. It uses gravity interactions only. It predicts the origin of inflaton
and its scalar potential. It is still {\it viable} and {\it consistent} with 
all known observations. Inflaton is not charged (singlet) under the SM gauge group.
The Starobinsky inflation has an end  (Graceful Exit), and gives the simple explanation 
to the WMAP-observed value of $n_s$. The  {\it key} difference of Starobinsky 
inflation from the other standard inflationary models (having $\ha m^2\f^2$ or 
$\l\f^4$ scalar potentials) is the very low value of $r$ --- see Fig.~2 for comparison 
and ref.~\cite{lnw} for more details; \\
(iii) the viable inflationary models, based on $\tilde{f}(R)=R+\hat{f}(R)$ 
gravity, turn out to be close to the simplest Starobinsky model (over the range
 of $R$ relevant to inflation), with $\hat{f}(R)\approx R^2A(R)$ and the 
slowly varying function $A(R)$  in the sense
\be \lb{slow}
 \abs{A'(R)}\ll \fracmm{A(R)}{R} \qquad  {\rm and}\qquad  
\abs{A''(R)}\ll \fracmm{A(R)}{R^2} 
\ee

\section{Supergravity and Superspace}

Supersymmetry (SUSY) is the leading proposal to new physics beyond the SM. Therefore,
it is quite natural to unify inflation with high-energy particle physics in the context 
of supersymmetry. 

SUSY is the symmetry between bosons and fermions. SUSY is the extension of 
Poincar\'e symmetry of spacetime, and is well motivated in HEP beyond the SM. 
Supersymmetry is also needed for consistency of strings. Supergravity (SUGRA) is the 
theory of {\it local} supersymmetry that automatically implies general coordinate 
invariance. Hence, considering inflation with supersymmetry necessarily leads to 
supergravity. As a matter of fact, most of studies of superstring- and brane-cosmology 
are also based on their effective description in the four-dimensional $N=1$ supergravity.

It is not our purpose to give a detailed account of SUSY and SUGRA, because of the 
existence of several good textbooks --- see eg., refs.~\cite{sb1,sb2,sb3}. In this 
Section we recall only the basic facts about $N=1$ supergravity in four spacetime 
dimensions, which are needed for our purposes.

A concise and manifestly supersymmetric description of SUGRA is provided  
by Superspace. In this section the natural units $c=\hbar=\kappa=M_{\rm Pl}=1$ 
are used for more simplicity.

Supergravity needs a {\it curved} superspace. However, they are not the same,
because one has to reduce the field content to the minimal one corresponding 
to an off-shell supergravity multiplet. It can be done by imposing certain off-shell
constraints on the supertorsion tensor in curved superspace \cite{sb1,sb2,sb3}.
 An off-shell supergravity multiplet has some extra (auxiliary) fields with
noncanonical dimensions, in addition to physical spin-2 field (metric) and 
spin-3/2 field (gravitino). It is worth mentioning that imposing the off-shell 
constraints is independent upon writing a supergravity action.
 
One may work either in a full (curved) superspace or in a chiral one. There are
some practical anvantages of using the chiral superspace, because it helps us to keep the
auxiliary fields unphysical (i.e. nonpropagating). The chiral superspace is more closely 
related to supergravity in components (in a Wess-Zumino gauge).

The chiral superspace density reads
\be \lb{den}
\ce(x,\theta) = e(x) \left[ 1 +i\theta\s^a\bar{\j}_a(x) -
\theta^2 \left( B^*(x) + \bar{\j}_a\bar{\s}^{ab}\bar{\j}_b\right) \right]~, \ee
where $e=\sqrt{-\det g_{\m\n}}$, $g_{\m\n}$ is a spacetime metric, 
$\j_a^{\a}=e_a^{\m}\j_{\m}^{\a}$ is a chiral gravitino, $B=S+iP$ 
is the complex scalar auxiliary field. We use the lower case middle greek 
letters $\m,\n,\ldots=0,1,2,3$ for curved spacetime vector indices, the 
lower case early latin letters $a,b,\ldots=0,1,2,3$ for flat (target) space 
vector indices, and the lower case early greek letters $\a,\b,\ldots=1,2$ for 
chiral spinor indices.

A solution to the superspace Bianchi identities together with the constraints 
defining the $N=1$ Poincar\'e-type {\it minimal} supergravity theory results 
in only {\it three} covariant tensor superfields $\car$, $\cg_a$ and 
$\cw_{\a\b\g}$, subject to the off-shell relations \cite{sb1,sb2,sb3}:
\be \lb{bi1}
 \cg_a=\bar{\cg}_a~,\qquad \cw_{\a\b\g}=\cw_{(\a\b\g)}~,\qquad
\bar{\de}_{\dt{\a}}\car=\bar{\de}_{\dt{\a}}\cw_{\a\b\g}=0~,\ee
and
\be \lb{bi2}
 \bar{\de}^{\dt{\a}}\cg_{\a\dt{\a}}=\de_{\a}\car~,\qquad
\de^{\g}\cw_{\a\b\g}=\frac{i}{2}\de\du{\a}{\dt{\a}}\cg_{\b\dt{\a}}+
\frac{i}{2}\de\du{\b}{\dt{\a}}\cg_{\a\dt{\a}}~~,\ee
where $(\de\low{\a},\bar{\de}_{\dt{\a}}.\de_{\a\dt{\a}})$ stand for the 
curved superspace $N=1$ supercovariant derivatives, and the bars denote 
complex conjugation.

The covariantly chiral complex scalar superfield $\car$ has the scalar
 curvature $R$ as the coefficient at its $\theta^2$ term, the real vector 
superfield $\cg_{\a\dt{\a}}$ has the traceless Ricci tensor, 
$R_{\m\n}+R_{\n\m}-\frac{1}{2}g_{\m\n}R$, as the coefficient at its 
$\theta\s^a\bar{\theta}$ term, whereas the covariantly chiral, complex, 
totally symmetric, fermionic superfield $\cw_{\a\b\g}$ has the self-dual part
of the Weyl tensor $C_{\a\b\g\d}$ as the coefficient at its linear 
$\theta^{\d}$-dependent term. 

A generic Lagrangian representing the supergravitational effective action in 
(full) superspace, reads
\be \lb{genc}
\Lag = \Lag(\car,\cg,\cw,\ldots) \ee
where the dots stand for arbitrary supercovariant derivatives of the 
superfields. 

The Lagrangian (\ref{genc}) it its most general form is, however, unsuitable 
for physical applications, not only because it is too complicated, but just 
because it generically leads to the ``propagating auxiliary'' fields whose physical
interpretation is unclear. The important physical condition of keeping the 
supergravity auxiliary fields to be truly auxiliary (ie. nonphysical or nonpropagating) 
in field theories with the higher derivatives was dubbed the {\it `auxiliary freedom'} 
in refs.~\cite{g1,g2}. To get the supergravity actions with the  `auxiliary freedom', we 
employ the chiral (curved) superspace.

\section{$F(\car)$ Supergravity in Superspace}

Here we focus on the scalar-curvature-sector of a generic 
higher-derivative supergravity (\ref{genc}), which is most relevant to the
FRLW cosmology, by ignoring the tensor curvature superfields $\cw_{\a\b\g}$ and
 $\cg_{\a\dt{\a}}$, as well as the derivatives of the scalar superfield $\car$. 
Then there is only one candidate for a locally supersymmetric action in the chiral 
curved superspace, 
\be \lb{action}
 S_F = \int d^4xd^2\theta\,\ce F(\car) + {\rm H.c.}
\ee
governed by a chiral {\it analytic} function $F(\car)$.~\footnote{The
field construction of this theory by using the 4D, $N=1$ superconformal tensor calculus 
was given in ref.~\cite{cec}.} 
 Besides having the manifest local $N=1$ supersymmetry, the action (\ref{action}) has the
 auxiliary freedom since the auxiliary field $B$ does not propagate. It distinguishes 
the action (\ref{action}) from other possible truncations of eq.~(\ref{genc}). 
The action (\ref{action}) gives rise to the spacetime {\it torsion} generated 
by gravitino, while its bosonic terms have the form 
\be \lb{mgrav}
 S_f = - \fracmm{1}{2} \int d^4x \,\sqrt{-g}\, \tilde{f}(R)  
\ee
Hence, eq.~(\ref{action}) can be considered as the locally $N=1$ 
supersymmetric extension of the $f(R)$-type gravity (Sec.~3). However, in
the context of supergravity, the `supersymmetrizable' bosonic functions 
$\tilde{f}(R)$ are very restrictive (see Secs.~9 and 10).

The superfield action (\ref{action}) is classically equivalent to 
\be \lb{lmult}
 S_V = \int d^4x d^2\theta\,\ce \left[ \cz\car -V(\cz)\right] + {\rm H.c.}
\ee
with the covariantly chiral superfield $\cz$ as the Lagrange multiplier
superfield.  Varying the action (\ref{lmult}) with respect to $\cz$  gives back
 the original action  (\ref{action}) provided that
\be \lb{lt1} 
F(\car) =\car\cz(\car)-V(\cz(\car)) \ee
where the function $\cz(\car)$ is defined by inverting the function
\be \lb{lt2}
\car =V'(\cz)  \ee
Equations (\ref{lt1}) and  (\ref{lt2}) define the superfield Legendre 
transform, and imply 
\be \lb{lt3}
F'(\car)=Z(\car)\qquad {\rm and}\qquad F''(\car)=Z'(\car)
=\fracmm{1}{V''(\cz(\car))}  \ee
where $V''=d^2V/d\cz^2$. The second formula (\ref{lt3}) is the {\it duality}
relation between the supergravitational function $F$ and the chiral 
superpotential $V$.

A supersymmetric (local) Weyl transform of the acton (\ref{lmult}) can be done 
entirely in superspace. In terms of the field components, the super-Weyl 
transform amounts to a Weyl transform, a chiral rotation and a (superconformal)
 $S$-supersymmetry  transformation \cite{howe}. The chiral density superfield 
$\ce$ appears to be the chiral {\it compensator} of the super-Weyl 
transformations,
\be \lb{swt}
\ce \to e^{3\F} \ce 
\ee
whose parameter $\F$ is an arbitrary covariantly chiral superfield,
$\bar{\de}_{\dt{\a}}\F=0$. Under the transformation (\ref{swt}) the 
covariantly chiral superfield $\car$ transforms as 
\be \lb{rwlaw}
\car \to e^{-2\F}\left( \car - \fracm{1}{4}\bar{\nabla}^2\right)
e^{\bar{\F}}
\ee
The super-Weyl chiral superfield parameter $\F$ can be traded for the chiral
Lagrange multiplier $\cz$ by using a generic gauge condition
\be \lb{ch2} \cz=\cz(\F) \ee
where $\cz(\F)$ is a holomorphic function of $\F$. It results in the  action   
\be \lb{chimat2}
S_{\F}=\int d^4x d^4\theta\, E^{-1} e^{\F+\bar{\F}}
\left[ \cz(\F) +{\rm H.c.} \right] 
- \int d^4x d^2\theta\, \ce e^{3\F}V(\cz(\F)) +{\rm H.c.} 
\ee

Equation (\ref{chimat2}) has the {\it standard} form of the action of 
a chiral matter superfield coupled to supergravity,
\be \lb{sstand}
S[\F,\bar{\F}]= \int d^4x d^4\theta\, E^{-1} \O(\F,\bar{\F}) 
+\left[ \int d^4x d^2\theta\, \ce P(\F) +{\rm H.c.} \right]
\ee  
in terms of the non-chiral potential $\O(\F,\bar{\F})$ and the chiral 
superpotential $P(\F)$. In our case (\ref{chimat2}) we find
\be \lb{spots}
\O(\F,\bar{\F}) =  e^{\F+\bar{\F}}
\left[ \cz(\F) +\bar{\cz}(\bar{\F}) \right]~,\quad
P(\F) =  -  e^{3\F} V(\cz(\F)) 
\ee 
The {\it K\"ahler} potential $K(\F,\bar{\F})$ is given by 
\be \lb{kaehler}
 K = -3\ln(-\fracmm{\O}{3})\quad {\rm or} \quad \O=-3e^{-K/3}
\ee
so that the action (\ref{sstand}) is invariant under the supersymmetric
(local) K\"ahler-Weyl transformations
\be \lb{swk}
K(\F,\bar{\F})\to  K(\F,\bar{\F}) +\L(\F) + \bar{\L}(\bar{\F})~, \qquad
P(\F) \to -  e^{-\L(\F)} P(\F)
\ee 
with the chiral superfield parameter $\L(\F)$. It follows that
\be \lb{csvt}
\ce \to e^{\L(\F)}\ce
\ee

The scalar potential in terms of the usual fields is given by the standard 
formula \cite{crem} 
\be \lb{cremv}
 \cv (\f,\bar{\f}) =\left. e^{K} \left\{ \abs{\fracmm{\pa P}{\pa\F}
+\fracmm{\pa K}{\pa\F}P}^2-3\abs{P}^2\right\} \right|  
\ee
where all the superfields are restricted to their  leading field 
components, $\left.\F\right|=\f(x)$, and we have introduced the notation
\be \lb{defc}
 \abs{\fracmm{\pa P}{\pa\F}+\fracmm{\pa K}{\pa\F}P}^2
\equiv \abs{D_{\F}P}^2=  D_{\F}P(K^{-1}_{\F\bar{\F}}) \bar{D}_{\bar{\F}}\bar{P}
\ee
with $K_{\F\bar{\F}}=\pa^2 K/\pa\F\pa\bar{\F}$. Equation (\ref{cremv}) can be 
simplified by making use of the K\"ahler-Weyl invariance (\ref{swk}) 
that allows one to choose a gauge
\be  \lb{setone}
P=1 \ee
It is equivalent to the well known fact that the scalar potential (\ref{cremv})
is actually governed by the {\it single} (K\"ahler-Weyl-invariant) potential 
\be\lb{spo}
G(\F,\bar{\F}) = \O +\ln\abs{P}^2 \ee
In our case  (\ref{spots}) we find
\be \lb{pot1p}
G=e^{\F+\bar{\F}}\left[ \cz(\F) +\bar{\cz}(\bar{\F}) \right]
+ 3(\F+\bar{\F}) + \ln(V(\cz(\F))+ \ln(\bar{V}(\bar{\cz}(\bar{\F}))
\ee
So let us choose a gauge by the condition
\be \lb{fix}  3\F + \ln(V(\cz(\F))=0\quad {\rm or}\quad
 V(\cz(\F))=e^{-3\F}
\ee
that is equivalent to eq.~(\ref{setone}). Then the $G$-potential (\ref{pot1p}) 
gets simplified to
\be \lb{center}
G = e^{\F+\bar{\F}}\left[ \cz(\F) +\bar{\cz}(\bar{\F}) \right] 
\ee
There is the correspondence between a holomorphic function $F(\car)$ in the 
supergravity action (\ref{action}) and a holomorphic function $\cz(\F)$ 
defining the scalar potential (\ref{cremv}),
\be \lb{cpot}
\cv =\left. e^G \left[ \left(\fracmm{\pa^2 G}{\pa\F\pa{\bar{\F}}}\right)^{-1}
\fracmm{\pa G}{\pa\F}\fracmm{\pa G}{\pa\bar{\F}} -3\right] \right|
\ee 
in the classically equivalent scalar-tensor supergravity. More simplifications are 
possible in a particular gauge and for a particular model --- see Sec.~13.

To the end of this section, we comment on the standard way of the inflationary model 
building by a choice of $K(\F,\bar{\F})$ and $P(\F)$ --- see eg., 
refs.~\cite{rev,yamag} for a review.

The factor $\exp(K/M^2_{\rm Pl})$ in the $F$-type scalar potential 
(\ref{cremv}) of the chiral matter-coupled supergravity, in the case of  the 
{\it canonical} K\"ahler potential, $K\propto \Bar{\F}\F$, results in the 
scalar potential $V\propto \exp(\abs{\F}^2/M^2_{\rm Pl})$ that is too steep to 
support chaotic inflation. Actually, it also implies $\eta\approx 1$ or, 
equivalently, $M^2_{\rm inflaton}\approx V_0/M^2_{\rm Pl}\approx H^2$. It is 
known as the  $\eta${\it -problem} in supergravity \cite{jap1}. 

As is clear from our discussion above, the $\eta$-problem is not really a 
supergravity problem, but it is the problem associated with the choice 
of the canonical K\"ahler potential for an inflaton superfield. The K\"ahler 
potential in supergravity is a (K\"ahler) gauge-dependent quantity, and its 
quantum renormalization is not under control. Unlike the one-field inflationary
 models, a generic K\"ahler potential is a function of at least two real scalar 
fields, so it implies a nonvanishing curvature in the target space of the 
{\it non-linear sigma-model} associated with the K\"ahler kinetic 
term.~\footnote{See eg., ref.~\cite{nlsm} for more about the non-linear 
sigma-models.} Hence, a generic K\"ahler potential cannot be brought 
to the canonical form by a field redefinion. 
  
To solve the $\eta$-problem associated with the simplest (naive) choice of the 
K\"ahler potential, on may assume that the K\"ahler potential $K$ possesses  
some shift symmetries (leading to its {\it flat} directions), and then 
choose inflaton in one such flat direction \cite{jap2}. However, in order 
to get inflation that way, one also has to add ({\it ad hoc}) the proper 
inflaton superpotential breaking the initially introduced shift symmetry, and 
then stabilize the inflationary trajectory with the help of yet another matter 
superfield.

The possible alternative is the $D$-term mechanism \cite{bdva}, where the
inflaton particle belongs to the matter {\it gauge} sector and, as a result, 
inflation is highly sensitive to gauge charges \cite{bdva}. This mechanism
is not related to spacetime and gravity.

It is worth mentioning that in the (perturbative) superstring cosmology one
gets the K\"ahler potential (see e.g., refs.~\cite{swe1,swe2}) 
\be \lb{scosm}
K\propto \log({\rm moduli~polynomial})_{\rm CY}
\ee over a Calabi-Yau (CY) space in the type-IIB superstring compactification, 
thus avoiding the $\eta$-problem but leading to a plenty of choices 
(``embarrassment of riches'') in the String Landscape and the associated high
unpredictability.

Finally, one still has to accomplish stability of a given inflationary model 
in supergravity against quantum corrections. Such corrections can easily spoil 
the flatness of the inflaton potential. The K\"ahler kinetic term is 
not protected against quantum corrections, because it is given by a full
superspace integral (unlike the chiral superpotential term). The $F(\car)$
supergravity action (\ref{action}) is given by a {\it chiral} superspace 
integral, so that it is protected against the quantum corrections given by 
full superspace integrals. 

To conclude this section, we claim that an $N=1$ locally supersymmetric 
extension of $f(R)$ gravity is possible. It is non-trivial  because  
the auxiliary freedom has to be preserved.  The new supergravity action 
(\ref{action}) is classically  equivalent to the standard $N=1$ Poincar\'e 
supergravity coupled to a {\it dynamical} chiral matter superfield, whose 
K\"ahler potential and the superpotential are dictated by a single 
{\it holomorphic} function. Inflaton can be identified with the real scalar 
field component of that chiral matter superfield originating from the 
supervielbein, and thus has the geometrical origin.

The action (\ref{action}) has yet another natural extension in the chiral curved 
superspace due to the last equation (\ref{bi1}), namely,
\be \lb{actionw}
 S_{\rm ext} = \int d^4xd^2\theta\,\ce F(\car,\cw) + {\rm H.c.}
\ee
where $\cw_{\a\b\g}$ is the $N=1$ covariantly-chiral Weyl superfield of 
the $N=1$ superspace supergravity \cite{us14}. In Superstring Theory the 
Weyl-tensor-dependence of the perturbative gravitational effective action is unambigously
 determined by the superstring scattering amplitudes or by the super-Weyl invariance of 
the corresponding non-linear sigma-model (see eg., ref.~\cite{nlsm}). However, the 
action of the type (\ref{actionw}) may only be generated from superstrings 
nonperturbatively. 

A possible connection of $F(\car)$ supergravity to the 
{\it Loop Quantum Gravity} was investigated in ref.~\cite{us3}.

\section{No-scale $F(\car)$ Supergravity}

In this section investigate a possibility of spontaneous supersymmetry breaking without 
fine tuning by imposing the condition of the vanishing scalar potential. Those no-scale 
supergravities are the starting point of many phenomenological applications of 
supergravity to HEP and inflationary theory, including superstring theory applications 
--- see eg., refs.~\cite{fluxes,kall} and the references therein. 

The no-scale supergravity arises by demanding the scalar potential 
(\ref{cremv}) to vanish. It results in the vanishing cosmological constant 
without fine-tuning \cite{noscale}. The no-scale  supergravity potential $G$ 
has to  obey the non-linear 2nd-order partial differential equation, which
follows from eq.~(\ref{cpot}),
\be \lb{nleq}
3\fracmm{\pa^2 G}{\pa\F\pa{\bar{\F}}}
=\fracmm{\pa G}{\pa\F}\fracmm{\pa G}{\pa\bar{\F}}
\ee 
A gravitino mass $m_{3/2}$ is given by the vacuum expectation value \cite{sb2}
\be \lb{ssb}
m_{3/2}= \VEV{e^{G/2}} \ee
 
The well known exact solution to eq.~(\ref{nleq}) is given by
\be \lb{wn}
G = -3\log(\F +\bar{\F}) +~const.\ee
In the recent literature the no-scale solution (\ref{wn}) is usually 
modified by other terms, in order to describe the universe with a positive 
cosmological constant --- see e.g., the KKLT mechanism \cite{kklt}.

To appreciate the difference between the standard no-scale supergravity 
solution and our `modified' supergravity, it is worth noticing that  
demanding eq.~(\ref{nleq}) gives rise to the first-order non-linear partial 
differential equation
\be \lb{first}
3\left( e^{\bar{\F}}X' + e^{\F}\bar{X}'\right) =\abs{e^{\bar{\F}}X' +
e^{\F}\bar{X}}^2
\ee
where we have introduced the notation
\be \lb{change}
\cz(\F) = e^{-\F}X(\F)~,\qquad X'=\fracmm{dX}{d\F}  \ee
in order to get the differential equation in its most symmetric and concise 
form. 

Accordingly, the gravitino mass (\ref{ssb}) is given by 
\be \lb{ssb2}
m_{3/2} = \VEV{\exp \ha\left( e^{\bar{\F}}X + e^{\F}\bar{X}\right)}
\ee

We are not aware of any non-trivial holomorphic exact solution to 
eq.~(\ref{first}). However, should it obey a holomorphic differential 
equation of the form
\be \lb{holde} 
X'= e^{\F}g(X,\F) 
\ee
with a holomorphic function $g(X,\F)$,  eq.~(\ref{first}) gives rise to the
functional equation
\be \lb{funce}
3\left( g + \bar{g}\right)= \abs{ e^{\bar{\F}}g +\bar{X}}^2
\ee 

Being restricted to the real variables $\F=\bar{\F}\equiv y$ and 
$X=\bar{X}\equiv x$, eq.~(\ref{first}) reads
\be \lb{real}
6x'=e^y(x'+x)^2~,\quad {\rm where}\quad x'=\fracmm{dx}{dy}
\ee
This equation can be integrated after a change of variables,
 \be \lb{chvar} 
 x=e^{-y}u~,
\ee
and it leads to a quadratic equation with respect to $ u'=du/dy$,
\be \lb{quade}
 (u')^2 - 6u' +6u=0
\ee
It follows
\be \lb{laste}
y=\int^u \fracmm{d\x}{3\pm \sqrt{3(3-2\x)}}=\mp \, \sqrt{1-\frac{2}{3}u} 
+\ln\left(\sqrt{3(3-2u)}\pm 3\right) + C~.
\ee

\section{Fields from Superfields in $F(\car)$ Supergravity}

For simplicity, now we set all fermionic fields to zero, and keep only bosonic 
field components of the superfields. It greatly simplies all equations but 
makes supersymmetry to be manifestly broken. Of course, SUSY is restored after 
adding back all the fermionic terms.
 
Applying the standard superspace chiral density formula \cite{sb1,sb2,sb3} 
\be \lb{chiden}
\int d^4xd^2\theta\,\ce \Lag =\int d^4x\, e\left\{ 
\Lag_{\rm last} +B\Lag_{\rm first}\right\} \ee 
to the action (\ref{action}) yields its bosonic part in the form
\be \lb{expa}
(-g)^{-1/2}L_{\rm bos}\equiv f(R,\tilde{R};X,\bar{X})=
F'(\bar{X}) \left[ \frac{1}{3}R_* +4\bar{X}X \right] +3X F(\bar{X})+{\rm 
H.c.}  \ee
where the primes denote differentiation with respect to the given argument.
We have used the notation
\be \lb{snot1} X=\frac{1}{3}B \qquad {\rm and} \qquad
R_*=R+\fracmm{i}{2}\ve^{abcd}R_{abcd} \equiv R + i\tilde{R} 
\ee
The $\tilde{R}$ does not vanish in $F(\car)$ supergravity, and it represents 
the pseudo-scalar superpartner of the real scalaron field in our construction.
 
Varying eq.~(\ref{expa}) with respect to the auxiliary fields $X$ and 
$\bar{X}$,
\be \lb{auxe}
 \fracmm{\pa L_{\rm bos}}{\pa X} =  \fracmm{\pa L_{\rm bos}}{\pa \bar{X}} 
= 0
\ee
 gives rise to the algebraic equations on the auxiliary fields,
\be\lb{aux1}
3\bar{F}+X(4\bar{F}'+7F')+4\bar{X}XF'' +\frac{1}{3}F''R_*=0
\ee
and its conjugate
\be \lb{aux2}
3F+\bar{X}(4F'+7\bar{F}')+4\bar{X}X\bar{F}'' +\frac{1}{3}\bar{F}''\bar{R}_*=0
\ee
where $F=F(X)$ and $\bar{F}=\bar{F}(\bar{X})$. The algebraic equations 
(\ref{aux1}) and (\ref{aux2}) cannot be explicitly solved for $X$ in a generic 
$F(\car)$ supergravity.

To  recover the standard (pure) supergravity in our approach, let us 
consider the simple special case when 
\be \lb{case1} F''=0 \qquad {\rm or,~equivalently,} \qquad 
F(\car)=f_0-\frac{1}{2}f_1\car \ee
with some complex constants $f_0$ and $f_1$,  where ${\rm Re}f_1>0$.  
Then eq.~(\ref{aux1}) is  easily solved as
\be \lb{sol1}
\bar{X} =\fracmm{3f_0}{5({\rm Re}f_1)} \ee
Substituting this solution back into the Lagrangian (\ref{expa}) yields
\be \lb{gract}
L= -\frac{1}{3}({\rm Re}f_1)R + \fracmm{9\abs{f_0}^2}{5({\rm Re}f_1)}
\equiv  -\frac{1}{2} M^2_{\rm Pl} R -\L \ee
where we have introduced the reduced Planck mass $M_{\rm Pl}$, and 
the cosmological constant $\L$ as 
\be \lb{fconst}
  {\rm Re}f_1= \frac{3}{2} M^2_{\rm Pl} \qquad {\rm and} \qquad
  \L = \fracmm{-6\abs{f_0}^2}{5M^2_{\rm Pl}} \ee
It is the standard pure supergravity with a {\it negative} cosmological 
constant \cite{sb1,sb2,sb3}.

\section{Generic $\car^2$ supergravity, and AdS Bound}

The simplest non-trivial $F(\car)$ supergravity is obtained by choosing 
$F''=const.\neq 0$
that leads to the $\car^2$-supergravity defined by a generic {\it quadratic} 
polynomial in terms of the scalar supercurvature \cite{us8}.

Let us recall that the stability conditions in $f(R)$-gravity are given by 
eqs.~(\ref{stab}) in the notation (\ref{fgra}). In the notation (\ref{expa})
used here, ie. when $f(R)=-\ha M^2_{\rm Pl} \tilde{f}(R)$, one gets the 
opposite signs,
\be \lb{stabc1} f'(R)<0 \ee
and
\be \lb{stabc2}  f''(R)>0
\ee 
The first (classical stability) condition (\ref{stabc1}) is related to the 
sign factor in front of the Einstein-Hilbert term (linear in $R$) in  the 
$f(R)$-gravity action, and it ensures that graviton is not a ghost. The 
second (quantum stability) condition (\ref{stabc2}) guarantees that scalaron 
is not a tachyon.

Being interested in the inflaton (scalaron) part of the bosonic $f(R)$-gravity 
action that follows from eq.~(\ref{expa}), we set gravitino to zero and the scalar $X$
to be real, which also implies the real $R$ or $R_*=R$. 

In $F(R)$ supergravity the stability condition (\ref{stabc1}) is now replaced by 
a stronger  condition,
\be \lb{cs}
F'(X) < 0 
\ee
It is easy to verify that eq.~(\ref{stabc1}) follows from eq.~(\ref{cs}) 
because of eq.~(\ref{auxe}). Equation (\ref{cs}) also ensures the classical 
stability of the bosonic $f(R)$ gravity embedding into the full $F(\car)$ 
supergravity against small fluctuations of the axion field.

In this Section we investigate a generic {\it quadratically generated}  Ansatz 
(with $F''=const.\neq 0$) that leads to the simplest non-trivial toy-model of 
$F(R)$ supergravity with the master function
\be \lb{an}
 F(\car) = f_0   -\fracmm{1}{2}f_1 \car + \fracmm{1}{2}f_2 \car^2 
\ee
having three coupling constants $f_0$, $f_1$ and $f_2$. We take all of them to 
be real, since we ignore this potential source of $CP$-violation here (see, 
however, the Outlook in Sec.~20).  As regards the mass dimensions 
of the quantities introduced, we have
\be \lb{dims}
[F]=[f_0]=3~, \quad [R]=[f_1]=2~, \quad{\rm and}\quad [\car]=[f_2]=1 
\ee

The bosonic Lagrangian (\ref{expa}) with the function (\ref{an}) reads
\be \lb{simple}
(-g)^{-1/2}L_{\rm bos}= 11 f_2 X^3 - 7f_1X^2 + \left( \frac{2}{3}f_2R 
+6f_0\right)X -\frac{1}{3}f_1R
\ee
Hence, the auxiliary field equation (\ref{auxe}) takes the form of a 
{\it quadratic} equation,
\be \lb{xquad}
\frac{33}{2}f_2 X^2 - 7f_1X + \frac{1}{3}Rf_2 + 3f_0 =0
\ee
whose solution is given by
\be \lb{xquads}
 X_{\pm} =  \fracmm{7}{3\cdot 11} \left[ \fracmm{f_1}{f_2} \pm 
\sqrt{ \fracmm{2\cdot 11}{7^2} (R_{\rm max}-R)}\right]
\ee
where we have introduced the {\it maximal\,} scalar curvature 
\be \lb{max}
R_{\rm max} = 
\fracmm{7^2}{2\cdot 11}\fracmm{f^2_1}{f^2_2} -3^2\fracmm{f_0}{f_2}
\ee
Equation (\ref{xquads}) obviously implies the automatic bound on the scalar
curvature (from one side only). In our notation, it corresponds to the (AdS) 
bound on the scalar curvature from above,
\be \lb{boun}
R<R_{\rm max}
\ee

The existence of the built-in maximal (upper) scalar curvature (or the AdS 
bound) is a nice bonus of our construction.  It is similar to the factor
$\sqrt{1-v^2/c^2}$ in Special Relativity. Yet another close analogy 
comes from the {\it Born-Infeld} non-linear extension of Maxwell 
electrodynamics, whose (dual) Hamiltonian is proportional to \cite{nlsm}
\be \lb{binf}
\left( 1-\sqrt {1- \vec{E}^2/E_{\rm max}^2-\vec{H}^2/H^2_{\rm max} 
+(\vec{E}\times \vec{H})^2/E_{\rm max}^2H_{\rm max}^2} \, \right)
\ee 
in terms of the electric and magnetic fields $\vec{E}$ and $\vec{H}$, 
respectively, with their maximal values. For instance, in String Theory
 one has $E_{\rm max}= H_{\rm max}=(2\p\a')^{-1}$ \cite{nlsm}.

Substituting the solution (\ref{xquads}) back into eq.~(\ref{simple}) yields 
the corresponding $f(R)$-gravity Lagrangian  
\be \lb{flag}
\eqalign{
f_{\pm}(R) = & \fracmm{2\cdot 7}{11}\fracmm{f_0f_1}{f_2} 
-\fracmm{2\cdot 7^3}{3^3\cdot 11^2}
\fracmm{f^3_1}{f^2_2} \cr
&  -\fracmm{19}{3^2\cdot 11} f_1R \mp \sqrt{ 
\fracmm{2}{11}}\left(\fracmm{2^2}{3^3}f_2\right)
 \left( R_{\rm max}-R \right)^{3/2}  \, } 
\ee
Expanding eq.~(\ref{flag}) into power series of $R$ yields
\be \lb{taylor}
f_{\pm}(R) = -\L_{\pm} - a_{\pm}R +b_{\pm}R^2 +{\cal O}(R^3)
\ee
whose coefficients are given by 
\be \lb{cosc}
\L_{\pm} = \fracmm{2\cdot 7}{3^2\cdot 11}f_1 \left(
R_{\rm max} - \fracmm{7^2}{2\cdot 3\cdot 11} \fracmm{f_1^2}{f^2_2}\right)
\pm \sqrt{\fracmm{2}{11}}\left(\fracmm{2^2}{3^3}f_2\right)R^{3/2}_{\rm max}
\ee
\be \lb{einhil}
a_{\pm} = \fracmm{19}{3^2\cdot 11}f_1 \mp \sqrt{\fracmm{2}{11}R_{\rm max}}
\left(\fracmm{2}{3^2}f_2\right)
\ee
and
\be \lb{2ndc}
b_{\pm} = \mp \sqrt{\fracmm{2}{11R_{\rm max}}} \left( \fracmm{f_2}{2\cdot 3^2}
\right) 
\ee

Those equations greatly simplify when $f_0=0$. One finds \cite{us5,us8}
\be \lb{fgro}
f^{(0)}_{\pm}(R) = \fracmm{-5\cdot 17 M^2_{\rm Pl} }{2\cdot 3^2\cdot 11} R
+ \fracmm{2\cdot 7}{3^2\cdot 11}M^2_{\rm Pl} 
\left(R - R_{\rm max} \right)\left[ 1\pm \sqrt{1-R/R_{\rm max} } \; \right] 
\ee 
where we have chosen
\be \lb{sol11}
f_1= \fracmm{3}{2}M^2_{\rm Pl}
\ee
in order to get the standard normalization of the Einstein-Hilbert term that 
is linear in $R$. Then, in the limit $R_{\rm max}\to +\infty$, both functions 
$f^{(0)}_{\pm}(R)$ reproduce General Relativity. In another limit $R\to 0$, one
finds a {\it vanishing} or {\it positive} cosmological constant,
\be \lb{cc}
\L^{(0)}_- =0 \qquad {\rm and}\qquad  
\L^{(0)}_+ = \fracmm{2^2\cdot 7}{3^2\cdot 11}M^2_{\rm Pl}R_{\rm max}
\ee

The stability conditions are given by eqs.~(\ref{stabc1}), (\ref{stabc2}) and 
(\ref{cs}), while the 3rd condition implies the 2nd one. In our case 
(\ref{flag}) we have
\be \lb{1der}
f'_{\pm}(R) = -\fracmm{19}{3^2\cdot 11} f_1 \pm \sqrt{\fracmm{2}{11}}\left(
\fracmm{2}{3^2}f_2 \right)\sqrt{R_{\rm max}-R}~ < 0
\ee
and
\be \lb{2der}
f''_{\pm}(R) = \mp \left( \fracmm{f_2}{3^2}\right)\sqrt{
\fracmm{2}{11(R_{\rm max}-R)}}~>0
\ee
while eqs.~(\ref{cs}), (\ref{an}) and  (\ref{xquads}) yield 
\be \lb{csf}
\pm \sqrt{ \fracmm{2\cdot 11}{7^2}(R_{\rm max}-R)}~ < \fracmm{19}{2\cdot 7}
\fracmm{f_1}{f_2}
\ee

It follows from eq.~(\ref{2der}) that 
\be \lb{lasth}
f_2^{(+)} < 0 \qquad {\rm and}\qquad f_2^{(-)} > 0 
\ee
Then the stability condition (\ref{stabc2}) is obeyed for any value of $R$.

{\it As regards the $(-)$-case,} there are {\it two} possibilities depending 
upon the sign of $f_1$. Should $f_1$ be {\it positive}, all the remaining 
stability conditions are automatically satisfied, ie. in the case of both 
$f_2^{(-)}>0$ and $f_1^{(-)}>0$.

Should $f_1$ be {\it negative,} $f_1^{(-)}<0$, we find that the remaining 
stability conditions (\ref{1der}) and (\ref{csf}) are {\it the same}, as they 
should, while they  are both given by
\be \lb{ssc}
R < R_{\rm max} -\fracmm{19^2}{2^3\cdot 11}\fracmm{f_1^2}{f_2^2} =
-\fracmm{3\cdot 5}{2^3\cdot 11}\fracmm{f_1^2}{f_2^2} -3^2\fracmm{f_0}{f_2}
\equiv R_{\rm max}^{\rm ins}
\ee

{\it As regards the $(+)$-case,} eq.~(\ref{csf}) implies that $f_1$ should
 be {\it negative}, $f_1<0$, whereas then eqs.~(\ref{1der}) and (\ref{csf}) 
result in {\it the same} condition (\ref{ssc}) again.

Since $R_{\rm max}^{\rm ins} < R_{\rm max}$, our results imply that the 
instability happens {\it before} $R$ reaches $R_{\rm max}$ in all cases with 
negative $f_1$.

As regards the particularly simple case (\ref{fgro}), the stability conditions 
allow us to choose the lower sign only.

A different example arises with a negative $f_1$. When choosing the lower 
sign (ie. a positive $f_2$) for definiteness, we find
\be \lb{mpc}
\eqalign{
f_{-}(R) = & -\fracmm{2\cdot 7}{11}f_0 \abs{\fracmm{f_1}{f_2}} +
\fracmm{2\cdot 7^3}{3^3\cdot 11^2}\abs{\fracmm{f^3_1}{f^2_2}} \cr
&  +\fracmm{19}{3^2\cdot 11} \abs{f_1}R + 
\sqrt{ \fracmm{2}{11}}\left(\fracmm{2^2}{3^3}f_2\right)
 \left( R_{\rm max}-R \right)^{3/2}  \, } 
\ee
Demanding the standard normalization of the Einstein-Hilbert term in this case 
implies
\be \lb{norm2}
R_{\rm max} = \fracmm{3^4\cdot 11}{2^3 f^2_2}
\left( \fracmm{M^2_{\rm Pl}}{2}+\fracmm{19}{3^2\cdot 11}\abs{f_1}\right)^2
\ee
where we have used eq.~(\ref{einhil}). It is easy to verify by using 
eq.~(\ref{cosc}) that the cosmological constant is always {\it negative} in 
this case, and the instability bound (\ref{ssc}) is given by
\be \lb{insb}
R_{\rm max}^{\rm ins} = \fracmm{3^4\cdot 11 M^2_{\rm Pl}}{2^3f_2^2}\left(
\fracmm{M^2_{\rm Pl}}{2^2} + \fracmm{19\abs{f_1}}{3^2\cdot 11}\right) 
< R_{\rm max}
\ee

The $f_{-}(R)$ function of eq.~(\ref{flag}) can be rewritten to the form
\be \lb{cho}
f(R)= \fracmm{7^3}{3^3\cdot 11^2}\fracmm{f_1^3}{f_2^2}
-\fracmm{2\cdot 7}{3^2\cdot 11}
f_1R_{\rm max}-\fracmm{19}{3^2\cdot 11}f_1R
 + f_2\sqrt{ \fracmm{2^5}{3^6\cdot 11}} ( R_{\rm max}-R)^{3/2}
\ee
where we have used eq.~(\ref{max}). There are {\it three} physically 
different regimes:

(i) the {\it high-curvature regime,} 
$R<0$ and $\abs{R}\gg R_{\rm max}$. Then eq.~(\ref{cho}) implies
\be \lb{hcre} 
f(R) \approx -\L_h - a_hR +c_h\abs{R}^{3/2}
\ee
whose coefficients are given by
\be \lb{hcrec}
\eqalign{
\L_h = &~ \fracmm{2\cdot 7}{3^2\cdot 11}f_1R_{\rm max} - 
\fracmm{7^3}{3^3\cdot 11^2}\fracmm{f_1^3}{f_2^2}~~, \cr
a_h = &~ \fracmm{19}{3^2\cdot 11}f_1~~, \cr
c_h = &~ \sqrt{ \fracmm{2}{ 11 } } \left( \fracmm{2^2}{3^3}f_2 \right) \cr}
\ee

(ii) the {\it low-curvature regime,} $\abs{R/R_{\rm max}}\ll 1$. Then 
eq.~(\ref{cho}) implies
\be \lb{lcre}
f(R) \approx -\L_l - a_lR~,
\ee
whose coefficients are given by
\be \lb{lcc}
\eqalign{
\L_l = &~ \L_h -\sqrt{ \fracmm{2R^3_{\rm max}}{11} }\left( 
\fracmm{2^2}{3^3}f_2\right)~~, 
\cr
a_l = &~ a_h +\sqrt{\fracmm{2R_{\rm max}}{11}}\left(\fracmm{2}{3^2}f_2\right)=
a_{-}=\fracmm{M^2_{\rm Pl}}{2}~~, \cr }
\ee
where we have used eq.~(\ref{einhil}).

(iii) the {\it near-the-bound regime} (assuming that no instability happens 
before it), $R=R_{\rm max}+\d R$, $\d R <0$, and $\abs{\d R/R_{\rm max}}\ll 1$.
 Then eq.~(\ref{cho}) 
implies 
\be \lb{nbre} f(R) \approx - \L_b +a_b\abs{\d R} +c_b \abs{\d R}^{3/2}
\ee
whose coefficients are
\be \lb{nbrec}
\eqalign{
\L_b = &~ \fracmm{1}{3}f_1R_{\rm max} - 
\fracmm{7^3}{3^3\cdot 11^2}\fracmm{f_1^3}{f_2^2}~~, \cr
a_b = &~ a_h~~, \cr
c_b = &~ \sqrt{ \fracmm{2}{11} }\left( \fracmm{2^2}{3^3}f_2\right)
\cr }
\ee

The cosmological dynamics may be either directly derived from the gravitational
equations of motion in the $f(R)$-gravity with a given function $f(R)$, or just
read off from the form of the corresponding scalar potential of a scalaron 
(see below). For instance, as was demonstrated in ref.~\cite{us5} for the 
special case $f_0=0$, a cosmological expansion is possible in the regime (i) 
towards the regime (ii), and then, perhaps, to the regime (iii) unless an 
instability occurs.

However, one should be careful since our toy-model (\ref{an}) does not pretend 
to be viable in the low-curvature regime, eg., for the present Universe. 
Nevertheless, if one wants to give it some physical meaning  there, by 
identifying it with General Relativity, then one should also fine-tune 
the cosmological constant $\L_l$ in eq.~(\ref{lcc})  to be ``small'' 
and positive. We find that it amounts to 
\be \lb{grb}
R_{\rm max} \approx \fracmm{3^4\cdot 7^2\cdot 11}{2^5\cdot 19^2}
\fracmm{M^4_{\rm Pl}}{f_2^2}
\equiv R\low{\L=0}
\ee 
with the actual value of $R_{\rm max}$ to be ``slightly'' above of that bound, 
$R_{\rm max}>R\low{\L=0}$. It is also posssible to have the vanishing 
cosmological constant, $\L_l=0$, when choosing $R_{\rm max}=R\low{\L=0}$. 
It is worth mentioning that it relates the values of $R_{\rm max}$ and $f_2$.

The particular $\car^2$-supergravity model (with $f_0=0$) was introduced in 
ref.~\cite{us5} in an attempt to get viable embedding of the Starobinsky model 
into $F(\car)$-supergravity. However, it failed because, as was found in 
ref.~\cite{us5}, the higher-order curvature terms cannot be ignored in 
eq.~(\ref{fgro}), ie. the $R^n$-terms with $n\geq 3$ are not small enough 
against the $R^2$-term. In fact, the possibility of destabilizing the 
Starobinsky inflationary scenario by the terms with higher powers of the scalar
 curvature, in the context of $f(R)$ gravity, was noticed earlier in 
refs.~\cite{ma2,bma}. The most general Ansatz (\ref{an}), which is mostly
{\it quadratic} in the supercurvature, does not help for that purpose either.

For example, the full $f(R)$-gravity function $f_-(R)$ in eq.~(\ref{fgro}), 
which we derived from our $\car^2$-supergravity, gives rise to the inflaton 
scalar potential
\be \lb{expot}
V(y) = V_0 \left( 11e^y +3\right)\left(e^{-y}-1\right)^2
\ee
where $V_0= (3^3/2^6)M^4_{\rm Pl}/f_2^2$. The corresponding inflationary  
parameters 
\be \lb{exeps}
\ve (y) = \fracmm{1}{3} \left[ \fracmm{ e^y\left( 11+ 11e^{-y} +6 e^{-2y}
\right) }{ (11e^y+3)
(e^{-y}-1)} \right]^2  \geq \fracmm{1}{3} \ee
and
\be \lb{exeta}
\eta(y)= \fracmm{2}{3} \fracmm{\left(11 e^y+5e^{-y}+12 e^{-2y}\right) }{ 
(11e^y +3) (e^{-y}-1)^2 } \geq \fracmm{2}{3}
\ee
are not small enough for matching the WMAP observational data. A solution to 
this problem is given in the next section. 

\section{Chaotic inflation in $F(\car)$ Supergravity}

Let us further generalize our Ansatz and consider a new $F(\car)$ function 
having the {\it cubic} form
\be \lb{cub}
F(\car)= -\frac{1}{2}f_1 \car + \frac{1}{2}f_2 \car^2 -\frac{1}{6}f_3\car^3
\ee
whose real (positive) coupling constants $f_{1,2,3}$ are of (mass) dimension 
$2$, $1$ and $0$, respectively.  Our conditions on the coefficients are
\be \lb{cco}
 f_3 \gg 1~,\qquad f_2^2 \gg f_1  
\ee 
The first condition is needed to have inflation at the curvatures much less 
than $M^2_{\rm Pl}$ (and to meet observations), while the second condition is 
needed to have the scalaron (inflaton) mass be much less than $M_{\rm Pl}$, in 
order to avoid large (gravitational) quantum loop corrections  after the end of
 inflation up to the present time.

The bosonic action is given by eq.~(\ref{expa}). For a real scalaron it reduces to
\be \lb{rede}
L/\sqrt{-g} = 2F'\left[ \fracmm{1}{3}R +4X^2\right] + 6XF
\ee 
so that the real auxiliary field is a solution to the algebraic equation
\be \lb{realx}
3F +11F'X + F''\left[ \fracmm{1}{3}R +4X^2\right] =0
\ee

Stability of the bosonic embedding in supergravity requires $F'(X)<0$ (Sec.~9).
 In the case (\ref{cub}) it gives rise to the condition $f_2^2 < f_1f_3$. 
For simplicity here, we will assume a stronger condition,
\be \lb{adcon}
f_2^2\ll f_1f_3 
\ee
Then the second term on the right-hand-side of eq.~(\ref{cub}) will not affect 
inflation, as is shown below. However, it will be quite important for reheating 
(see Secs.~13 and 14).

Equation (\ref{rede}) with the Ansatz (\ref{cub}) reads
\be \lb{bos3}
L = -5f_3X^4 + 11f_2 X^3 - (7f_1 +\frac{1}{3}f_3R)X^2 +\frac{2}{3}f_2RX 
-\frac{1}{3}f_1R
\ee 
and gives rise to a cubic equation on $X$,
\be \lb{aux3} 
X^3 -\left( \fracmm{33f_2}{20f_3}\right)X^2 +\left( \fracmm{7f_1}{10f_3} 
+\fracmm{1}{30}R\right)X - \fracmm{f_2}{30f_3}R =0
\ee
We find three consecutive (overlapping) regimes.
\begin{itemize}
\item The high curvature regime including inflation  is given by 
\be \lb{reg1}
\d R<0 \quad {\rm and} \quad 
\fracmm{\abs{\d R}}{R_0}\gg \left(\fracmm{f^2_2}{f_1f_3}\right)^{1/3}
\ee
where we have introduced the notation $R_0=21f_1/f_3>0$ and $\d R = R+R_0$. 
With our sign conventions we have $R<0$ during the de Sitter and matter 
dominated stages. In the regime (\ref{reg1})  the $f_2$-dependent terms in 
eqs.~(\ref{bos3}) and  (\ref{aux3}) can be neglected, and we get
\be \lb{aux31}
X^2 = -\frac{1}{30} \d R
\ee
and
\be \lb{lag1}
L = -\fracmm{f_1}{3}R + \fracmm{f_3}{180}(R+R_0)^2
\ee
It closely reproduces the Starobinsly inflationary model (Sec.~2) since 
inflation occurs at $\abs{R}\gg R_0$. In particular, we can identify
\be \lb{f3m}
f_3=\fracmm{15M^2_{\rm Pl}}{M^2_{\rm inf}}
\ee
It is worth mentioning that we cannot simply set $f_2=0$ in 
eq.~(\ref{cub}) because it would imply $X=0$ and $L=-\frac{f_1}{3}R$ 
for $\d R>0$. As a result of that the scalar degree of freedom would disappear 
that would lead to the breaking of a regular Cauchy evolution. Therefore, the 
second term in eq.~(\ref{cub}) is needed to remove that degeneracy.

\item The intermediate (post-inflationary) regime is given by
\be \lb{reg2} 
 \fracmm{\abs{\d R}}{R_0}\ll 1
\ee
In this case $X$ is given by a root of the cubic equation
\be \lb{root2}
 30 X^3 +(\d R)X +\fracmm{f_2R_0}{f_3}=0
\ee
It also implies that the 2nd term in eq.~(\ref{aux3}) is always small.
Equation (\ref{root2}) reduces to eq.~(\ref{aux31}) under the conditions 
(\ref{reg1}).
\item The low-curvature regime (up to $R=0$) is given by
\be \lb{reg3}
\d R>0 \quad {\rm and} \quad 
\fracmm{\d R}{R_0}\gg \left(\fracmm{f^2_2}{f_1f_3}\right)^{1/3}
\ee
It yields
\be \lb{alow}
X = \fracmm{f_2R}{f_3(R+R_0)}
\ee
and
\be \lb{lag3}
L = -\fracmm{f_1}{3}R + \fracmm{f_2^2R^2}{3f_3(R+R_0)} 
\ee
It is now clear that $f_1$ should be equal to $3M_{Pl}^2/2$
in order to obtain the correctly normalized Einstein gravity 
at $|R|\ll R_0$. In this regime the scalaron mass squared 
 is given by
\be \lb{smass}
 \fracmm{1}{3\abs{f''(R)}}=\fracmm{f_3R_0M_{\rm Pl}^2}{4f_2^2}
= \fracmm{21f_1}{4f_2^2}M_{\rm Pl}^2 = \fracmm{63M_{\rm Pl}^4}{8f_2^2}
\ee
in agreement with the case of the absence of the ${\cal R}^3$ term,
studied in the previous section. The scalaron mass squared (\ref{smass}) is
much less than $M_{Pl}^2$ indeed, due to the second inequality in 
eq.~(\ref{cco}), but it is  much more than one at the end of inflation 
$(\sim M^2)$.
\end{itemize}

It is worth noticing that the corrections to the Einstein action in
eqs.~(\ref{lag1}) and (\ref{lag3})  are of ther same order (and small) 
at the borders of the intermediate region (\ref{reg2}).

The roots of the cubic equation (\ref{aux3}) are given by the textbook 
(Cardano) formula \cite{abs}, though that formula is not very illuminating 
in a generic case. The Cardano formula greatly simplifies in the most 
interesting (high curvature) regime where inflation takes place, and the 
Cardano discriminant is
\be \lb{disc}
D \approx \left( \fracmm{R}{90}\right)^3 < 0
\ee
It implies that all three roots are real and unequal. The Cardano formula 
yields the 
roots
\be \lb{card}
X_{1,2,3}\approx \fracmm{2}{3}\sqrt{ \fracmm{-R}{10} } \cos
\left( \fracmm{27}{4f_3 \sqrt{-10R/f^2_2}} + C_{1,2,3} \right) 
+\fracmm{11f_2}{20f_3}
\ee 
where the constant $C_{1,2,3}$ takes the values $(\p/6,5\p/6,3\p/2)$. 

As regards the leading terms, eqs.~(\ref{bos3}) and (\ref{card}) result in the 
$(-R)^{3/2}$ correction to the $(R+R^2)$-terms in the effective Lagrangian in 
the  high-curvature regime $|R|\gg f_2^2/f_3^2$. In order to verify that this 
correction does not change our results under the conditions 
(\ref{reg1}), let us consider the $f(R)$-gravity model with
\be  \lb{try}  \tilde{f}(R) = R - b(-R)^{3/2} - aR^2 \ee
whose parameters $a>0$ and $b>0$ are subject to the conditions $a\gg 1$ and  
$b/a^2\ll 1$. It is easy to check that $\tilde{f}'(R)>0$ 
for $R\in (-\infty,0]$, as  is needed for (classical) stability.

Any $f(R)$ gravity model is classically equivalent to the scalar-tensor 
gravity with certain scalar potential (Sec.~3). The scalar potential can be 
calculated from a given function $f(R)$ along the standard lines (Sec.~3). 
We find (in the high curvature regime)
\be \lb{poten}
V(y) = \fracmm{1}{8a} \left( 1- e^{-y}\right)^2 +
\fracmm{b}{8\sqrt{2a}}e^{-2y}\left(e^y-1\right)^{3/2}
\ee
in terms of the inflaton field $y$. The first term of this equation is the 
scalar potential associated with the pure $(R+R^2)$ model, and the 2nd term is 
the correction due to the $R^{3/2}$-term in eq.~(\ref{try}). It is now clear 
that for large positive $y$ the vacuum energy in  the first term dominates and 
drives inflation until the vacuum energy is compensated by the $y$-dependent 
terms near  $e^y=1$.   

It can be verified along the lines of ref.~\cite{mchi} that 
the formula for scalar perturbations remains the same as that for
the model (\ref{qua}), ie. $\Delta_{\cal R}^2\approx 
N^2M^2_{\rm inf}/(24\pi^2M_{\rm Pl}^2)$, where $N$ is the number of e-folds 
from the end of inflation. So, to fit the observational data, one 
has to choose
\be \lb{f3v}
 f_3 \approx 5N_e^2/(8\pi^2 \Delta_{\cal R}^2)\approx
6.5\cdot 10^{10} (N_e/50)^2
\ee
Here the value of $\Delta_{\cal R}$ is taken from ref.~\cite{wmap} and the 
subscript ${\cal R}$ has a different meaning from the rest of this review.

We conclude that the model (\ref{cub}) with a sufficiently small $f_2$ obeying 
the conditions (\ref{cco}) and (\ref{adcon}) gives a viable realization of the
 chaotic $(R+R^2)$-type inflation in supergravity. The only significant 
difference with respect to the original $(R+R^2)$ inflationary model is the 
scalaron mass that becomes much larger than $M$ in supergravity, soon after the 
end of inflation when $\d R$ becomes positive. It makes the scalaron decay faster 
and creation of the usual matter (reheating) more effective.

The whole series in powers of ${\car}$ may also be considered, instead of the 
limited Ansatz (\ref{cub}). The only necessary condition for embedding 
inflation is that $f_3$ should be anomalously large. When the curvature grows,
 the $\car^3$-term should become important much earlier than the convergence 
radius of the whole series without that term. Of course, it means that viable
 inflation does not occur for any function $F(\car)$ but only inside a small 
region of measure zero in the space of all those functions. However, the 
same is true for all known inflationary models, so the very existence of 
inflation has to be taken from the observational data, not from a pure thought.

The results of this section can be considered as the viable alternative to the 
earlier proposals \cite{jap2,bdva} for realization of chaotic 
inflation in supergravity. But inflation is not the only target of our 
construction. As is well known \cite{star1,star2,star3}, the scalaron decays 
into pairs of particles and anti-particles of quantum matter fields, while its 
decay into gravitons is strongly suppressed \cite{s81}. It thus represents 
the {\it universal mechanism of viable reheating} after inflation and provides 
a transition to the subsequent hot radiation-dominated stage of the universe 
evolution. In its turn, it leads to the standard primordial nucleosynthesis 
(BBN) after. In $F(R)$ supergravity the scalaron has a pseudo-scalar 
superpartner that may be the source of a strong $CP$-violation and then, subsequently, 
lepto- and baryo-genesis that may lead to baryon (matter-antimatter) asymmetry 
\cite{fy,dineku,buchm,desy} --- see Sec.~20 for more.

\section{More about Inflationary Dynamics in our Model}

%%%%%%%%%%%%%%%%%%%%%%%%%%%%%%%%%%%%%%%%%%%%%%%%%%%%%%%%%

\newcommand{\op}{\`}
\def\be{\begin{equation}}
\def\ee{\end{equation}}
\def\ba{\begin{eqnarray}}
\def\ea{\end{eqnarray}}
\def\bs{\begin{subequations}}
\def\es{\end{subequations}}
\def\tX{\tilde{X}}
\def\tv{\tilde{v}}
\def\tT{\tilde{T}}
\def\ce{{\cal E}}
\def\cy{{\cal Y}}
\def\l{\lambda}
\def\j{\psi}
\def\slpa{\slash{\pa}}       
\def\fracmm#1#2{{{#1}\over{#2}}}
\def\dt#1{{\buildrel {\hbox{\LARGE .}} \over {#1}}}   

\renewcommand{\S}{{\text{\tiny $\phi$}}}
\newcommand{\rd}{{\rm d}}
\newcommand{\rr}{{\rm rad}}
\def\d{{\rm d}}
\newcommand{\T}{{\text{\tiny $T$}}}
\newcommand{\sH}{{\text{\tiny $\phi H$}}}
\newcommand{\sV}{{\text{\tiny $\phi V$}}}
\newcommand{\tH}{{\text{\tiny $TH$}}}
\newcommand{\tV}{{\text{\tiny $TV$}}}
\renewcommand{\H}{{\text{\tiny $H$}}}
\newcommand{\V}{{\text{\tiny $V$}}}
\newcommand{\eff}{{\text{\tiny eff}}}
\newcommand{\vp}{\varphi}
\newcommand{\tk}{\tilde{k}}

\def\R{\cal{R}}

\def\cob{\color{blue}}
\def\cor{\color{red}}

\def\car{{\cal R}}
\def\cy{{\cal Y}}

\renewcommand{\a}{\alpha}
\renewcommand{\b}{\beta}
\renewcommand{\c}{\gamma}
\renewcommand{\d}{\delta}
\newcommand{\la}{\lambda}
\newcommand{\nn}{\nonumber\\}
\newcommand{\pp}[1]{(\ref{#1})}

%%%%%%%%%%%%%%%%%%%%%%%%%%%%%%%%%%%%%%%%%%%%%%%%%%%%%%%%%%%%%%%

The supersymmetric extension of the simplest $R^2$-type inflationary model
in the previous section has some important improvements against the original 
Starobinsky's model, because it is characterized by {\it two} mass scales of a 
scalar degree of freedom (scalaron): $M$ (associated with the inflationary era) 
and $m$ (associated with the preheating era).~\footnote{Compared to the earlier sections, 
we rescale $M$ by the factor of $\sqrt{6}$ here, in order to make it equal to the scalaron 
mass during inflation.} They correspond to two free real parameters  $f_2$ and $f_3$ in our 
Ansatz \pp{cub}. The allowed values of the masses $M$ and $m$ can be derived from the 
amplitude of the CMB temperature anisotropies. In the previous section the viability of our 
model was established only in certain limit of its parameter space. Here we show that our 
model is consistent with the joint observational constraints of the WMAP and the PLANCK in 
the 
regime where a sufficient amount of inflation (with the number of e-foldings larger than 50) 
is realized. We also find observational bounds on the parameter values. In the low-energy 
regime 
relevant to preheating, we derive the effective scalar potential in the presence of a 
pseudo-scalar field $\chi$ coupled to the inflaton (scalaron) field 
$\phi$ (the field $\chi$ was ignored in the previous section). This potential is employed for 
numerical analysis of the preheating stage after inflation. If $m$ is much larger than $M$, 
we find that there exists the preheating stage in which the field perturbations $\delta \chi$ 
and $\delta \phi$ rapidly grow by a broad parametric resonance by which the both field 
perturbations $\delta \chi$ and $\delta \phi$ are amplified (Sec.~14). The dynamics of 
reheating appears to be different from that in the original Starobinsky's $f(R)$ model and, 
in fact, more efficient.

In order to recover the standard behaviour of General Relativity in the low-energy regime 
we require that $f_1=3M_{\rm Pl}^2/2$. The mass squared of the scalar degree of 
freedom is given by $m^2=1/(3f''(R))$, where $f(R)$ is related to the Lagrangian $L(R)$ as 
$L(R)=-M_{\rm Pl}^2 f(R)/2$. According to Sec.~10 in the limit $|R| \ll R_0$ we have  
\be
m^2=\fracmm{21f_1 M_{\rm Pl}^2}{4f_2^2}
=\fracmm{63M_{\rm Pl}^4}{8f_2^2}
\label{masslow}
\ee
In the high-curvature regime the scalaron mass squared is given by
\be
M^2=\fracmm{15M_{\rm Pl}^2}{f_3}
\ee
Hence, the constants $f_{1,2,3}$ can be expressed by using 
the three mass scales $M_{\rm Pl}$, $m$, and $M$, as follows:
\be \label{repar}
f_1=\fracmm{3}{2} M_{\rm pl}^2\,,\quad
f_2=\sqrt{\fracmm{63}{8}} \fracmm{M_{\rm Pl}^2}{m}\,,\quad
f_3=\fracmm{15M_{\rm Pl}^2}{M^2}\
\ee

The conditions $f_2^2<f_1f_3$, $f_3\gg 1$ and $f^2_2\gg f_1$ of Sec.~10
translate into
\be
m>\sqrt{\frac{7}{20}}\,M\,,\qquad
M  \ll M_{\rm Pl}\,,\qquad
m \ll M_{\rm Pl}\,~,
\label{mcon}
\ee
respectively.

The high-energy regime (A) satisfies the condition $|R| \gg R_0$ with
the flat FLRW background described by the line element $ds^2=dt^2-a^2(t) d{\bm x}^2$.
It is convenient to introduce the following dimensionless functions:
\be
\alpha \equiv \fracmm{M^2}{mH}\,,\qquad
\beta \equiv \fracmm{M^2}{H^2}
\label{albe}
\ee
and represent $R_0$ as $R_0=21M^2/10$. During inflation the functions (\ref{albe}) 
should satisfy the conditions $\alpha \ll 1$ and $\beta \ll 1$ (see below).
In eq.~(\ref{lag1}) the term $f_3 R^2/180$ is the dominant 
contribution during inflation. Hence, we neglect the higher-order terms
beyond that of the first (linear) order in $\alpha$ and $\beta$.
Then the Lagrangian following from eq.~(\ref{lag1}) is given by
\be
f(R) \simeq \fracmm{3}{10}R-\fracmm{R^2}{6M^2}
-\fracmm{3\sqrt{105}}{100} \fracmm{(-R)^{3/2}}{m}\,.
\label{fRinf}
\ee
We assume that the Lagrangian (\ref{fRinf}) is valid by the end of inflation.

In the flat FLRW spacetime the field equations of motion are
\ba
3 {\cal F} H^2 &=& (f-R {\cal F})/2-3H \dot{\cal F}\,,
\label{be1} \\
-2{\cal F} \dot{H} &=& \ddot{{\cal F}}-H \dot{\cal F}\,,
\label{be2}
\ea
where ${\cal F} \equiv f'(R)$.
It is useful to define the new slow-roll parameters as \cite{fgrev2}
\be
\epsilon_1 \equiv -\fracmm{\dot{H}}{H^2}\,,\qquad
\epsilon_2 \equiv \fracmm{\dot{{\cal F}}}{2H{\cal F}}\,,\qquad
\epsilon_3 \equiv \fracmm{\ddot{{\cal F}}}{H \dot{{\cal F}}}
\ee
which satisfy $|\epsilon_i| \ll 1$ ($i=1,2,3$).
It follows from eq.~(\ref{be2}) that 
\be
\epsilon_1=-\epsilon_2 (1-\epsilon_3)
\label{ep123}
\ee
In what follows we carry out the linear expansion in terms of 
the variables $\epsilon_i$ ($i=1,2,3$), $\alpha$, $\beta$, and 
$s \equiv \ddot{H}/(H \dot{H})$.

For the Lagrangian (\ref{fRinf}) we have
\ba
{\cal F} &=& \fracmm{4H^2}{M^2} \left( 1+\frac{27 \sqrt{35}}{400} \alpha
+\frac{3}{40} \beta -\frac12 \epsilon_1 \right)\,,\label{Fva} \\
\dot{\cal F} &=& -\fracmm{8H^3}{M^2} \epsilon_1
\left( 1+\frac{27 \sqrt{35}}{800} \alpha
+\frac14 s \right)
\ea
Then the variable $\epsilon_2$ is given by 
\be
\epsilon_2=-\epsilon_1 \left( 1-\frac{27 \sqrt{35}}{800} \alpha
-\frac{3}{40} \beta+\frac12 \epsilon_1 +\frac14 s \right)
\ee
Comparing this with eq.~(\ref{ep123}) we obtain
\be
\epsilon_3=-\frac{27 \sqrt{35}}{800} \alpha
-\frac{3}{40} \beta+\frac12 \epsilon_1 +\frac14 s
\label{ep3d}
\ee
Similarly, eq.~(\ref{be1}) gives the following relations:
\be
\epsilon_1=\frac{3\sqrt{35}}{200} \alpha+\frac{1}{20} \beta
\label{ep1}
\ee
and 
\be
\epsilon_2=-\frac{3\sqrt{35}}{200} \alpha-\frac{1}{20} \beta
\label{ep2}
\ee

Equation (\ref{ep1}) is equivalent to 
\be
\dot{H}=-\fracmm{3\sqrt{35}}{200} \fracmm{M^2}{m}
\left( H+\fracmm{10m}{3\sqrt{35}} \right)
\ee
This differential equation can be easily integrated.  It yields 
\be
H(t)=\left( H_i +\fracmm{10m}{3\sqrt{35}} \right)
\exp \left[ \fracmm{3\sqrt{35}}{200} \fracmm{M^2}{m}
(t_i-t) \right]-\fracmm{10m}{3\sqrt{35}}
\label{Hana}
\ee
where $H_i$ is the initial value of $H$ at $t=t_i$.
So we find
\be
s=-\frac{3\sqrt{35}}{200}\alpha
\label{s}
\ee
Substituting eqs.~(\ref{ep1}) and (\ref{s}) into 
eq.~(\ref{ep3d}) we obtain
\be
\epsilon_3=-\frac{3\sqrt{35}}{100} \alpha
-\frac{1}{20} \beta
\label{ep3}
\ee

The end of inflation ($t=t_f$) is identified
by the condition $\epsilon_1=1$. By using the solution (\ref{Hana}), we have
\ba
\hspace{-0.6cm}
t_i-t_f &=& \fracmm{200m}{3\sqrt{35}M^2}
\ln \biggl(
\fracmm{63M^2}{80m (3\sqrt{35}H_i+10m)}
\nonumber \\
\hspace{-0.6cm}  &\times&
\biggl[ 1+\frac{800}{63} \left( \frac{m}{M} \right)^2
+\sqrt{1+\frac{1600}{63} \left( \frac{m}{M} \right)^2}
\biggr] \biggr)
\label{tfi}
\ea

We define the number of e-foldings from the onset of inflation ($t=t_i$)
to the end of inflation ($t=t_f$) as $N (t_i) \equiv \int_{t_i}^{t_{f}}H\,dt$.
{}From eqs.~(\ref{Hana}) and (\ref{tfi}) we can express $N(t_i)$ in 
terms of $H_i$, $M$, and $m$. The number of e-foldings $N$ corresponding to the time $t$ 
can be derived by replacing $H_i$ in the expression of $N(t_i)$ 
for $H$. It follows that 
\ba
N &=& \fracmm{1}{126\alpha^2}
\biggl[ 3\alpha (80\sqrt{35}-21\alpha-\sqrt{7(63\alpha^2+1600\beta)})
\nonumber \\
&-&400 \beta (8 \ln 2+3 \ln 5)+800\beta 
\nonumber \\
&\times& \ln \biggl( \fracmm{\sqrt{7} (63\alpha^2+800\beta)+
21\alpha\sqrt{63\alpha^2+1600\beta}}{21\alpha+2\sqrt{35} \beta}
\biggr) \biggr]\nonumber \\
\label{efold}
\ea

In the limit $\alpha \to 0$ one has $N \to 10/\beta-1/2$, ie.
$\beta \to 20/(2N+1)$. In this case the $R^2/(6M^2)$ term in the Lagrangian (\ref{fRinf})
dominates over the dynamics of inflation, which corresponds to the Starobinsky's $f(R)$ model.
In another limit $\beta \to 0$ it follows that  $N \to 40\sqrt{35}/(21\alpha)-1$, ie.
$\alpha \to 40\sqrt{35}/[21(N+1)]$. Then we obtain the following bounds on 
$\alpha$ and $\beta$:
\be
0<\alpha<\fracmm{40\sqrt{35}}{21(N+1)}\,~,
\qquad
0<\beta<\fracmm{20}{2N+1}
\label{alcon}
\ee

In order to realize inflation with eg., $N=60$, the two variables need to be in the range
$0<\alpha<0.185$ and $0<\beta<0.165$. For the number of e-foldings relevant to the CMB 
temperature anisotropies ($50 \lesssim N \lesssim 60$) the slow-roll parameters
given in eqs.~(\ref{ep1}), (\ref{ep2}), (\ref{ep3}) are much  smaller than unity, so that 
the slow-roll approximation employed above is justified.

\section{Facing Observational Tests}

In this section we study more closely whether the $f(R)$ model (\ref{fRinf}) satisfies
the observational constraints of the CMB temperature anisotropies. The power spectra of scalar 
and tensor perturbations generated during inflation based on $f(R)$ theories were calculated 
in ref.~\cite{mchi}.

The scalar power spectrum of the curvature perturbation is given by \cite{fgrev2}
\be
{\cal P}_{\rm s}=\fracmm{1}{24\pi^2 {\cal F}}
\left( \fracmm{H}{M_{\rm pl}} \right)^2
\fracmm{1}{\epsilon_2^2}
\label{Ps}
\ee
Using eqs.~(\ref{Fva}), (\ref{ep1}), and (\ref{ep2}), 
it follows that
\be
{\cal P}_{\rm s} \simeq \fracmm{1250}{3\pi^2}
\left( \fracmm{M}{M_{\rm pl}} \right)^2
\left( 3\sqrt{35} \alpha+10\beta \right)^{-2}
\ee
where in the expression of ${\cal F}$ we have neglected 
the terms $\alpha$ and $\beta$ relative to 1. Using the WMAP7 normalization 
${\cal P}_{\rm s}=2.4 \times 10^{-9}$ at the pivot wave number 
$k_0=0.002$~Mpc$^{-1}$ \cite{wmap}, the mass $M$ is constrained to be 
\be
M \simeq 7.5 \times 10^{-6} \left( 3\sqrt{35} \alpha+10\beta
\right) M_{\rm Pl}
\label{WMAPnor}
\ee

In the limit $\alpha \to 0$ and $\beta \to 20/(2N+1)$
we have $M/M_{\rm Pl}=7.5 \times 10^{-4}/(N+1/2)$.
In another limit $\alpha \to 40\sqrt{35}/[21(N+1)]$ 
and $\beta \to 0$ it follows that 
$M/M_{\rm Pl}=1.5 \times 10^{-3}/(N+1)$.
In the intermediate regime characterized by eq.~(\ref{alcon})
we can numerically find the values of $\alpha$ and $\beta$ 
for given $N$ satisfying the constraint (\ref{efold}), 
which allows us to evaluate $M$ from eq.~(\ref{WMAPnor}).
{}From eq.~(\ref{albe}) the mass scale $m$ is also
known by the relation $m=(\sqrt{\beta}/\alpha)M$.

\begin{figure}
\includegraphics[height=3.2in,width=3.3in]{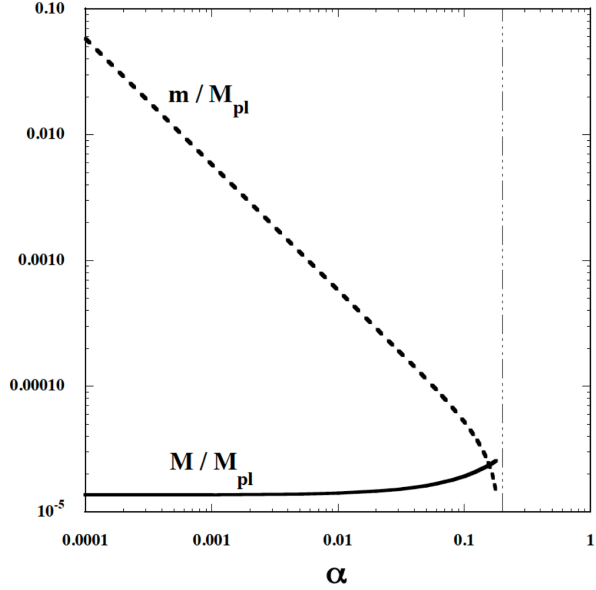}
\caption{\small The two masses $M$ and $m$ 
versus the variable $\alpha$ in the regime 
$10^{-4} \le \alpha \le 0.18$ 
for the number of e-foldings $N=55$. 
We also show the upper bound $\alpha_{\rm max}=0.201$
determined by Eq.~(\ref{alcon}).
$M$ is weakly dependent on $\alpha$
with the order of $10^{-5}M_{\rm Pl}$, whereas
$m$ strongly depends on $\alpha$.
The condition $m>\sqrt{7/20}\,M$ is satisfied
for $\alpha<0.178$.
\label{fig3}}
\end{figure}

In Fig.~\ref{fig3} we plot $M$ and $m$ versus $\alpha$
in the regime $10^{-4} \le \alpha \le 0.18$ for $N=55$. 
In this case $\alpha$ is bounded to be 
$0<\alpha<0.201$ from Eq.~(\ref{alcon}).
The mass $M$ weakly depends on $\alpha$ with 
the order of $10^{-5}M_{\rm Pl}$, while $m$ changes 
significantly depending on the values of $\alpha$.
For $\alpha$ much smaller than 1 we have $m \gg M$, 
while $m$ is of the same order as $M$ for $\alpha \gtrsim 0.1$.
We recall that there is the condition $m>\sqrt{7/20}\,M$.
For $N=55$ this condition gives the upper bound $\alpha<0.178$.

The scalar spectral index $n_{\rm s}$ can be defined by 
$n_{\rm s}=1+d \ln {\cal P}_{\rm s}/d \ln k$, which is evaluated at the Hubble radius 
crossing $k=aH$ (where $k$ is a comoving wave number) \cite{Lidsey,Lidsey2,Bassett}.
In $f(R)$ gravity it is given by \cite{fgrev2}
\be
n_{\rm s}=1-4\epsilon_1+2\epsilon_2-2\epsilon_3
\ee
By using eqs.~(\ref{ep1}), (\ref{ep2}) and (\ref{ep3}), 
we obtain
\be
n_{\rm s} =
1-\frac{3\sqrt{35}}{100}\alpha-\frac15 \beta
\label{nsab}
\ee

The tensor power spectrum is given by \cite{fgrev2}
\be
{\cal P}_{\rm t}=\fracmm{2}{\pi^2 {\cal F}}
\left( \fracmm{H}{M_{\rm pl}} \right)^2
\label{Pt}
\ee
{}From Eqs.~(\ref{Ps}) and (\ref{Pt}) the tensor-to-scalar
ratio is 
\be
r \equiv \frac{{\cal P}_{\rm t}}{{\cal P}_{\rm s}}
=48\epsilon_2^2=\frac{3}{2500}
\left( 3\sqrt{35} \alpha+10 \beta \right)^2\,.
\label{rab}
\ee

In the limit $\alpha \to 0$ and $\beta \to 20/(2N+1)$
the observables (\ref{nsab}) and (\ref{rab}) reduce to 
\ba
n_{\rm s} (\alpha \to 0) &=& 1-\fracmm{4}{2N+1}\,,
\label{asy1n} \\
r (\alpha \to 0) &=& \fracmm{48}{(2N+1)^2}~~,
\label{asy1}
\ea
which agree with those in the Starobinsky's
$f(R)$ model \cite{mchi}.  For $N=55$ one has $n_{\rm s} 
(\alpha \to 0)=0.964$ and $r (\alpha \to 0)=3.896 \times 10^{-3}$.
In another limit $\alpha \to 40\sqrt{35}/[21(N+1)]$ 
and $\beta \to 0$ it follows that 
\ba
n_{\rm s} (\beta \to 0) &=& 1-\fracmm{2}{N+1}\,,
\label{asy2n} \\
r (\beta \to 0) &=& \fracmm{48}{(N+1)^2}
\label{asy2}
\ea
For $N=55$ one has $n_{\rm s} (\beta \to 0)=0.964$
and $r (\beta \to 0)=1.531 \times 10^{-2}$.
While the scalar spectral indices (\ref{asy1n}) and 
(\ref{asy2n}) are practically identical for $N \gg 1$, 
$r (\beta \to 0)$ is about four times as large as $r(\alpha \to 0)$.
For the intermediate values of $\alpha$ between 0 and 
$40\sqrt{35}/[21(N+1)]$ we need to numerically derive $\beta$ 
satisfying eq.~(\ref{efold}) for given $N$.
Then $n_{\rm s}$ and $r$ are known from 
eqs.~(\ref{nsab}) and (\ref{rab}).

%%%%%
\begin{figure}[t]
\includegraphics[height=3.3in,width=3.5in]{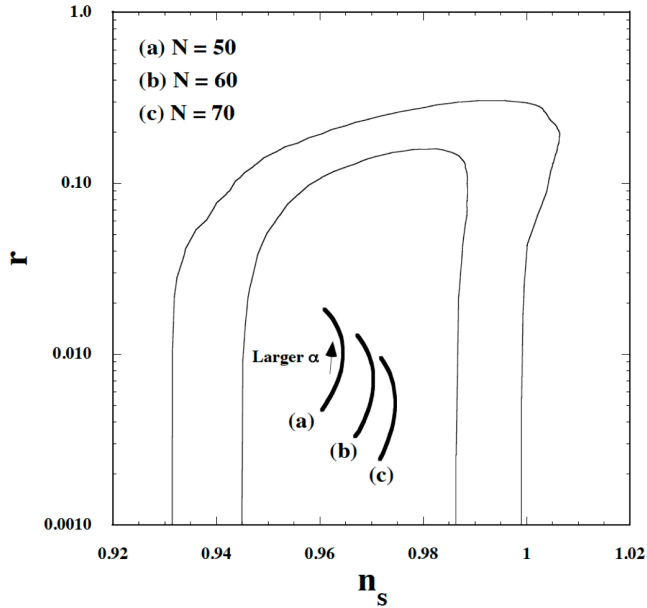}
\caption{\small The three thick lines show the theoretical values 
of $n_{\rm s}$ and $r$ for $N=50, 60, 70$ with
$\alpha$ ranging in the region (\ref{alcon}).
The thin solid curves are the 1$\sigma$ (inside)
and 2$\sigma$ (outside) observational contours 
constrained by the joint data analysis of 
WMAP7, BAO, and HST.
For $\alpha \to 0$,  $n_{\rm s}$ and $r$ are given by 
Eqs.~(\ref{asy1n}) and (\ref{asy1}).
In the limit $\beta \to 0$,
$n_{\rm s}$ and $r$ approach the values given in 
Eqs.~(\ref{asy2n}) and (\ref{asy2}).
\label{fig4}}
\end{figure}
%%%%%

In Fig.~\ref{fig4} we plot the theoretical values of $n_{\rm s}$
and $r$ in the $(n_{\rm s}, r)$ plane for $N=50, 60, 70$ together 
with the $1\sigma$ and $2\sigma$ observational contours
constrained by the joint data analysis of WMAP7 \cite{wmap}, 
Baryon Acoustic Oscillations (BAO) \cite{BAO}, 
and the Hubble constant measurement (HST) \cite{HST}.
The observational bounds are derived by using the 
standard consistency relation $r=-8n_{\rm t}$ \cite{Bassett}, 
where $n_{\rm t}=d \ln {\cal P}_{\rm t}/d \ln k$ is the tensor 
spectral index. In $f(R)$ gravity this relation also holds
by using the equivalence of the power spectra between 
the Jordan and Einstein frames \cite{fgrev2}.

The Starobinsky's $f(R)$ model, which corresponds to the 
limit $\alpha \to 0$ with the observables 
given in eqs.~(\ref{asy1n}) and (\ref{asy1}), 
is well within the current observational 
bound. In the regime $\alpha \ll \beta$ one has 
$\beta \simeq 20/(2N+1)-\sqrt{35}\alpha/5$, so that 
\ba
\hspace{-0.3cm}
n_{\rm s} (\alpha \ll \beta) &=& 1-\fracmm{4}{2N+1}
+\fracmm{\sqrt{35}}{100} \alpha\,,\\
\hspace{-0.3cm}
r (\alpha \ll \beta) &=& \fracmm{48}{(2N+1)^2}
\left[ 1+\fracmm{\sqrt{35}(2N+1)}{200} \alpha \right]^2
\ea
This shows that both $n_{\rm s}$ and $r$ increase for
larger $\alpha$ satisfying the condition $\alpha \ll \beta$. 
As we see in Fig.~\ref{fig4}, $n_{\rm s}$ switches to 
decrease at some value of $\alpha$, whereas
$r$ continuously grows toward the asymptotic value 
given in Eq.~(\ref{asy2}).

{}From Fig.~\ref{fig4} we find that the $f(R)$ model (\ref{fRinf})
in which $\alpha$ is in the range (\ref{alcon}) is inside the 
$1\sigma$ observational contour.
The condition $m>\sqrt{7/20}\,M$ provides the 
constraints $\alpha<0.194$, $\alpha<0.165$, $\alpha<0.143$
for $N=50, 60, 70$ respectively, while the bound (\ref{alcon})
in each case corresponds to 
$\alpha<0.221$, $\alpha<0.185$, $\alpha<0.159$.
When $N=60$ the scalar spectral index and the tensor-to-scalar ratio 
are $n_{\rm s}=0.969$, $r=0.0110$ for $\alpha=0.165$ and 
$n_{\rm s}=0.967$, $r=0.0129$ for $\alpha=0.185$, which
are not very different from each other.
For the background in which inflation is sustained with the 
number of e-foldings $N>50$ the model is consistent 
with the current observations.

Note that the nonlinear parameter $f_{\rm NL}$ of the
scalar non-Gaussianities is of the order of the slow-roll 
parameters in $f(R)$ gravity \cite{Defe} --- see also Sec.~20.
Hence, in current observations, this does not provide additional constraints 
to those studied above.

\section{Effective Scalar Potential for Preheating}

In this section we derive the effective scalar potential and the kinetic terms 
of a {\it complex} scalaron field in the low-energy regime (B) characterized by $|R| \ll R_0$. 
In doing so, let us return to the original $F(\car)$  supergravity action 
(\ref{action}) and perform the superfield Legendre transformation --- see Sec.~6. 
As is usual, we temporarily set $M_{\rm Pl}= 1$ to simplify our calculations.  The Legendre
transform yields the equivalent action 
\be 
S=\int d^4 x\,d^2 \theta\, {\cal E}\, \left[ -\cy \car +Z(\cy) \right] 
+{\rm H.c.}\,,
\label{frsup2}
\ee
where we have introduced the new covariantly chiral superfield $\cy$ and the 
new holomorphic function $Z(\cy)$ related to the function $F$ as 
\be \label{zfun}
F(\car) = -\car \cy(\car) + Z(\cy(\car))
\ee

The equation of motion of the superfield $\cy$, which follows from the variation 
of the action (\ref{frsup2}) with respect to $\cy$, has the algebraic form
\be \label{yem}
\car = Z'(\cy) \, ,
\ee
so that the function $\cy(\car)$ is obtained by inverting the function $Z'$.  
Substituting the solution $\cy(\car)$ back into the action (\ref{frsup2}) yields 
the original action (\ref{action}) because of eq.~(\ref{zfun}). 
We also find 
\be \label{leg}
\cy =- F'(\car) 
\ee
The inverse function $\car(\cy)$ always exists under the physical condition  
$F'(\car) \neq 0$. As regards the $F$-function (\ref{cub}), eq.~(\ref{leg}) yields
a quadratic equation with respect to $\car$, whose solution is
\be \label{quads}
\car(\cy) =  \fracmm{\sqrt{14}M^2}{20m} \left[ 
1 - \sqrt{ 1 + \fracmm{80m^2}{21M^2}(Y-3/4) }
\right]\,,
\ee 
where we have used the parametrization (\ref{repar}). 
Equation (\ref{quads}) is also valid for the leading complex scalar field 
components $\left.\car\right|=\bar{B}/3=\bar{X}$ and 
$\left.\cy\right|\equiv Y$, where $Y$ is the complex scalaron field.

The kinetic terms of $\cy$ are obtained by using the identity 
\be \label{siegel}  
\int d^4 x\,d^2 \theta\, {\cal E}\, \cy \car  +{\rm H.c.}=
\int d^4 x \, d^4 \theta \, E^{-1} (\cy +\bar{\cy} ) \, ,
\ee
where $E^{-1}$ is the full curved superspace density \cite{sb1,sb2,sb3}.
 Therefore, the K\"ahler potential reads 
\be \label{kahl}
K = -3 \ln \left( \cy + \bar{\cy} \right) 
\ee
up to an additive constant. It gives rise to the kinetic terms
\ba \label{kinc}
{\cal L}_{\rm kin} &=& \left. \fracmm{\pa^2 K}{\pa\cy \pa\bar{\cy}}\right|_{\cy=Y}
\pa_{\mu}Y\pa^{\mu}\bar{Y} \nonumber \\
&=& 3 \fracmm{ \pa_{\mu}Y\pa^{\mu}\bar{Y} }
{(Y +\bar{Y})^2}=
3 \fracmm{ (\pa_{\mu}y)^2 +  (\pa_{\mu}z)^2}{4y^2} \; ,
\ea
where we have used the notation $Y=y +iz$ in terms of the two real fields 
$y$ and $z$. The imaginary component $z$ corresponds to 
a pseudo-scalar field. The kinetic terms (\ref{kinc}) represent 
the {\it non-linear sigma model} \cite{nlsm}  with the hyperbolic target space of 
(real) dimension two, whose metric is known as the standard Poincar\'e metric. 
The kinetic terms are invariant under arbitrary rescalings 
$Y \to AY$ with constant parameter $A \neq 0$. 

The effective scalar potential $V(Y,\bar{Y})$ of a complex scalaron $Y$ 
in the regime (B), where supergravity decouples (it corresponds to rigid supersymmetry) 
is easily derived from eq.~(\ref{frsup2}) when keeping only scalars 
(i.e. ignoring their spacetime derivatives together with all fermionic contributions) and 
eliminating the auxiliary fields, near the minimum of the scalar potential. We find 
\be \label{scals}
V = \frac{21}{2} \left|  Z'(Y) \right|{}^2 =  \frac{21}{2} \left|  
\car(Y) \right|{}^2 \, ,
\ee
which gives rise to the chiral superpotential
\be \label {chipot} 
W (\cy) = \sqrt{\frac{21}{2}} Z(\cy)
\ee

The superfield equations (\ref{kahl}) and (\ref{chipot}) are {\it model-independent}, ie. 
they apply to any function $F({\cal R})$ in the large $M_{\rm Pl}$ limit, near the minimum of 
the scalar potential with the vanishing cosmological constant. The exact scalar potential 
including the supergravity effects is derived in Appendix A, but it is not very illuminating. 

There is no field redefinition that would bring all the kinetic terms (\ref{kinc}) to 
the free form. The canonical (free) kinetic term of a {\it real} scalaron $y$ alone can be 
obtained via the field redefinition 
\be \label{repara}
y = A \exp( -\sqrt{2/3}\,\phi)
\ee
The scalaron potential vanishes at $y=3/4$.
Demanding that this minimum corresponds to $\phi=0$, we have  
$A=3/4$ and hence $y=(3/4) \exp( -\sqrt{2/3}\,\phi)$.
Defining a rescaled field $\chi$ as $\chi=\sqrt{8/3}\,z$, the 
kinetic term (\ref{kinc}) can be written as 
\be \label{kine2}
{\cal L}_{\rm kin}=\frac12 (\partial_{\mu} \phi)^2+
\frac12 e^{2\sqrt{2/3}\,\phi/M_{\rm pl}} (\partial_{\mu} \chi)^2\,.
\ee
Here and in what follows we restore the reduced Planck
mass $M_{\rm Pl}$. 

The total potential (\ref{scals}) including both the fields
$\phi$ and $\chi$ is given by  
\be \label{Vpc}
V(\phi, \chi)=\fracmm{147M^4M_{\rm pl}^2}{400m^2} \left|
\sqrt{{\cal B}(\phi)+i{\cal C}(\chi)}-1 \right|^2\,, 
\ee
where 
\ba  \label{BC}
{\cal B}(\phi) &=& 1+\fracmm{20m^2}{7M^2}
\left( e^{-\sqrt{2/3}\,\phi/M_{\rm pl}}-1  \right)\,,\\
{\cal C}(\chi) &=& \fracmm{80m^2}{21M^2}
\sqrt{\fracmm{3}8}\,\fracmm{\chi}{M_{\rm pl}}\,.
\ea
In order to express (\ref{Vpc}) in a more convenient form
we write $\sqrt{{\cal B}(\phi)+i{\cal C}(\chi)}=p+iq$, where 
$p$ and $q$ are real. This gives the relations $p^2-q^2={\cal B}(\phi)$ and 
$2pq={\cal C}(\chi)$. Solving these equations for $p$,  we find
\be 
p=\frac{1}{\sqrt{2}} \left[ {\cal B}(\phi)+
\sqrt{{\cal B}^2(\phi)+{\cal C}^2(\chi)}
\right]^{1/2}\,,
\ee
where we have chosen the solution $p>0$ to recover
$p=\sqrt{{\cal B}(\phi)}$ for ${\cal B}(\phi)>0$
in the limit ${\cal C}(\chi) \to 0$.
Then the field potential (\ref{Vpc}) reads 
\ba 
\hspace{-0.3cm}V(\phi, \chi) &=& 
\fracmm{147M^4 M_{\rm pl}^2}{400m^2} \biggl[1+
\sqrt{{\cal B}^2 (\phi)+{\cal C}^2 (\chi)} \nonumber \\
\hspace{-0.3cm}& &-\sqrt{2} \left\{ {\cal B}(\phi)
+\sqrt{{\cal B}^2 (\phi)+{\cal C}^2 (\chi)} \right\}^{1/2} \biggr].
\label{Vtotal}
\ea

In the absence of the pseudo-scalar $\chi$
the potential (\ref{Vtotal}) reduces to 
\be  \label{scalpot}
V(\phi)= \fracmm{147M^4 M_{\rm pl}^2}{400m^2}
\left[ 1+|{\cal B}(\phi)|-\sqrt{2} \left\{ {\cal B}(\phi)
+|{\cal B} (\phi)| 
\right\}^{1/2} \right]
\ee
For the field $\phi$ satisfying the condition 
${\cal B}(\phi)<0$ it follows that 
\be  \label{scalpot1}
V(\phi)=\frac{21}{20}M^2 M_{\rm pl}^2
\left( 1-e^{-\sqrt{2/3}\,\phi/M_{\rm pl}}
\right)\,,
\ee
which approaches the constant 
$V(\phi) \to 21M^2M_{\rm pl}^2/20$
in the limit $\phi \to \infty$. Defining the slow-roll parameter 
$\epsilon_V=(M_{\rm pl}^2/2)(V_{,\phi}/V)^2$,  we have
\be  
\epsilon_V=\fracmm{x^2}{3(1-x)^2}\,,\qquad
x=e^{-\sqrt{2/3}\,\phi/M_{\rm pl}}
\ee
The end of inflation is characterized by the criterion $\epsilon_V=1$. 
This gives $x_f=e^{-\sqrt{2/3}\,\phi_f/M_{\rm pl}}=(3-\sqrt{3})/2$ 
and hence $\phi_f=0.558M_{\rm pl}$. For $m>M$ the condition ${\cal B}(\phi_f)<0$
is satisfied, so that the potential (\ref{scalpot1}) is valid 
at the end of inflation. If $m$ is close to the border value $\sqrt{7/20}\,M$, 
then the potential (\ref{scalpot1}) is already invalid at the end of inflation.

For small $\phi$ satisfying the condition ${\cal B}(\phi)>0$ the potential (\ref{scalpot}) reads 
\be  \label{scalpot4}
V(\phi)=\fracmm{147M^4M_{\rm pl}^2}{400m^2} \left[ 1-\sqrt{1+
\fracmm{20m^2}{7M^2} \left( e^{-\sqrt{2/3}\phi/M_{\rm pl}}-1
\right)} \right]^2
\ee
In this case Taylor expansion around $\phi=0$ gives 
rise to the leading-order contribution $V(\phi)=m^2 \phi^2/2$.
Reheating occurs around the potential minimum
through the oscillation of the canonical field $\phi$.

The full effective potential involving the interaction between 
the fields $\phi$ and $\chi$ is given by eq.~(\ref{Vtotal}).
Expanding the potential (\ref{Vtotal}) around $\phi=\chi=0$
and picking up the terms up to fourth-order in the fields, 
we obtain 
\be \eqalign{
V(\phi, \chi) 
\simeq~ &
\fracmm{1}{2}m^2 \phi^2+
\fracmm{\sqrt{6}m^2 (10m^2-7M^2)}{42M^2M_{\rm pl}}\phi^3
+\fracmm{(1500m^4-1260m^2M^2+343M^4)m^2}
{1764M^4 M_{\rm pl}^2}\phi^4 \cr
&  +\fracmm{1}{2}m^2 \chi^2-\fracmm{25 m^6}{49M^4 M_{\rm pl}^2}\chi^4 
+\fracmm{5\sqrt{6}m^4}{21M^2M_{\rm pl}} \phi \chi^2
+\fracmm{5m^4 (10m^2-7M^2)}{147M^4M_{\rm pl}^2}\phi^2 \chi^2 \cr}
\label{Vre}
\ee 
The scalaron $\phi$ is coupled to the pseudo-scalar $\chi$
through the interaction given in the second line of eq.~(\ref{Vre}).

\section{Preheating after Inflation}

Here we study the dynamics of preheating for the two-field system 
described by the kinetic term (\ref{kine2}) and 
the effective potential (\ref{Vtotal}). The background equations 
of motion on the flat FLRW background are 
\ba 
& & 3M_{\rm pl}^2 H^2=\dot{\phi}^2/2+e^{2b}\dot{\chi}^2/2+V\,,
\label{Hubeq}\\
& & \ddot{\phi}+3H \dot{\phi}+V_{,\phi}-b_{,\phi}e^{2b} \dot{\chi^2}=0\,,
\label{phieqm} \\
& & \ddot{\chi}+(3H+2b_{,\phi} \dot{\phi})\dot{\chi}+e^{-2b}V_{,\chi}=0\,,
\label{ddotchi}
\ea
where $b(\phi)=\sqrt{2/3}\,\phi/M_{\rm pl}$ and 
``${}_{,\phi}$'' represents a partial derivative with 
respect to $\phi$.

In Fourier space the field perturbations $\delta \phi_k$ 
and $\delta \chi_k$ with the comoving wave number 
$k$ obey the following equations:
\ba 
& &\ddot{\delta \phi}_k+3H \dot{\delta \phi}_k+[k^2/a^2+V_{,\phi \phi}
-(2b_{,\phi}^2+b_{,\phi \phi}) e^{2b} \dot{\chi}^2]\delta \phi_k \nonumber \\
& &=-V_{,\phi \chi} \delta \chi_k+2b_{,\phi}e^{2b} \dot{\chi}
\dot{\delta \chi}_k \,,\label{delphieq} \\
& &\ddot{\delta \chi}_k+(3H+2b_{,\phi} \dot{\phi}) 
\dot{\delta \chi}_k
+(k^2/a^2+e^{-2b} V_{,\chi \chi})\delta \chi_k \nonumber \\
& &=
-e^{-2b} (V_{,\phi \chi}-2b_{,\phi}V_{,\chi}+2b_{,\phi \phi}
e^{2b} \dot{\phi} \dot{\chi})\delta \phi_k
-2b_{,\phi} \dot{\chi} \dot{\delta \phi}_k \nonumber \\
\label{delchieq}
\ea

\begin{figure}
\includegraphics[height=2.6in,width=3.2in]{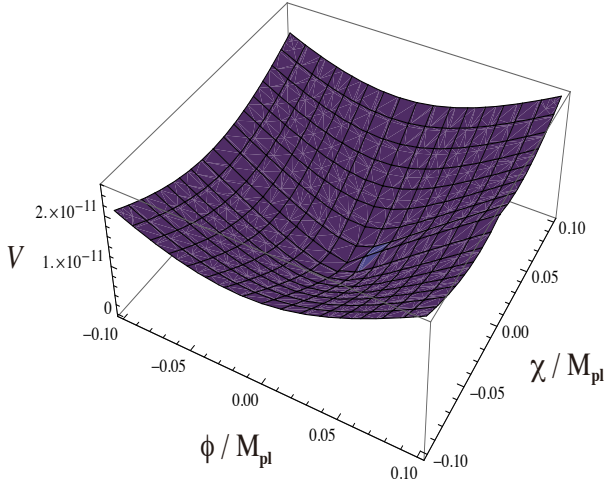}
\caption{\small
Effective potential (\ref{Vtotal}) for $m=1.14 \times 10^{-4}M_{\rm pl}$
and $M=1.62 \times 10^{-5} M_{\rm pl}$ in the region
$-0.1<\phi/M_{\rm pl}<0.1$ and $-0.1<\chi/M_{\rm pl}<0.1$.
\label{fig5}}
\end{figure}
%%%%%

%%%%%
\begin{figure}
\includegraphics[height=3.0in,width=3.1in]{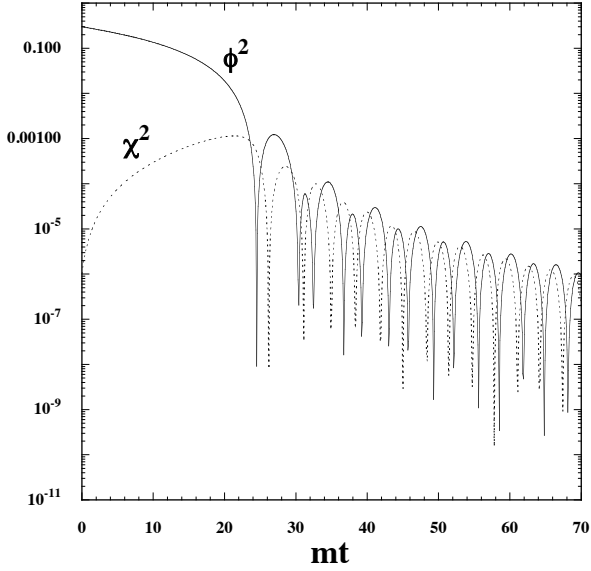}
\caption{
Evolution of the background fields $\phi^2$ and $\chi^2$
(both are normalized by $M_{\rm pl}^2$) for 
$m=1.14 \times 10^{-4} M_{\rm pl}$ and 
$M=1.62 \times 10^{-5} M_{\rm pl}$
with the initial conditions $\phi=0.55 M_{\rm pl}$, 
$\chi=10^{-3} M_{\rm pl}$, 
$\dot{\phi}=-1.6 \times 10^{-2}mM_{\rm pl}$, and 
$\dot{\chi}=1.5 \times 10^{-3}mM_{\rm pl}$.
\label{fig6}}
\end{figure}

The derivative $V_{,\chi}$ of the potential (\ref{Vtotal})
vanishes at $\chi=\pm \chi_c$, where
\be \label{chic}
\chi_c=\frac{\sqrt{210}M}{20m}
\left[ 1-e^{-\sqrt{2/3}\,\phi/M_{\rm pl}}-\frac{21}{80}
\left( \frac{M}{m} \right)^2 \right]^{1/2}M_{\rm pl}
\ee

The local minima exist in the $\chi$ direction 
provided that 
\be \label{phic}
\phi>\sqrt{\frac32} \ln \left[ 1-\frac{21}{80}
\left( \frac{M}{m} \right)^2 \right]^{-1} M_{\rm pl}
\equiv \phi_c~~,
\ee
whereas they disappear for $\phi<\phi_c$.
In Fig.~\ref{fig5} we plot the potential (\ref{Vtotal})
with respect to $\phi$ and $\chi$ for $m=1.14 \times 10^{-4}M_{\rm pl}$
and $M=1.62 \times 10^{-5} M_{\rm pl}$.
Since $\phi_c=6.5 \times 10^{-3}M_{\rm pl}$ in this case, 
the potential has the local minima in the $\chi$ direction 
for $\phi>6.5 \times 10^{-3}M_{\rm pl}$.
{}From eq.~(\ref{chic}) the field value $\chi_c$ increases
for larger $\phi$. For the model parameters used 
in Fig.~\ref{fig5}, for example, one has
$\chi_c=0.028 M_{\rm pl}$ at $\phi=0.1 M_{\rm pl}$ and  
$\chi_c=0.059 M_{\rm pl}$ at $\phi=0.5M_{\rm pl}$.

If the initial conditions of the fields are $0<\chi<\chi_c$ 
and $\phi>\phi_c$, the field $\chi$ grows 
toward the local minimum at $\chi=\chi_c$.
After $\phi$ drops below $\phi_c$, the field $\chi$
approaches the global minimum at $\chi=0$.
In Fig.~\ref{fig6} we show one example for the
evolution of the background fields $\phi$ and $\chi$
with the same values of $m$ and $M$ as those 
in Fig.~\ref{fig5}.
The energy density of the field $\chi$ catches up to 
that of the inflaton around the onset of reheating.

As we see in eq.~(\ref{phic}), the critical field value $\phi_c$
gets smaller for increasing $m/M$. 
Hence, for larger $m/M$, the potential (\ref{Vtotal}) 
possesses the local minima at $\chi=\pm \chi_c$ 
for a wider range of $\phi$.
The potential in the region $|\chi|<\chi_c$ 
can be flat enough to lead to inflation by the slow-roll 
evolution of the field $\chi$, even if $\phi$ is smaller 
than $\phi_f=0.558M_{\rm pl}$.
For larger ratio $m/M$ inflation ends with the field value 
much smaller than $\phi_f$.
If $m/M=20$ and $m/M=83$, for example, the amplitudes of 
the field $\phi$ at the onset of oscillations are 
$\phi_i =1.5 \times 10^{-2} M_{\rm pl}$ and 
$\phi_i = 5.0 \times 10^{-3} M_{\rm pl}$, respectively.

Let us consider the regime where the condition 
\be \label{phicon}
\left( \fracmm{m}{M} \right)^2 \fracmm{|\phi|}{M_{\rm Pl}} \ll 1
\ee
is satisfied. 
Then the potential (\ref{Vre}) is approximately
given by $V(\phi, \chi) \simeq m^2 \phi^2/2+m^2 \chi^2/2$, 
in which case both $\phi$ and $\chi$ have the 
same mass $m$.
This gives rise to the matter-dominated epoch 
(where $H=2/(3t)$) driven by the oscillations 
of two massive scalar fields.
{}From eq.~(\ref{phieqm}) we have that  
$\ddot{\phi}+(2/t)\dot{\phi}+m^2 \phi \simeq 0$, 
whose solution is
\be
\phi(t) \simeq \frac{\pi}{2mt} \phi_i \sin (mt)\,.
\label{infosc}
\ee
Here the initial field value $\phi_i$ corresponds to the
time $t_i=\pi/(2m)$.

In order to discuss the dynamics of the field perturbations 
in eqs.~(\ref{delphieq}) and (\ref{delchieq}) 
we define the two frequencies $\omega_{\phi}$ 
and $\omega_{\chi}$, as 
$\omega_{\phi}^2=k^2/a^2+V_{,\phi \phi}
-(2b_{,\phi}^2+b_{,\phi \phi}) e^{2b} \dot{\chi}^2$ and 
$\omega_{\chi}^2=k^2/a^2+e^{-2b}V_{,\chi \chi}$.
As long as the condition (\ref{phicon}) is satisfied, 
it is sufficient to pick up the terms up to cubic order in fields.
It then follows that 
\ba \label{omephi}
\omega_{\phi}^2 & \simeq &
\fracmm{k^2}{a^2}+m^2+\fracmm{\sqrt{6}m^2 (10m^2-7M^2)}
{7M^2 M_{\rm pl}} \phi\,,\\
\omega_{\chi}^2 &=&
\fracmm{k^2}{a^2}+m^2e^{-2b}+\fracmm{10\sqrt{6}m^4}
{21M^2 M_{\rm pl}}e^{-2b} \phi\,,
\ea
where, in eq.~(\ref{omephi}), we have neglected the contribution 
of the term $-(2b_{,\phi}^2+b_{,\phi \phi}) e^{2b} \dot{\chi}^2$.

\begin{figure}
\includegraphics[height=3.0in,width=3.1in]{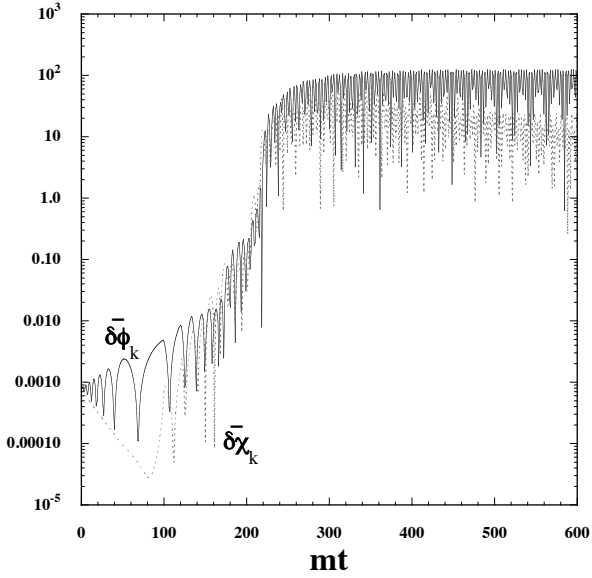}
\caption{\small
Evolution of the field perturbations
$\bar{\delta \phi}_k=k^{3/2}\delta \phi_k/M_{\rm pl}$
and $\bar{\delta \chi}_k=k^{3/2}\delta \chi_k/M_{\rm pl}$
with the wave number $k=m$
for $m=1.16 \times 10^{-3} M_{\rm pl}$ and 
$M=1.39 \times 10^{-5} M_{\rm pl}$.
We choose the background initial conditions 
$\phi=0.1 M_{\rm pl}$, $\chi=1.0 \times 10^{-3} M_{\rm pl}$, 
$\dot{\phi}=-8.48 \times 10^{-4}mM_{\rm pl}$, and 
$\dot{\chi}=1.18 \times 10^{-5}mM_{\rm pl}$.
\label{fig7}}
\end{figure}

\begin{figure}
\includegraphics[height=3.0in,width=3.1in]{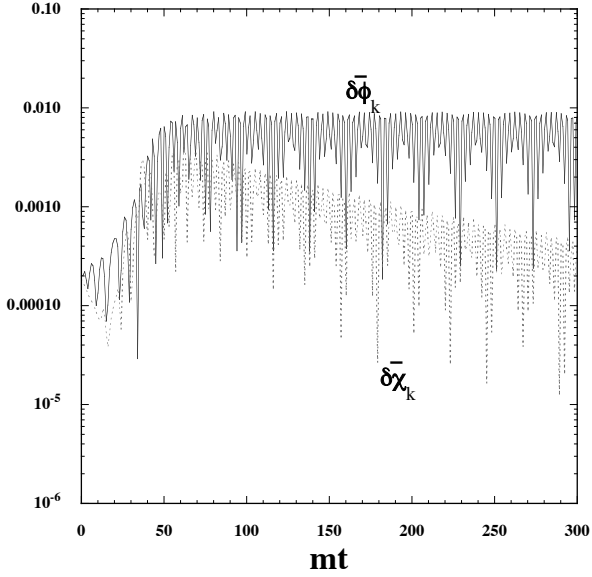}
\caption{\small
Evolution of the field perturbations
with the wave number $k=m$
for $m=2.89 \times 10^{-4}  M_{\rm pl}$ and 
$M=1.46 \times 10^{-5}M_{\rm pl}$.
We choose the background initial conditions 
$\phi=0.1M_{\rm pl}$, $\chi=1.0 \times 10^{-3}M_{\rm pl}$, 
$\dot{\phi}=-7.35 \times  10^{-3} mM_{\rm pl}$, and 
$\dot{\chi}=6.85 \times 10^{-4} mM_{\rm pl}$.
\label{fig8}}
\end{figure}

We introduce the rescaled fields $\delta \varphi_k=a^{3/2} \delta \phi_k$
and $\delta X_k=a^{3/2}e^{b} \delta \chi_k$ to estimate the growth of 
perturbations in the regime (\ref{phicon}).
Neglecting the contributions of the r.h.s. of eqs.~(\ref{delphieq})
and (\ref{delchieq}) and also using the approximation 
$e^{-2b} \simeq 1$ in the regime $H \ll m$, the field perturbations 
$\delta \varphi_k$ and $\delta X_k$
obey the following equations
\ba
& &\fracmm{d^2}{dz^2} \delta \varphi_k+
\left[ A_k-2q_{\phi} \cos (2z) \right]  \delta \varphi_k \simeq 0\,,
\label{delvarphi} \\
& &\fracmm{d^2}{dz^2} \delta X_k+
\left[ A_k-2q_{\chi} \cos (2z) \right]  \delta X_k \simeq 0\,,
\label{delX} 
\ea
where $2z=mt+\pi/2$.
The quantities $A_k$, $q_{\phi}$, and $q_{\chi}$ are given by 
\ba
A_k &=& 4+4 \fracmm{k^2}{m^2 a^2}\,,\\
q_{\phi} &=& 
\frac{20\sqrt{6}}{7} 
\left(1-\fracmm{7M^2}{10m^2}
\right) \left( \frac{m}{M} \right)^2  
\fracmm{\phi_i}{M_{\rm pl}} \fracmm{\pi/2}{mt}\,,\\
q_{\chi} &=& \fracmm{20\sqrt{6}}{21} 
\left( \fracmm{m}{M} \right)^2 
\fracmm{\phi_i}{M_{\rm pl}} \fracmm{\pi/2}{mt}\,,
\ea
which are time-dependent.

Equations (\ref{delvarphi}) and (\ref{delX}) are the so-called
{\it Mathieu equations} describing the parametric resonance caused by
oscillations of the field $\phi$ \cite{Robert,Robert2,KLS,KLS2}. In the regime 
(\ref{phicon}) both $q_{\phi}$ and $q_{\chi}$ are smaller than 1 for 
$t \ge t_i=\pi/(2m)$. In this case the resonance occurs in narrow bands 
near $A_k=l^2$, where $l=1, 2, \cdots$ \cite{KLS,KLS2,Tsuji}.
As the physical momentum $k/a$ redshifts away, the field
perturbations approach the instability band at $A_k=4$.
Although $\delta \varphi_k$ and $\delta X_k$ can be amplified
for $A_k \simeq 4$ and $q_{\phi} \lesssim 1$, $q_{\chi} \lesssim 1$, 
this narrow parametric resonance is not efficient enough to 
lead to the growth of $\delta \phi_k$ and $\delta \chi_k$
against the Hubble friction \cite{KLS,KLS2}.

If the initial field $\phi_i$ satisfies the condition $(m/M)^2|\phi_i|/M_{\rm pl} \gg 1$, 
the quantities $q_{\phi}$ and $q_{\chi}$ are much larger than 1 at the onset 
of reheating. This corresponds to the so-called {\it broad} resonance regime \cite{KLS,KLS2}
in which the perturbations $\delta \phi_k$ and $\delta \chi_k$ can grow even against the 
Hubble friction. We caution, however, that eqs.~(\ref{delvarphi}) and (\ref{delX}) are 
no longer valid because the background solution (\ref{infosc}) is subject to change due to 
the effect of higher-order terms in the potential (\ref{Vtotal}).
Still, the non-adiabatic particle production occurs around the potential minimum ($\phi=0$) 
\cite{KLS,KLS2}. In this region the dominant contribution to the potential 
is the quadratic term $m^2 \phi^2/2$.
Hence, it is expected that preheating can be efficient for the values of $q_{\phi}$ and $q_{\chi}$ 
much larger than 1 at the onset of the field oscillations.

We numerically solve the perturbations equations (\ref{delphieq}) 
and (\ref{delchieq}) together with the background equations
(\ref{Hubeq}), (\ref{phieqm}), and (\ref{ddotchi}) 
for the full potential (\ref{Vtotal}) without using the approximate 
expression (\ref{Vre}). In Figs.~\ref{fig7} and \ref{fig8} we plot the evolution of 
the field perturbations $\delta \phi_k$ and $\delta \chi_k$
with the wave number $k=m$ for two different choices of 
the parameters $m$ and $M$ (which are constrained by 
the WMAP normalization in Fig.~\ref{fig3}).
The initial conditions of the perturbations are chosen to 
recover the vacuum state characterized by 
$\delta \varphi_k(t_i)=e^{-i \omega_{\phi} t_i}/\sqrt{2\omega_{\phi}}$ 
and 
$\delta X_k(t_i)=e^{-i \omega_{\chi} t_i}/\sqrt{2\omega_{\chi}}$.

Figure \ref{fig7} corresponds to the mass scales
$m=1.16 \times 10^{-3}M_{\rm pl}$ and 
$M=1.39 \times 10^{-5}M_{\rm pl}$, i.e., 
the ratio $m/M=83$.
The field value at the onset of oscillations is found 
to be $\phi_i = 5.0 \times 10^{-3} M_{\rm pl}$, in which case
$q_{\phi}(t_i)=244$ and $q_{\chi}(t_i)=81$.
Figure \ref{fig7} shows that both $\delta \phi_k$ and $\delta \chi_k$ 
 rapidly grow by the broad parametric resonance.
The growth of the field perturbations ends when $q_{\phi}$
and $q_{\chi}$ drop below 1.

Figure \ref{fig8} corresponds to the ratio $m/M=20$, 
in which case $\phi_i = 1.5 \times 10^{-2} M_{\rm pl}$, 
$q_{\phi}(t_i)=41$, and $q_{\chi}(t_i)=4.6$.
Compared to the evolution in Fig.~\ref{fig7}, 
preheating is less efficient because of the smaller
values of $q_{\phi}(t_i)$ and $q_{\chi}(t_i)$.
The parameter to control the efficiency of preheating 
is the mass ratio $m/M$.
For larger $m/M$ the creation of particles tends to be 
more significant.
For the mass $m$ smaller than $10^{-4} M_{\rm pl}$
the field perturbations $\delta \phi_k$ and $\delta \chi_k$
hardly grow against the Hubble friction 
because they are not in the broad resonance regime.

In our numerical simulations we did not take into account 
the rescattering effect between different modes of the 
particles. The lattice simulation \cite{lattice,lattice1,lattice2,Felder} is required 
to deal with this problem. It will be of interest to see, how the 
non-linear effect can affect the evolution of perturbations 
at the final stage of preheating.

\section{Current Status of our Model}

In the preceeding sections we studied the viability of the $f(R)$ inflationary scenario in the 
context of $F({\cal R})$ supergravity. In the high-energy regime characterized by the condition 
$|R| \gg R_0$ there is a correction of the form $(-R)^{3/2}/m$ to the function 
$f(R)=3R/10-R^2/(6M^2)$. Introducing the dimensionless functions $\alpha$ and $\beta$
in eqs.~(\ref{albe}), we showed that these are constrained to be
in the range (\ref{alcon}) to realize inflation with the number of 
e-foldings $N$.

The masses of the scalaron field in the regimes $|R| \gg R_0$
and $|R| \ll R_0$ are approximately given by $M$ and $m$, respectively. 
{}From the WMAP normalization of the CMB temperature anisotropies we derived 
$M$ and $m$ as a function of $\alpha$ in Fig.~\ref{fig3}.
The weak dependence of $M$ with respect to $\alpha$
means that the term $-R^2/(6M^2)$ needs to dominate
over the correction $(-R)^{3/2}/m$ during inflation.
We also showed that the model is within the 1$\sigma$ 
observational contour constrained from the joint data
analysis of WMAP7, BAO, and HST, 
by evaluating the scalar spectral index $n_{\rm s}$
and the tensor-to-scalar ratio $r$.

In the presence of the pseudo-scalar field $\chi$ coupled to 
the scalaron field $\phi$ we derived the effective potential (\ref{Vtotal}) 
and their kinetic energies (\ref{kine2}) 
in the low-energy regime ($|R| \ll R_0$).
Provided that the condition (\ref{phic}) is satisfied, the effective potential 
has two local minima at $\chi=\pm \chi_c$.
Around the global minimum at $\phi=\chi=0$ 
the system is described by two massive scalar fields
with other interaction terms given in Eq.~(\ref{Vre}).
Even if $\chi$ is initially close to 0, $\chi$ typically catches up 
to $\phi$ around the onset of the field oscillations (see Fig.~\ref{fig6}).

In the regime where the field $\phi$ is in the range (\ref{phicon})
we showed that both the field perturbations 
$\delta \varphi_k=a^{3/2} \delta \phi_k$ and 
$\delta X_k=a^{3/2}e^b \delta \chi_k$ obey the Mathieu 
equations (\ref{delvarphi}) and (\ref{delX}).
This corresponds to the narrow resonance regime in which 
$q_{\phi}$ and $q_{\chi}$ are smaller than the order of unity.
The broad resonance regime is characterized by the condition
$(m/M)^2 |\phi|/M_{\rm pl} \gg 1$, but in this case the expansion 
(\ref{Vre}) of the effective potential around the minimum is
no longer valid. In order to confirm the presence of 
the broad resonance we numerically solved the perturbation equations
(\ref{delphieq}) and (\ref{delchieq}) for the full potential (\ref{Vtotal}). 
Indeed we found that preheating of the both perturbations 
$\delta \phi_k$ and $\delta \chi_k$ is efficient in this regime.
As we see in Figs.~\ref{fig7} and \ref{fig8}, the broad parametric 
resonance is more significant for larger values of $m/M$.

Our results lend compelling support to the phenomenological 
viability of the bosonic sector of $F(\car)$ supergravity, 
in addition to its formal consistency. It is also worthwhile to
recall that supergravity unifies bosons and fermions with General Relativity,
highly constrains particle spectrum and interactions, has the ideal candidate
for a dark matter particle such as the lightest super-particle (see Sec.~20). 
It may also be deduced from quantum gravity such as superstring theory. The $F(\car)$ 
supergravity action (\ref{action}) is truly chiral in superspace, so that it is
expected to be protected against quantum corrections, which is important for stabilizing 
the masses $M$ and $m$ in quantum theory. 

\section{Cosmological Constant in $F(\car)$ Supergravity} 

The Standard ($\L$-CDM) Model in cosmology gives a phenomenological description of
the observed {\it Dark Energy} (DE) and {\it Dark Matter} (DM). It is based on the
use of a small positive cosmological constant $\L$ and a {\it Cold Dark Matter} (CDM), 
and is consistent with all observations coming from the existing cosmological, Solar
system and ground-based laboratory data. However, the $\L$-CDM  Model cannot be the 
ultimate answer to DE, since it implies its time-independence. For example, the 
`primordial' DE responsible for inflation in the early Universe was different from $\L$ 
and unstable.  The {\it dynamical}  (ie. time-dependent) models of DE can be easily 
constructed by using the $f(R)$ gravity theories, defined via replacing the scalar 
curvature $R$ by a function $f(R)$ in the gravitational action. The $f(R)$ gravity 
provides the self-consistent non-trivial alternative to the $\L$-CDM Model. The viable 
$f(R)$-gravity-based models of the current DE are known \cite{cde1,cde2,cde3}, 
and the combined inflationary-DE models are possible too \cite{fgrev2}. 

The natural question arises, whether $F(\car)$ supergravity is also capable to describe
the present DE and eg., a {\it positive} cosmological constant. It is non-trivial because 
the standard (pure) supergravity can only have a zero or negative cosmological constant. 
In this section we further extend the Ansatz used in Sec.~10 for the $F$-function, and
apply it to get a positive cosmological constant in the regime of a {\it low} spacetime 
curvature.

Throughout this section we again use the units $c=\hbar=M_{\rm Pl}=1$. We recall that an 
AdS-spacetime has a positive scalar curvature, and a dS-spacetime has a negative scalar 
curvature in our notation.

The embedding of $f(R)$ gravity into $F(\car)$ supergravity is given by (Sec.~8)
\be \lb{cemb1}
f(R) = f(R,X(R))
\ee
where the function $f(R,X)$ (or the gravity Lagrangian $\Lag$) is defined by
\be \lb{cemb2}
\Lag= f(R,X) = 2F'(X) \left[ \fracmm{1}{3} R +4 X^2 \right] + 6XF(X)   
\ee
and the function $X=X(R)$ is determined by solving an algebraic equation,
\be \lb{cemb3}
 \fracmm{\pa f(R,X)}{\pa X} =0 
\ee

The cosmological constant in $F(\car)$ supergravity is thus given by
\be \lb{ccc1}
\L = - f(0,X_0) 
\ee
where $X_0=X(0)$. It should be mentioned that $X_0$ represents the vacuum expectation
value of the auxiliary field $X$ that determines the scale of the supersymmetry 
breaking. Both inflation and DE imply $X_0\neq 0$.

To describe DE in the present Universe, ie. in the regime of a {\it low} spacetime 
curvature $R$, the function $f(R)$ should be close to the Einstein-Hilbert (linear) 
function $f_{\rm EH}(R)$ with a small positive $\L$,
\be \lb{cdep}
\abs{f(R)-f_{\rm EH}(R) } \ll \abs{f_{\rm EH}(R)},\quad 
\abs{f'(R)-f'_{\rm EH} } \ll 1,\quad
\abs{Rf''(R)} \ll 1  
\ee   
ie. $f(R)\approx -\fracmm{1}{2}R -\L$ for small $R$ with the very small and positive 
$\L\approx 10^{-118}(M^4_{\rm Pl})$.

Equations (\ref{cemb2}) and (\ref{ccc1}) imply
\be \lb{ccc2} 
\L =-8F'(X_0)X_0^2 -6X_0F(X_0) 
\ee
where $X_0$ is a solution to the algebraic equation
\be \lb{ccc3}
4X_0^2F''(X_0) + 11X_0F'(X_0) + 3F(X_0) =0
\ee
As is clear from eq.~(\ref{ccc2}), to have $\L\neq 0$, one must have $X_0\neq 0$, ie.
a (spontaneous) supersymmetry breaking. However, in order to proceed further, we need 
a reasonable Ansatz for the $F$-function.

The simplest opportunity is given by expanding the function $F(\car)$ in Taylor 
series with respect to $\car$. Since the $N=1$ chiral superfield $\car$ has
$X$ as its leading field component (in $\q$-expansion), one may expect that the Taylor
expansion is a good approximation as long as $\abs{X_0}\ll 1(M_{\rm Pl})$. As was
demonstrated in Sec.~10, a viable (successful) description of inflation is
possible in $F(\car)$ supergravity, when keeping the {\it cubic} term $\car^3$ in the
Taylor expansion of the $F(\car)$ function. It is, therefore, natural to expand the
function $F$ up to the cubic term with respect to $\car$, and use it as our Ansatz 
here,
\be \lb{can}
 F(\car) = f_0  - \fracmm{1}{2}f_1\car + \fracmm{1}{2}f_2\car^2 
-\fracmm{1}{6}f_3\car^3 
\ee
with some real coeffieints $f_0,f_1,f_2,f_3$. The Ansatz (\ref{can}) differs from the
one used in eq.~(\ref{cub}) by the presence of the new parameter $f_0$ only. It is
worth emphasizing here that $f_0$ is {\it not} a cosmological constant because one 
still has to eliminate the auxiliary field $X$. The stability conditions (Sec.~9) imply
\be \lb{csa1}
f_1>0~~, \qquad f_2>0~~,\qquad f_3 > 0 
\ee
and
\be \lb{sa2}
f_2^2 < f_1 f_3
\ee
Inflation requires $f_3\gg 1$ and $f_2^2\gg f_1$.~\footnote{The stronger condition 
$f^2_2\ll f_1f_3$ was used in Sec.~10 for simplicity.} As was already found in 
Sec.~10, in order to meet the WMAP observations, the parameter$f_3$ should be 
approximately equal to $6.5\cdot 10^{10}(N_e/50)^2$. The cosmological constant in the 
high-curvature regime does not play a significant role in early universe, so it can be ignored.

In the low curvature regime, in order to recover the Einstein-Hilbert term, one has
to fix $f_1=3/2$ (Sec.~10). Then the Ansatz (\ref{can}) leads to the gravitational 
Lagrangian 
\be 
\lb{cfrx}
f(R,X) = -5f_3X^4 +11f_2X^3 - \fracmm{1}{3}f_3\left(R+\fracmm{63}{2f_3}\right)
X^2 + \left(6f_0 + \fracmm{2}{3}f_2R\right)X - \fracmm{1}{2}R 
\ee
and the auxiliary field equation
\be \lb{cafe}
X^3 - \fracmm{33f_2}{20f_3}X^2 +\fracmm{1}{30}\left( R + \fracmm{63}{2f_3}\right)
X - \fracmm{1}{30f_3}\left( f_2R +9f_0\right)=0
\ee
whose formal solution is available via the standard Cardano-Vi\`ete  
formulae \cite{abs}.

In the low-curvature regime we find a cubic equation for $X_0$ in the form
\be \lb{ccub}
 X_0^3 - \left(\fracmm{33f_2}{20f_3}\right)X_0^2 
+\left(\fracmm{21}{20f_3}\right)X_0 -\left(\fracmm{3f_0}{10f_3}\right)=0
\ee

`Linearizing' eq.~(\ref{ccub}) with respect to $X_0$ brings the solution $X_0=2f_0/7$
whose substitution into the action (\ref{cfrx}) gives rise to a {\it negative} 
cosmological constant, $\L_0=-6f_0^2/7$. This way we recover the standard supergravity 
case.

Equations (\ref{cfrx}) and (\ref{ccub}) allow us to write down the exact eq.~(\ref{ccc1})
for the cosmological constant in the {\it factorized} form
\be \lb{ccce}
\L(X_0) = - \fracmm{11f_2}{4}X_0 ( X_0-X_-)(X_0-X_+)
\ee
where $X_{\pm}$ are the roots of the quadratic equation $x^2-\fracmm{21}{11f_2}x
+\fracmm{18f_0}{11f_2}=0$, ie.
\be \lb{croots}
X_{\pm} = \fracmm{21}{22f_2}\left[ 1\pm \sqrt{ 1- \fracmm{2^3\cdot 11}{7^2}f_0f_2}\right]
\ee
Since $f_0f_2$ is supposed to be very small, both roots  $X_{\pm}$ are real and positive.

Equation (\ref{ccce}) implies that $\L>0$ when either (I) $X_0<0$, or (II)  $X_0$ is inside
the interval $(X_-,X_+)$.

By using {\it Matematica} we were able to numerically confirm the existence of solutions to
eq.~(\ref{ccub}) in the region (I) when $f_0<0$, but not in the region (II). So, to this 
end, we continue with the region (I) only.  All real roots of eq.~(\ref{ccub}) are given by
\be \lb{c3roots}
\eqalign{
(X_0)_1 = ~&~ 2\sqrt{-Q} \cos\left(\fracmm{\vq}{3}\right) +\fracmm{11f_2}{20f_3}~~,\cr
(X_0)_2 = ~&~ 2\sqrt{-Q} \cos\left(\fracmm{\vq+2\p}{3}\right) +\fracmm{11f_2}{20f_3}~~,\cr
(X_0)_3 = ~&~ 2\sqrt{-Q} \cos\left(\fracmm{\vq+4\p}{3}\right) +\fracmm{11f_2}{20f_3}~~,
\cr }
\ee
in terms of the Cardano-Vi\`ete parameters
\be \lb{ccvp} 
\eqalign{
Q= ~&~ -\fracmm{11f_2}{2^2\cdot 5f_3} -\fracmm{7^2}{2^4\cdot5^2 f_3^2}\approx 
-\fracmm{11f_2}{20f_3}~~,\cr
\hat{R}= ~&~ -\fracmm{3\cdot 7\cdot 11f_2}{2^5\cdot 5^2 f^2_3} + \fracmm{3f_0}{2^2\cdot 5f_3}
+\fracmm{11^3f_2^3}{2^6\cdot 5^3f_3^3} \approx -\fracmm{1}{20f_3}\left( -\fracmm{21}{2}Q
+3f_0\right)~~\cr}
\ee
and the angle $\vq$ defined by  
\be \lb{cvq}
\cos\vq =\fracmm {\hat{R}}{\sqrt{-Q^3}}
\ee
The Cardano discriminant reads $D=\hat{R}^2+Q^3$. All three roots are real provided that $D<0$. 
It is known to be the case in the high-curvature regime (Sec.~10), and it is also the case 
when $f_0$ is extremely small. Under our requirements on the parameters the angle $\vq$ 
is very close to zero, so the relevant solutions $X_0<0$ are given by the 2nd and 3rd lines
 of eq.~(\ref{c3roots}), with  $X_0\approx f_0/10$. 

We thus demonstrated that it is possible to have a {\it positive} cosmological constant 
(at low spacetime curvature) in the particular $F(\car)$ supergravity without its coupling
 to super-matter, as described by the Ansatz (\ref{can}). The same Ansatz is applicable for 
describing a viable chaotic inflation in supergravity 
(at high spacetime curvature). A positive cosmological constant was achieved as the 
{\it non-linear} effect (with respect to the superspace curvature and spacetime curvature) in the 
narrow part of the parameter space (it is, therefore, highly constrained). It also implies 
the apparent violation of the Strong Energy Condition in our model.

Of course, describing the DE in the present Universe requires enormous fine-tuning of our 
parameters in the $F$-function. However, it is the common feature of all known approaches 
to the DE. Our analysis does not contribute to `explaining' the smallness of the 
cosmological constant. Yet another attempt for describing DE by an $F(R)$ supergravity model
with spontaneous breaking of supersymmetry was proposed in ref.~\cite{us13}.

\section{Nonminimal Scalar-Curvature Coupling in Gravity and Supergravity, and Higgs 
inflation}

One can pursue different strategies in a theoretical search for inflaton. For
instance, inflaton may be either a new exotic particle or something that we 
already know `just around the corner'. In this review we advocate the second 
``economical'' approach. Besides the Starobinsky inflation another ``economical''
approach is given by the so-called {\it Higgs} inflation \cite{f1,f2,f3}.

According to the cosmology textbooks, a Higgs particle of the Standard Model 
{\it cannot} serve as inflaton because the SM parameters are $\lambda\approx 1$, 
$m_{\rm H}\approx 10^2~GeV$, and $(\d T/T)\approx 1$, whereas inflation 
requires (see Sec.~4) $\lambda\approx 10^{-13}$, $m_{\rm inf}\approx 
10^{13}~GeV$, and $(\d T/T)\approx 10^{-5}$. Nevertheless, it is possible to 
reach the
required values when assuming that Higgs particle is {\it nonminimally} 
coupled to gravity  \cite{f1,f2,f3}. For instance, adding the nonminimal 
coupling of the Higgs field to the scalar spacetime curvature is natural in 
curved spacetime because it is required by renormalization \cite{bdev}. 

In this section we compare the inflationary scalar potential, derived by the 
use of the nonminimal coupling \cite{f1,f2,f3}, with the scalar potential that 
follows from the $(R+R^2)$ inflationary model (Sec.~2), and confirm that they
are {\it the same}. Then we also upgrade that equivalence to supergravity. In 
this section we set $M_{\rm Pl}=1$ too.

The original motivation of Refs.~\cite{f1,f2,f3} is based on the assumption 
that there is no new physics beyond the Standard Model up to the Planck scale. 
Then it is natural to search for the most economical mechanism of inflation 
by identifying inflaton with Higgs particle. We assume that there is the new physics 
beyond the Standard Model, and it is given by supersymmetry. Then it is quite natural 
to search for the most economical mechanism 
of inflation in the context of supergravity. Moreover, we do not have to identify 
our inflaton with a Higgs particle of the Minimal Supersymmetric Standard Model.
Let us begin with the 4D Lagrangian 
\be \lb{1bl}
\Lag_{\rm J} = \sqrt{-g_{\rm J}} \left[ -\ha (1+\x \phi_{\rm J}^2)R_{\rm J} + 
\ha g^{\m\n}_{\rm J}\pa_{\m}\phi_{\rm J}\pa_{\n}\phi_{\rm J} -V(\phi_{\rm J}) \right]
\ee
where we have introduced the real scalar field $\phi_{\rm J}(x)$, nonminimally
coupled to gravity (with the coupling constant $\x$) in Jordan frame, with the 
Higgs-like scalar potential  
\be \lb{1hpot}
V(\phi_{\rm J}) = \fracmm{\l}{4}(\phi_{\rm J}^2-v^2)^2 
\ee

The action (\ref{1bl}) can be rewritten to Einstein frame by redefining the 
metric via a Weyl transformation,
\be \lb{1me}
g^{\m\n} = \fracmm{g_{\rm J}^{\m\n}}{(1+\x\phi_{\rm J}^2)}
\ee
It gives rise to the standard Einstein-Hilbert term $(-\ha R)$ for gravity in
the Lagrangian. However, it also leads to a nonminimal (or noncanonical) 
kinetic term of the scalar field $\phi_{\rm J}$. To get the canonical kinetic 
term, a scalar field redefinition is needed, $\phi_{\rm J}\to \vf(\phi_{\rm J})$, 
subject  to the condition
\be \lb{1fre}
 \fracmm{d\vf}{d\phi_{\rm J}} = \fracmm{ 
\sqrt{1+\x(1+6\x)\phi_{\rm J}^2}}{1+\x\phi_{\rm J}^2}
\ee
As a result, the non-minimal theory (\ref{1bl}) is classically equivalent to
the standard (canonical) theory of the scalar field $\vf(x)$ minimally coupled 
to gravity,
\be \lb{1mina}
\Lag_{\rm E} = \sqrt{-g}\left\{ -\ha R + \ha g^{\m\n}\pa_{\m}\vf\pa_{\n}\vf
 -V(\vf) \right\}
\ee
with the scalar potential 
\be \lb{1hpote}
V(\vf) = \fracmm{V(\phi_{\rm J}(\vf))}{[1+\x\phi_{\rm J}^2(\vf)]^2}
\ee

Given a large positive $\x\gg 1$, in the small field limit one finds from
eq.~(\ref{1fre}) that $\f_{\rm J}\approx \vf$, whereas in the large $\vf$ 
limit one
gets
\be \lb{1fres}
 \vf \approx \sqrt{\fracmm{3}{2}} \log (1+\x\f_{\rm J}^2) 
\ee

Then eq.~(\ref{1hpote}) yields the scalar potential:\\
(i) in the {\it very small} field limit, $\vf<\sqrt{\fracmm{2}{3}}\x^{-1}$, as 
\be \lb{1vsp}
V_{\rm vs}(\vf) \approx   \fracmm{\l}{4}\vf^4
\ee
(ii) in the {\it small} field limit, $\sqrt{\fracmm{2}{3}}\x^{-1}<\vf\ll
\sqrt{\fracmm{3}{2}}$, as
\be \lb{1smp}
V_{\rm s}(\vf) \approx   \fracmm{\l}{6\x^2}\vf^2,
\ee
(iii) and in the {\it large} field limit, $
\vf\gg\sqrt{\fracmm{2}{3}}\x^{-1}$, as 
\be \lb{1lp}
V(\vf) \approx \fracmm{\l}{4\x^2}\left(
1-\exp\left[ -\sqrt{\fracmm{2}{3}} \vf\right] \right)^2
\ee
We have assumed here that $\x\gg 1$ and $v\x\ll 1$. 

Identifying inflaton with Higgs particle requires the parameter $v$ to be of 
the order of weak scale, and the coupling $\l$ to be the Higgs boson 
selfcoupling at the inflationary scale. The scalar potential (\ref{1lp}) is  
perfectly suitable to support a slow-roll inflation, while its consistency 
with the WMAP normalization condition (Sec.~4) for the observed CMB amplitude 
of density perturbations at the e-foldings number $N_e=55$  gives rise to the 
relation $\x/\sqrt{\l}\approx 5\cdot 10^4$ \cite{f1,f2,f3}.

The scalar potential (\ref{1smp}) corresponds to the post-inflationary 
matter-dominated epoch described by the oscillating inflaton field $\vf$ with 
the frequency
\be \lb{1freq}
\o = \sqrt{\fracmm{\l}{3}}\,\x^{-1} =M_{\rm inf}
\ee  

When gravity is extended to 4D, $N=1$ supergravity, any physical real scalar 
field should be complexified, becoming the leading complex scalar field 
component of a chiral (scalar) matter supermultiplet. In a curved 
superspace of $N=1$ supergravity, the chiral matter supermultiplet is described
 by a covariantly chiral superfield $\F$ obeying the constaraint  
$\Bar{\nabla}_{\dt{\a}}\F=0$.  The standard (generic and minimally coupled) 
matter-supergravity action is given by in superspace by eqs.~(\ref{sstand}) and
and (\ref{kaehler}), namely, 
\be \lb{1msg}
S_{\rm MSG}= -3\int d^4x d^4\q E^{-1}\exp \left[ 
-\fracmm{1}{3}K(\F,\Bar{\F})\right] +\left\{ 
\int d^4x d^2\q \ce W(\F) +{\rm H.c.} \right\}
\ee
in terms of the K\"ahler potential $K=-3\log (-\fracmm{1}{3}\O)$ and the 
superpotential $W$ of the chiral supermatter, and the full density $E$ and 
the chiral density $\ce$ of the superspace supergravity (Sec.~5).

The non-minimal matter-supergravity coupling in superspace reads
\be \lb{1nma}
S_{\rm NM}= \int d^4x d^2\q \ce X(\F)\car +{\rm H.c.}
\ee
in terms of the chiral function $X(\F)$ and the N=1 chiral scalar 
supercurvature superfield $\car$ obeying  $\Bar{\nabla}_{\dt{\a}}\car=0$.
In terms of the field components of the superfields the non-minimal action 
(\ref{1nma}) is given by 
\be \lb{1bnma}
\int d^4x d^2\q \ce X(\F)\car +{\rm H.c.} = -\fracmm{1}{6}\int d^4x \sqrt{-g}
X(\f_c)R + {\rm H.c.} +\ldots
\ee
 stand for the fermionic terms, and $\f_c=\left. \F\right|=
\f+i\c$ is the leading complex scalar field component of the superfield 
$\F$. Given $X(\F)=-\x\F^2$ with the real coupling constant $\x$, we find 
the bosonic contribution 
\be \lb{1sbnm}
S_{\rm NM,bos.}=  \fracmm{1}{6}\x\int d^4x \sqrt{-g}
\left( \f^2-\c^2\right) R 
\ee
It is worth noticing that the supersymmetrizable (bosonic) non-minimal 
coupling reads
$\left[ \f_c^2 +(\f_c^{\dg})^2\right]R$, not $(\f_c^{\dg}\f_c)R$.

Let us now introduce the manifestly supersymmetric nonminimal action 
(in Jordan frame) as
\be \lb{1nmact}
S = S_{\rm MSG} + S_{\rm NM} 
\ee

In curved superspace of $N=1$ supergravity the (Siegel's) chiral integration 
rule \be \lb{1sieg}
\int d^4x d^2\q \ce \Lag_{\rm ch}  = \int d^4x d^4\q E^{-1}
\fracmm{\Lag_{\rm ch}}{\car} 
\ee
applies to any chiral superfield Lagrangian $\Lag_{\rm ch}$ with
$\Bar{\nabla}_{\dt{\a}}\Lag_{\rm ch}=0$. It is, therefore, possible to 
rewrite eq.~(\ref{1nma}) to the equivalent form
\be \lb{1nmeq}
S_{\rm NM} = \int d^4x d^4\q E^{-1} \left[ X(\F) +\Bar{X}(\Bar{\F})\right]
\ee
We conclude that adding $S_{\rm NM}$ to $S_{\rm MSG}$ is equivalent to
the simple change of the $\O$-potential as ({\it cf}. ref.~\cite{jones})
\be \lb{1chan}
\O \to \O_{\rm NM} = \O +X(\F) + \bar{X}(\Bar{\F})
\ee
It amounts to the change of the K\"ahler potential as 
\be \lb{1kchan}
K_{\rm NM} = -3 \ln \left[ e^{-K/3} -\fracmm{X(\F)+\Bar{X}(\Bar{\F})}{3}
\right]
\ee
The scalar potential in the matter-coupled supergravity (\ref{1msg})
 is given by eq.~(\ref{cpot}),
\be \lb{cremv2}
V(\f,\bar{\f})=e^G \left[  
\left( \fracmm{\pa^2 G}{\pa\f\pa\bar{\f}}\right)^{-1}\fracmm{\pa G}{\pa\f}
\fracmm{\pa G}{\pa\bar{\f}} -3\right]
\ee
in terms of the K\"ahler-gauge-invariant function (\ref{spo}), ie.
\be \lb{1gfun}
G = K +\ln\abs{W}^2
\ee
Hence, in the nonminimal case (\ref{1nmact}) we have
\be \lb{1gfunnm}
G_{\rm NM} = K_{\rm NM} + \ln\abs{W}^2
\ee

Contrary to the bosonic case, one gets a nontrivial K\"ahler potential
$K_{\rm NM}$, ie. a {\it Non-Linear Sigma-Model} (NLSM) as the kinetic term of
$\f_c=\f+i\c$ (see ref.~\cite{nlsm} for more about the NLSM). Since the 
NLSM target space in general has a nonvanishing curvature, no field 
redefinition generically exist that could bring the kinetic term to the free 
(canonical) form with its K\"ahler potential $K_{\rm free}=\Bar{\F}\F$.    

Let's now consider the full action (\ref{1nmact}) under the slow-roll 
condition, ie. when the contribution of the kinetic term is negligible. Then 
eq.~(\ref{1nmact}) takes the truly chiral form
\be \lb{1slowra}
S_{\rm ch.}= \int d^4x d^2\q \ce\left[ X(\F)\car +W(\F)\right] +{\rm H.c.} 
\ee
When choosing $X$ as the independent chiral superfield, $S_{\rm ch.}$ can
be rewritten to the form
\be \lb{1slowra1}
S_{\rm ch.}= \int d^4x d^2\q \ce\left[ X\car -\cz (X)\right] 
+{\rm H.c.} 
\ee
where we have introduced the notation
\be \lb{1nota}
\cz(X) = - W(\F(X))
\ee

In its turn, the action (\ref{1slowra1}) is equivalent to the chiral $F(\car)$ 
supergravity action (\ref{action}), whose function $F$ is related to the 
function $\cz$ via Legendre transformation (Sec.~6)
\be \lb{1ltr}
 \cz= X\car -F~,\quad F'(\car)=X\quad {\rm and}\quad \cz'(X)=\car
\ee
It implies the equivalence between the reduced action (\ref{1slowra}) and the 
corresponding $F(\car)$ supergravity whose $F$-finction obeys 
eq.~(\ref{1ltr}).

Next, let us consider the special case of eq.~(\ref{1slowra}) 
when the superpotential 
is given by 
\be \lb{1wz}
W(\F) = \fracmm{1}{2}m\F^2 +\fracmm{1}{6}\tilde{\l} \F^3
\ee
with the real coupling constants $m>0$ and $\tilde{\l}>0$. The model (\ref{1wz}) 
is known as the {\it Wess-Zumino} (WZ) model in 4D, $N=1$ rigid supersymmetry. 
It has the most general {\it renormalizable} scalar superpotential in the 
absence of supergravity. In terms of the field components, it gives rise 
to the Higgs-like scalar potential.

For simplicity, let us take a cubic superpotential,
\be \lb{1wz3}
W_3(\F) = \fracmm{1}{6}\tilde{\l} \F^3
\ee
or just assume that this term dominates in the superpotential (\ref{1wz}), and
choose the $X(\F)$-function in eq.~(\ref{1slowra}) in the form
\be  \lb{1xcho}
 X(\F) = -\x \F^2
\ee
with a large positive coefficient $\x$, $\x>0$ and $\x\gg 1$, in accordance 
with eqs.~(\ref{1bnma}) and (\ref{1sbnm}).

Let us also simplify the $F$-function of eq.~(\ref{cub}) by keeping only the
most relevant cubic term, 
\be \lb{1ks3}
F_3(\car) = - \fracmm{1}{6}f_3\car^3
\ee

It is straightforward to calculate the $\cz$-function for the $F$-function
(\ref{1ks3}) by using eq.~(\ref{1ltr}). We find
\be \lb{1eks}
-X =\fracmm{1}{2}f_3\car^2 \quad {\rm and}\quad 
\cz'(X)=\sqrt{\fracmm{-2X}{f_3}}
\ee
Integrating the last equation with respect to $X$  yields
\be \lb{1cz}
\cz(X) = -\fracmm{2}{3}\sqrt{\fracmm{2}{f_3}}(-X)^{3/2}=
-\fracmm{2\sqrt{2}}{3}\fracmm{\x^{3/2}}{f^{1/2}_3}\F^3
\ee
where we have used eq.~(\ref{1xcho}). In accordance to eq.~(\ref{1nota}),
the $F(\car)$-supergravity $\cz$-potential (\ref{1cz}) implies 
the superpotential 
\be \lb{1sp}
W_{\rm KS}(\F)= \fracmm{2\sqrt{2}}{3}\fracmm{\x^{3/2}}{f^{1/2}_3}\F^3 
\ee
It coincides with the superpotential (\ref{1wz3}) of the WZ-model, 
provided that we identify the couplings as
\be \lb{1cid}
 f_3 = \fracmm{32\x^3}{\tilde{\l}^2}
\ee

We conclude that the original nonminimally coupled matter-supergravity
theory (\ref{1nmact}) in the slow-roll approximation with the superpotential
(\ref{1wz3}) is classically equivalent to the $F(\car)$-supergravity theory
with the $F$-function given by eq.~(\ref{1ks3}) when the couplings
are related by eq.~(\ref{1cid}).  

The inflaton mass $M$ in the supersymmertic case, according to 
eqs.~(\ref{f3m}) and (\ref{1cid}), is given by
\be \lb{1minfs}
M_{\rm inf}^2 = \fracmm{15\tilde{\l}^2}{32\x^3}
\ee
Since the value of $M_{\rm inf}$ is fixed by the WMAP normalization (Sec.~4),
the value of $\x$ in the supersymmetric case is $\x_{\rm susy}^3
=(45/32)\x_{\rm bos}^2$, 
or $\x_{\rm susy}\approx 10^3$, ie. is {\it lower} than that in the bosonic case. 
We have asssumed here that $\tilde{\l}\approx {\cal O}(1)$.

The established equivalence begs for a fundamental reason. In the high-curvature 
(inflationary) regime the $R^2$-term dominates over the $R$-term in the Starobinsky
action (\ref{qua}), while the coupling constant in front of the $R^2$-action is
dimensionless (Sect.~2). The Higgs inflation is based on the Lagrangian (\ref{1bl}) 
with the relevant scalar potential $V_4=\frac{1}{4}\l\f_J^4$ (the parameter $v$ is 
irrelevant for inflation), whose coupling constants $\x$ and $\l$ are also dimensionless. 
Therefore, both relevant actions are {\it  scale invariant}. Inflation breaks that 
symmetry spontaneously.

The supersymmetric case is similar: the nonminimal action (\ref{1slowra}) with the 
$X$-function (\ref{1xcho}) and the superpotential (\ref{1wz3}) also have only dimensionless
 coupling constants $\x$ and $\tilde{\l}$, while the same it true for the 
$F(\car)$-supergravity 
action with the $F$-function (\ref{1ks3}), whose coupling constant $f_3$ is dimensionless 
too. Therefore, those actions are both {\it scale invariant}, while inflation spontaneously
 breaks that invariance.

A spontaneous breaking of the scale invariance {\it necessarily} leads to a Goldstone 
particle (or dilaton) associated with spontaneously broken dilatations. So, perhaps, 
Starobinsky scalaron (inflaton) may be identified with the Goldstone dilaton!

The basic field theory model, describing both inflation {\it and} the 
subsequent reheating, reads (see eg., eq.~(6) in ref.~\cite{KLS,KLS2})
\be \lb{1basm}
\eqalign{
 L/\sqrt{-g}  = &~  \ha \pa_{\m}\f\pa^{\m}\f - V(\f) +\ha \pa_{\m}\c\pa^{\m}\c
-\ha m^2_{\c}\c^2 +\ha \tilde{\x} R\c^2 +\bar{\j}(i\g^{\m}\pa_{\m}-m_{\j})\j
 \cr
 &~ -\ha g^2\f^2\c^2 - h(\bar{\j}\j)\f \cr}
\ee
with the inflaton scalar field $\f$ interacting with another scalar field 
$\c$ and a spinor field $\j$. The nonminimal supergravity theory (\ref{1nmact})
 with the Wess-Zumino superpotential (\ref{1wz}) can be considered as the $N=1$
 locally supersymetric extension of the basic model (\ref{1basm}) after rescaling  
$\f_c$ to $(1/\sqrt{2})\f_c$ and identifying $\tilde{\x}=-\fracm{1}{3}\x$ because of
  eq.~(\ref{1sbnm}). Therefore, {\it pre-heating} (ie. the nonperturbative enhancement 
of particle production due to a broad parametric resonance \cite{KLS,KLS2}) is a generic 
feature of our supergravity models.

The axion $\c$ and fermion $\j$ are both requred by supersymmetry, being 
in the same chiral supermultiplet with the inflaton $\f$. The scalar interactions are 
\be \lb{1sint}
V_{\rm int}(\f,\c) =m\hat{\l}\f(\f^2 +\c^2) +\fracmm{\hat{\l}^2}{4}(\f^2+\c^2)^2 
\ee  
whereas the Yukawa couplings are given by
\be \lb{1yu}
 L_{\rm Yu}= \ha \hat{\l} \f (\bar{\j}\j) +\ha \hat{\l} \c (\bar{\j}i\g_5\j)  
\ee
Supersymmetry implies the unification of couplings since $h=-\ha \hat{\l}$ and
$g^2=\hat{\l}^2$ in terms of the single coupling constant $\hat{\l}$. If 
supersymmetry is unbroken, the masses of $\f$, $\c$ and $\j$ are all the same. However, 
inflation already breaks supersymmetry, so the spontaneously broken 
supersymmetry is appropriate here.

To conclude, inflationary slow-roll dynamics in Einstein gravity theory with 
a nonminimal scalar-curvature coupling can be equivalent to that in the certain
$f(R)$ gravity theory. We just extended that correspondence to $N=1$ supergravity. 
The nonminimal coupling in supergravity can be rewritten in terms of the standard 
(`minimal') $N=1$ matter-coupled supergravity, by using their manifestly 
supersymmetric formulations in curved superspace. The equivalence relation 
between the supergravity theory with the nonminimal scalar-curvature coupling 
and the $F(\car)$ supergravity during slow-roll inflation is, therefore, established.

The equivalence is expected to hold even after inflation, during {\it initial}
reheating with harmonic oscillations. In the bosonic case the equivalence holds
until the inflaton field value is higher than $\o\approx M_{\rm Pl}/\x_{\rm bos}\approx
10^{-5} M_{\rm Pl}$. In the superymmetric case we have the same bound 
$\o\approx M_{\rm Pl}/\x_{\rm susy}^{3/2}\approx 10^{-5}M_{\rm Pl}$.  

The Higgs inflation and the renormalization group can be used to compute the mass of a
Higgs particle in the Standard Model by descending from the inflationary scale to the
electro-weak scale. For example, in the {\it two-loop} approximation one finds \cite{msh}  
\be \lb{mshhm}
 129~GeV < m_H < 194~GeV 
\ee
with the theoretical uncertainty of about $\pm 2~GeV$. It is to be compared to the observed
Higgs mass at the Linear Hadron Collider (LHC) in 2012 \cite{LHC} 
\be \lb{dsearch}
{\rm LHC~(ATLAS):} ~~m_H = 126\pm 0.8~GeV 
\ee
Therefore, the bosonic Higgs inflation is (almost) ruled out. It is worth noticing that in a 
{\it supersymmetric} extension of the SM (like the MSSM and NMSSM) there are more particles, 
when compared to the bosonic SM. Hence, the SUSY renormalization group trajectory is going to 
be steeper, while the theoretical SUSY bounds on the Higgs mass at the electro-weak scale are 
going to be {\it lower} than those in eq.~(\ref{mshhm}).

\section{Quantum Particle Production (Reheating)}

Reheating is a transfer of energy from inflaton to ordinary particles and fields. It took place
after inflation but before BBN and hot radiation domination. All particles in the universe are
believed to be created via the inflaton decay soon after the inflation. The leading channel of 
the particle production is preheating (due to the nonperturbative parametric resonance). The 
resonance eventially disappeared when the inflaton field became sufficiently small, and it was 
replaced by perturbative decay. The reheating provided {\it initial conditions} for the 
BBN that began after the first 3 minutes (such as the initial temperature of baryogenesis, 
DM abundance, relic monopoles and gravitinos, etc.). Both preheating and reheating are highly 
{\it model-dependent}. In our approach we advocate the (super)gravitational preheating and 
reheating due to the universal coupling of (super)inflaton to conformally noninvariant fields
(see also Ref.~\cite{yy}).

The classical solution (neglecting particle production) near the minimum 
of the inflaton scalar potential reads
\be \lb{zmin} 
 a(t)\approx a_0\left(\fracmm{t}{t_0}\right)^{2/3}\qquad {\rm and} \qquad
\vf(t)\approx\left( \fracmm{M_{\rm Pl}}{3M_{\rm inf}} \right)
\fracmm{\cos\[M_{\rm inf}(t-t_0)]}{t-t_0} \ee
A {\it time-dependent} classical spacetime background leads to {\it 
quantum} production of particles with masses $m<\o =M_{\rm inf}$ \cite{bdev}.
Actually, the amplitude of $\vf$-oscillations decreases much faster
\cite{KLS,KLS2}, namely, as
\be \lb{klsa}
\exp[-\ha (3H+\G)t]
\ee 
via inflaton decay and the universe expansion, as the solution to the
inflaton equation 
\be \lb{ppue}
\ddt{\vf}+3H\dt{\vf}+(m^2+\P)\vf=0
\ee 

Here $\P$ denotes the polarization operator that effectively describes particle
production. Unitarity (optical theorem) requires Im$(\P)=m\G$. The assumption 
$m\gg H$ was used here \cite{KLS,KLS2}.

The Starobinsky model (in Jordan frame) with the action 
\be \lb{starjo}
 S = \int d^4x \sqrt{-g\low{J}}f_S(R\low{J}) + S_{\rm SM}(g^{\m\n}\low{J},
\j)
\ee
after the conformal transformation to Einstein frame takes the form
\be \lb{starei}
S = S_{\rm scalar-tensor~gravity}(g\low{\m\n},\vf) 
+ S_{\rm SM}(g^{\m\n}e^{-\s\vf},\j)
\ee 
so that the inflaton $\vf$ couples to all {\it non-conformal} terms and fields
$\j$, due to the universality of gravitational interaction. Therefore, the 
Starobinsky inflation automatically leads to the {\it universal} mechanism of particle 
production.

For example, let us consider the scalar and spinor fields in the Jordan frame, with the action
\be\lb{2ex}
 S = -\fracmm{M_{\rm Pl}^2}{2}\int d^4x \sqrt{-g}f(R) + S_ {\rm matter}\ee
where the matter is represented by the the standard Klein-Gordon and Dirac actions, 
$S_ {\rm matter}=S_{\rm KG} + S_{\rm Dirac}$, with the minimal coupling to gravity,
\be \lb{kgac}
S_{\rm KG} = \int d^4x \sqrt{-g} \left( \ha g^{\m\n}\pa_{\m}\vf\pa_{\n}\vf-
\ha m_{\vf}\vf^2\right) \ee
and
\be
\lb{dac} 
S_{\rm Dirac }= \int d^4x \sqrt{-g} \left( i\bar{\j}\Hat{D}\j- m_{\j}\bar{\j}\j\right)
\ee

After rewriting the full action to the Einstein frame by a Weyl transformation of the metric 
with the scalaron field $\f$,
\be\lb{wtra}
 g_{\m\n}\to \O\tilde{g}_{\m\n},\quad \O(\f) = exp\left[ \sqrt{2/3}~\f/M_{\rm Pl}\right]~,\ee
and rescaling the matter scalar and spinor fields to get their canonical kinetic terms as
\be \lb{resca}
 \vf\to \tilde{\vf}=\O^{-1/2}\vf, \quad \j\to\tilde{\j}=\O^{-3/4}\j~,\ee
where we have used $\Hat{D}=\g^{\m}D_{\m}=e^{\m}_a\g^a D_{\m}$ and $\Hat{D}\to\O^{-1/2}\Hat{D}$, 
one finds 
\be\lb{resact}
S = S_{\rm quintessence}[\f,\tilde{g}] +  S_{\rm KG}[\tilde{\vf},\tilde{g},{ \f}] 
+ S_{\rm Dirac}[\tilde{\j},\tilde{g},{ \f}] \ee
where  \cite{gpan}
\be \lb{kgresac}
\eqalign{  S_{\rm KG}[\tilde{\vf},\tilde{g},\f] = & \int d^4x\sqrt{-\tilde{g}}
\left(   \ha \tilde{g}^{\m\n}\pa_{\m}\tilde{\vf}\pa_{\n}\tilde{\vf} - \ha \O^{-1}
 m_{\vf}^2\tilde{\vf}^2  \right. \cr 
& \left.  + \fracmm{\tilde{\vf}^2}{12M^2_{\rm Pl}} \tilde{g}^{\m\n}\pa_{\m}\f\pa_{\n}\f
+ \fracmm{\tilde{\vf}}{\sqrt{6}M_{\rm Pl}} \tilde{g}^{\m\n}\pa_{\m}\tilde{\vf}\pa_{\n}\f\right)
\cr}\ee
and  \cite{gpan}
\be\lb{rescdir}  S_{\rm Dirac}[\tilde{\j},\tilde{g},{ \f}]
= \int d^4x \sqrt{-\tilde{g}} \left( i\Bar{\tilde{\j}}~\Hat{\tilde{D}}~\tilde{\j}
- \O^{-1/2} m_{\j}\Bar{\tilde{\j}}\tilde{\j}\right)\ee

As is clear from those equations, all interactions with inflaton are suppressed by the factors 
of $M_{\rm Pl}$. Hence, they are only relevant for the large $\f$-values comparable to 
$M_{\rm Pl}$. Those interactions (and decay rates) are sensitive to the mass and spin of the 
created particles. The conformal couplings to not contribute to the inflaton decay \cite{gpan}. 
In particular, the domimant contribution to the inflaton  decay rate in the {\it scalar} channel 
comes from the 3rd term in the action $S_{\rm KG}[\tilde{\vf},\tilde{g},{ \f}]$ of 
eq.~(\ref{kgresac}). The only contribution to the inflaton decay rate in the {\it spinor} 
channel comes from the mass term in $S_{\rm Dirac}[\tilde{\vf},\tilde{g},{ \f}]$ of
eq.~(\ref{rescdir}).

The perturbative decay rates of the inflaton into a pair of scalars $(s)$ or into a pair of 
spin-1/2 fermions $(f)$ are given by \cite{star1,star2,vilen}
\be \lb{drates}
 \G_{\vf\to ss} =\fracmm{M^3_{\rm inf}}{192\p M^2_{\rm Pl}} \qquad
{\rm and} \qquad 
\G_{\vf\to ff} =\fracmm{M_{\rm inf}M^2_f}{48\p M^2_{\rm Pl}}~~, 
\ee
respectively. The perturbative decay rate of the inflaton into a pair of 
{\it gravitino} is \cite{eht6}
\be \lb{grates}
 \G_{\vf\to 2\j_{3/2}} =\fracmm{\abs{G_{,\vf}}^2}{288\p}
\fracmm{M^5_{\rm inf}}{m^2_{3/2}M^2_{\rm Pl}}
\ee
Being proportional to $M^5_{\rm inf}$, eq.~(\ref{grates}) may lead to the 
cosmologically disastrous gravitino overproduction in early universe \cite{ety3}, 
if the gravitino mass is relatively small (under $100~GeV$). In the case of the 
large-field inflation, when the inflaton expectation value has the order of the Planck 
mass (it includes the Starobinsky inflation), one can demonstrate that 
eq.~(\ref{grates}) reduces to the scalar decay rate (\ref{drates}) proportional
to $M^3_{\rm inf}$  \cite{ety3}. 

The energy transfers by the time $t_{\rm reh} \geq \left(\sum_{s,f}\G_{s,f}\right)^{-1}$. 
The {\it reheating temperature} is given by \cite{chi2,gpan}
\be \lb{rehte}
 T_{\rm reh} \propto \sqrt{ \fracmm{M_{\rm Pl}\G}{(\# d.o.f.)^{1/2}} }
\approx 10^9 ~{\rm GeV} \ee
that gives the maximal temperature of the primordial plasma. 

In the context of supergravity coupled to the supersymmetric matter (like MSSM)
gravitino can be either LSP (= the lightest sparticle) or NLSP (= not LSP). In the
LSP case (that usually happens with gauge mediation of supersymmetry breaking and
$m_{3/2}\ll 10^2~$ GeV)  gravitino is stable due to the R-parity conservation. If 
gravitino is NLSP, then it is unstable (it usually happens with gravity- or anomaly-
mediation of supersymmetry breaking, and $m_{3/2}\gg 10^2~$ GeV). Unstable gravitino
can decay into LSP. See ref.~\cite{media} for a review of mediation of supersymmetry
breaking from the hidden sector to the visible sector.

Stable gravitino may be the dominant part of {\it Cold Dark Matter} (CDM) \cite{buchm1}.
There exist severe Big Bang Nucleosynthesis (BBN)~\footnote{See ref.~\cite{bbn} for 
a review of BBN.} constraints on the overproduction of 
${}^3He$ in that case, which give rise to the upper bound on the reheating temperature 
of thermally produced gravitinos, $T_{\rm reh}<10^{5\div 6}$ GeV \cite{jap8,kkmy}. The 
reheating temperature (\ref{rehte}) is unrelated to that bound because it corresponds to
the much earlier time in the history of the Universe.

When gravitino is NLSP of mass $m_{3/2}\gg 10^2$ GeV, the BBN constraints are drastically
relaxed because the gravitino lifetime becomes much shorter than the BBN time 
(about 1 sec) \cite{jap8,kkmy}. In that case the most likely CDM candidate is MSSM 
neutralino, while the reheating temperature may be as high as $10^{10}$ GeV \cite{kkmy}. 

An overproduction of gravitinos from inflaton decay and scattering processes should be 
avoided, in order to prevent overclosure of the universe. The cosmological constraints on 
gravitino abundances were formulated in ref.~\cite{ety3}. Those constraints are very 
model-dependent.

The rate of decay changes with time, along with the decreasing amplitude of inflaton 
oscillations. It stops when the decay rate becomes smaller than the production 
rate. The reheating transfers most of energy to radiation, and leads to a radiation-dominated 
universe with $a \propto t^{1/2}$.

In the matter-coupled $F(\car)$ supergravity with the action
\be \lb{mcsug}
 S = \left[ \int d^4x d^2\q \, \ce F(\car) + {\rm H.c.} \right] 
+S\low{\rm SSM}(E,\J) \ee
after the super-Weyl transformation, $\ce\to \ce e^{3\F}$, we get
\be \lb{mcsug2}
 S = S_{\rm scalar-tensor~supergravity}(E,\F) + 
S\low{\rm SSM}(e^{\F + \Bar{\F}}E,\J) \ee
so that the superscalaron $\F$ is {\it universally} coupled to the SSM matter 
superfields $\J$.

\section{Conclusion}

${}\quad\bullet\quad$ A manifestly $4D$, $N=1$ supersymmetric {\it extension} 
of $f(R)$ gravity exist, it is {\it chiral} and is parametrized by a 
holomorphic function. An $F(\car)$ supergravity is classically {\it equivalent}
 to the scalar-tensor theory of a chiral scalar superfield (with certain
 K\"ahler potential and superpotential) miminally coupled to the $N=1$ 
Poincar\'e supergravity in four spacetime dimensions (with nontrivial $G$ 
and $K$), ie. the $N=1$ supersymmetric quintessence.

The {\it classical} equivalence between the $F(\car)$ supergravity and the 
quintessence N=1 supergravity has the same physical contents as the classical 
equivalence between $f(R)$ gravity and scalar-tensor gravity, ie. 
{\it the same} inflaton scalar potential and, therefore, the same inflationary 
dynamics. However, the physical nature of inflaton in the $f(R)$ gravity and 
the scalar-tensor gravity is very different. In the $f(R)$ gravity the inflaton
field is the spin-$0$ part of metric, whereas in a {\it generic} scalar-tensor 
gravity inflaton is a matter particle. The inflaton interactions with other matter 
fields are, in general, different in both theories. It gives rise to the 
different inflaton decay rates and different reheating, ie. implies different 
physics in the post-inflationary universe.

Similar remarks apply to the equivalence between Higgs inflation and
Starobinsky inflation (Sec.~12). The equivalence does not have to be valid 
{\it after} inflation. For example, the {\it reheating} temperature 
$T_{\rm reh}$ after the Higgs inflation is about $10^{13}~GeV$ \cite{f1,f2,f3},
 whereas after the Starobinsky inflation one has $T_{\rm reh}\approx 10^9~GeV$ 
\cite{star2}, or the one order more in the supersymmetric case.

${}\quad\bullet\quad$ It is expected that the classical equivalence is {\it broken} 
in quantum theory because the classical equivalence is achieved via a non-trivial 
field redefinition (Secs. 3 and 6). When doing that field redefinition in the quantum 
path integrals defining those quantum theories (unter their unitarity bounds), 
it gives rise to a non-trivial Jacobian that already implies the {\it quantum 
inequivalence}, even before taking into account renormalization.~\footnote{See
ref.~\cite{kmy} for the first steps of quantization with a higher time derivative.}

In the supergravity case, there is one more reason for the quantum
inequivalence between the $F(\car)$ supergravity and the clasically equivalent 
quintessence supergravity. The K\"ahler potential of the scalar superfield is 
described by a {\it full} superspace integral and, therefore, it receives quantum 
corrections that can easily spoil classical solutions describing an accelerating 
universe (those corrections are not under control). It was the reason for introduction
 of flat directions in the K\"ahler potential and popular realizations of inflation in 
supergravity by the use of a chiral scalar superpotential along the flat directions 
\cite{jap1,jap2,yamag}. The $F(\car)$ supergravity action is truly chiral, so that the
 function $F(\car)$ is already protected against the quantum perturbative corrections 
given by full superspace integrals. It is the important part of physical motivation 
for $F(\car)$ supergravity. It also explains why we consider $F(\car)$ supergravity 
as the viable and self-consistent alternative to the K\"ahler flat directions for 
realizing slow-roll inflation in supergravity. Of course, one can also consider both ways
together \cite{yy}.

${}\quad\bullet\quad$ The Starobinsky model of chaotic inflation can be 
embedded into $F(\car)$ supergravity. It is the viable realization of chaotic inflation 
in supergravity, and gives a simple solution to the $\eta$-problem.

${}\quad\bullet\quad$ A simple extension of our inflationary model (Sec.~16) has 
a {\it positive} cosmological constant in the regime of low spacetime curvature 
(Secs.~10 and 11). It is non-trivial because the standard supergravity with usual matter 
can only have a negative or vanishing cosmological constant \cite{gibb}. It happens 
because the usual (known) matter does not violate the {\it Strong Energy Condition} 
(SEC) \cite{hawe}. A violation of the SEC is required for an accelerating universe, and 
 is easily achieved in $f(R)$ gravity due to the fact that the quintessence field in 
$f(R)$ gravity is part of metric (ie. the unusual matter). Similarly, the quintessence
 scalar superfield in $F(\car)$ supergravity is part of super-vielbein, and also gives 
rise to a violation of the SEC.

In the $F(\car)$ supergravity model we considered (Secs.~10 and 11), the effective 
$f(R)$ gravity function in the high-curvature regime is essentially given by the 
Starobinsky function 
$(-\fracmm{M^2_{\rm Pl}}{2} R+\fracmm{M^2_{\rm Pl}}{12M^2_{\rm inf}} R^2)$. 
In the low-curvature regime it is essentially given by the Einstein-Hilbert function 
with a cosmological constant, $(-\fracmm{M^2_{\rm PL}}{2} R-\Lambda)$. Therefore, our
  model has a cosmological solution describing an inflationary universe of the 
quasi-dS type with $H(t)= (M^2_{\rm inf}/6)(t_{end}-t)$ at early times $t< t_{end}$, 
and an accelerating universe of the dS-type with $H=\Lambda$ at late times. 

The dynamical chiral superfield in $F(\car)$ supergravity may be identified 
with the dilaton-axion chiral superfield in quantum 4D Superstring Theory, 
when demanding the $SL(2,{\bf Z})$ symmetry of the effective action. As is
well known, String Theory supports the higher-derivative gravity. In particular,
the required $R^2A(R)$ terms may appear in the (nonperturbative) gravitational 
effective action after superstring compactification (with fluxes, after moduli
stabilization). The problem is how to get the anomalously {\it large} coefficient in 
front of the $\car^3$-term in the effective $F(\car)$ supergravity theory that would be 
consistent with the superstring dynamics.

Supersymmetry in $F(R)$ supergravity is broken by inflation but is restored near
the minimum of the scalar potential. The anomaly- or gravitationally-mediated supersymmetry 
breaking (in the hidden sector) may serve as the important element for the new particle 
phenomenology (beyond the Standard Model) based on the matter-coupled  $F(R)$ supergravity 
theory.

\section{Outlook: $CP$-violation, Baryonic Asymmetry, \newline
Lepto- and Baryo-genesis, Non-Gaussianity, Tests} 

The observed part of our Universe is highly $C-$ and $CP-$asymmetric
(no antimatter). Inflation naturally implies a {\it dynamical} origin of 
the baryonic matter predominance due to a nonconserved baryon number. The main 
conditions for the dynamical generation of the cosmological baryon asymmetry 
in early universe were formulated in Ref.~\cite{saha}:
\begin{enumerate}
\item nonconservation of baryons ({\it cf.} SUSY, GUT, EW theory),
\item  $C-$ and $CP-$symmetry breaking (confirmed experimentally),
\item deviation from thermal equilibrium in initial hot universe.
\end{enumerate}

The first condition is clearly necessary. And (in theory) there is no 
fundamental reason for the baryon number conservation. The baryon asymmetry
should have originated from spontaneous breaking of the $CP$ symmetry that
was present at very early times, so is the need for the second condition. Then 
the third condition is required by the $CPT$ symmetry, when the $CP$-violation 
is compensated by the $T$-violation, so it has to be no thermal equilibrium. 

There exist many scenarios of baryogenesis (see ref.~\cite{dineku} 
for a review), all designed to explain the observed asymmetry (BBN, CMB): 
\be \lb{baras}
 \b= \fracmm{n_B -n_{\Bar{B}}}{n_{\g}} = (6.0 \pm 0.5) \cdot 10^{-10} \ee
Here $n_B$ stands for the concentration of baryons, $n_{\Bar{B}}$ for the concentration
of anti-baryons, and $n_{\g}$ for the concentration of photons.

Perhaps, the most popular scenario is the nonthermal 
{\it baryo-through-lepto-genesis} \cite{buchm,fy}, ie. a creation of lepton asymmetry
 by L-nonconserving decays of a heavy  ($m\approx 10^{10}$ GeV) Majorana 
neutrino, and a subsequent transformation of the lepton asymmetry into the 
baryonic asymmetry by $CP$-symmetric, B-nonconserving and (B-L)-conserving 
electro-weak processes. 

The thermal leptogenesis requires the high reheating temperature,
$T_{\rm reh}\geq 10^9$ GeV \cite{buchm2}, which is consistent with eq.~(\ref{rehte}). 
 
The matter-coupled $F(\car)$ supergravity theory may contribute towards 
the {\it origin} and the {\it mechanism} of $CP$-violation and baryon 
asymmetry, because
 
${}\quad\bullet\quad$ {\it complex} coefficients of $F(\car)$-function and the
complex nature of the $F(\car)$ supergravity are the simple {\it source} of 
explicit $CP$-violation and complex Yukawa couplings; 

${}\quad\bullet\quad$ the nonthermal leptogenesis is possible via decay of 
heavy sterile neutrinos (FY-mechanism) {\it universally produced} by 
(super)scalaron decays, or via neutrino oscillations in  early 
universe \cite{akh3};

${}\quad\bullet\quad$ the existence of the natural Cold Dark Matter candidates 
({\it gravitino}, {\it axion}, {\it inflatino} or, maybe, {\it inflaton} itself!) 
in $F(\car)$ supergravity;

${}\quad\bullet\quad$ as is well known, {\it non-Gaussianity} is a measure of inflaton 
{\it interactions} described  by its 3-point functions and higher -- cf. eq.~(\ref{tempr}). 
The non-Gaussianity parameter $f_{\rm NL}$ is defined in terms of the (gauge-invariant) 
comoving curvature perturbations as 
\be \lb{nonga}
 \hat{\car} = \hat{\car}_{\rm gr} + \fracmm{3}{5}f_{\rm NL}
\hat{\car}^2_{\rm gr}
\ee
The non-Gaussianity was not observed yet, though it is expected. As regards 
the single-field inflationary models, they predict  \cite{malda} 
\be \lb{mald} 
f_{\rm NL} = \fracmm{5}{12}(1-n_s) \approx 0.02
\ee
The Starobinsky inflation is known to yield highly Gaussian fluctuations, which is
consistent with the recent Planck (2013) data \cite{planck2}.

Finally, we would like to comment on possible testing of $f(R)$ gravity and 
$F(\car)$ supergravity in Solar system and ground-based experiments.

As regards the large-scale structure of the present universe, the scalaron (ie. 
the dynamical spin-0 part of metric) may be responsible for its acceleration or 
Dark Energy. However, since scalaron is universally coupled to all matter
with gravitational strength, it may lead to an unacceptable violation of the 
equivalence principle. To avoid it, the scalaron should be ``screened off''
on the Solar system scales, because of the strong observational constraints from 
experimental tests of the equivalence principle \cite{eptest1,eptest2}. Moreover, 
it should not give rise to a large violation of the equivalence principle in 
ground-based (on Earth) laboratories, because of the tight constraints on the 
fifth fundamental force in Nature \cite{fifthf}. 

A natural solution to both problems is provided by {\it Chameleon Cosmology}
\cite{cham1,cham2}, because the effective scalaron mass is dependent upon a local 
matter density $\r$ (see also refs.~\cite{dampol1,dampol2}). The effective scalar potential 
of the scalaron (Chameleon) field takes the form 
\be \lb{chameleon}
V _{\rm eff} (\vf) = V(\vf) + \r \exp\left( \b\vf/M_{\rm Pl}\right)  
\ee
where the parameter $\b$ is of the order $1$. The exponential factor here arises
due to the universal coupling of the scalaron to the  matter of density $\r$ ---
see eq.~(\ref{starei}). As a result, the effective Chameleon mass is about $\r$, 
so that in a sufficiently dense environment one can evade the observational constraints 
on the equivalnce principle and the fifth force.

\section*{Acknowledgements}

I am grateful to my collaborators: S.J. Gates Jr., A.A. Starobinsky, S. Tsujikawa,  T. Terada
and N. Yunes and my students: S. Kaneda and N. Watanabe, for their efforts. 
I wish to thank the Theory Division of CERN in Geneva, the Institute of Theoretical Physics
in Hannover, the Heisenberg Institute of Physics in Munich, the Einstein Institute of 
Gravitational Physics in Potsdam, the DESY Theory Group in Hamburg and the Center of 
Theoretical Physics in Marseille for their kind hospitality extended to me during preparation of 
this paper. I also thank Hiroyuki Abe, Joseph Buchbinder, Gia Dvali, Antonio De Felice, Richard
Grimm, Koichi Hamaguchi, Artur Hebecker, Simeon Hellerman, Norihiro Iizuka, Satoshi Iso, Renata
Kallosh, Sergey Kuzenko, Kazunori Kohri, Kei-ichi Maeda, Mihail Shaposhnikov, Liam McAllister, 
Stefan Theisen, Roland Triay, Alexander Westphal, Bernard de Wit, Masahide Yamaguchi, Tsutomu 
Yanagida and Norimi Yokozaki for useful discussions.

This work was supported in part by the TMU Graduate School of Science and Engineering in Tokyo,
the World Premier International Research Center Initiative of MEXT in Japan, the German Academic
Exchange Service (DAAD), the Max-Planck Institute of Physics in Munich, the Max-Planck Institute of
Gravitational Physics in Potsdam, and the SFB 676 of the University of Hamburg and DESY in Germany.

\section*{Appendix A: Scalar Potential in $F(\car)$ Supergravity}
\vglue.2in

The exact K\"ahler potential and the superpotential in a {\it generic} $F(\car)$ Supergravity 
described by the action (\ref{action}) with the fixed chiral compensator are found in Sec.~13 --- 
see eqs.~(\ref{kahl}) and (\ref{frsup2}), respectively. It is, therefore, straightforward to 
compute the full scalar potential by the use of eqs.~(\ref{cremv}) or (\ref{cpot}), with all
gravitational corrections included.

Equation (\ref{cpot}) in the units with $M_{\rm Pl}=1$ for the chiral superpotential $Z(\cy)$ 
reads
\be \label{cremm}
\cv = e^G \left[ \fracmm{\pa G}{\pa\cy} \left( \fracmm{\pa^2G}{\pa\cy\pa\Bar{\cy}}\right)^{-1}
 \fracmm{\pa G}{\pa\Bar{\cy}} -3 \right]_{\cy=Y} \tag{A.1}
\ee
in terms of the K\"ahler {\it gauge-invariant} function  
$G(\cy,\Bar{\cy})= K(\cy,\Bar{\cy}) + \ln\abs{Z(\cy)}^2$. Substituting the K\"ahler potential of
eq.~(\ref{kahl}) yields the scalar potential in the form
\be \label{fullpo}
\cv = \fracmm{1}{3(\cy+\Bar{\cy})}\left\{ \abs{ \fracmm{\pa Z}{\pa\cy}}^2 
-\fracmm{3}{\cy+\Bar{\cy}}\left( \Bar{Z} \fracmm{\pa Z}{\pa\cy} 
+Z \fracmm{\pa\Bar{Z}}{\pa\bar{\cy}}\right)\right\} \tag{A.2}
\ee
In the case of the cubic {\it Ansatz} (\ref{cub}) for the $F(\car)$ function, we find
$$  \lb{zfunc}
Z(\cy)  = \fracmm{\sqrt{14}}{60} \fracmm{M^2}{m} \left\{ 
3(\cy -3/4) - 2(\cy - 3/4) \sqrt{ 1 + \fracmm{80m^2}{21M^2}(\cy - 3/4)}\right.$$ 
$$\left. + \fracmm{21M^2}{40m^2} \left( 1-  \sqrt{ 1 + \fracmm{80m^2}{21M^2}(\cy - 3/4)}\right)
\right\} \eqno(A.3)
$$
When substituting it into eq.~(\ref{fullpo}) one arrives at a very lengthy formula for the
scalar potential with many square roots, which is not very illuminating. It is therefore,
no surprise that such scalar potentials were not investigated earlier. 

We would like to emphasize that in our approach there is no need to use the scalar potential because 
it is much easier to work in the original picture with the $F$-function. See recent 
refs.~\cite{afour,atwo} for the different (non-minimal) approaches to the Starobinsky inflation in 
supergravity by the use of two or three chiral superfields.

\end{document}